\documentclass[11pt,a4paper]{article}
\usepackage{jheppub}                       
\usepackage{abb}                           
\usepackage{amsbsy,bm,mathrsfs}
\usepackage{array,booktabs,longtable,multirow}
\usepackage{xcolor}
\usepackage{tikz}\usetikzlibrary{calc,arrows.meta}
\hypersetup{colorlinks=true,linkcolor=blue!60!black,citecolor=blue!60!black,urlcolor=blue!60!black}
\usepackage[capitalise]{cleveref}
\crefname{equation}{Eq.}{Eqs.}\Crefname{equation}{Eq.}{Eqs.}
\crefname{section}{Sec.}{Secs.}\Crefname{section}{Sec.}{Secs.}
\crefname{subsection}{Sec.}{Secs.}\Crefname{subsection}{Sec.}{Secs.}
\crefname{figure}{Fig.}{Figs.}\Crefname{figure}{Fig.}{Figs.}
\crefname{table}{Table}{Tables}\Crefname{table}{Table}{Tables}
\crefname{appendix}{App.}{Apps.}\Crefname{appendix}{App.}{Apps.}
\newcommand{\pic}{\pi_c}
\newcommand{\la}{\lambda}
\newcommand{\n}{\nu_{\rm eff}}
\newcommand{\nbare}{\nu}
\newcommand{\muf}{\mu_{\rm eff}}
\newcommand{\meff}{m_{\rm eff}}
\newcommand{\mpl}{M_{\rm Pl}}
\newcommand{\fpi}{f_\pi}

\newcommand{\dgz}{\delta g^{00}}
\newcommand{\Dz}{\Delta_\zeta}
\newcommand{\Dzo}{\Delta_{\zeta,0}}

\newcommand{\gthree}{\mu}
\newcommand{\Wn}{\mathsf{W}}
\newcommand{\Vv}{\mathsf{V}}
\newcommand{\Pp}{\mathsf{P}}
\newcommand{\bW}{\mathbb{W}}
\newcommand{\bV}{\mathbb{V}}
\newcommand{\bP}{\mathbb{P}}
\newcommand{\Ma}{M}
\newcommand{\Wcal}{\mathcal{W}}
\newcommand{\Ccal}{\mathcal{C}}
\newcommand{\Acal}{\mathcal{A}}
\newcommand{\Scal}{\mathcal{S}}
\newcommand{\Ncal}{\mathcal{N}}
\newcommand{\Hcal}{\mathcal{H}}

\newcommand{\Tcal}{\mathcal{T}}

\newcommand{\Qcal}{\mathcal{Q}}
\newcommand{\Ycal}{\mathcal{Y}}

\newcommand{\Dw}{D}          
\newcommand{\Gsk}{G}         
\newcommand{\Kb}{K}          
\newcommand{\dd}{\mathrm{d}}
\newcommand{\ii}{i}
\newcommand{\Real}{\mathrm{Re}}
\newcommand{\Imag}{\mathrm{Im}}
\newcommand{\hyp}{{}_2F_1}
\newcommand{\sech}{\operatorname{sech}}
\newcommand{\csch}{\operatorname{csch}}

\title{Exact bispectra in strongly mixed multifield inflation}
\author{Lucas Pinol}
\affiliation{Laboratoire de Physique de l'\'Ecole Normale Sup\'erieure, ENS, CNRS, Universit\'e PSL,\\
Sorbonne Universit\'e, Universit\'e Paris Cit\'e, F-75005 Paris, France}
\emailAdd{lucas.pinol@phys.ens.fr}

\abstract{Using the effective field theory of inflationary fluctuations extended by a massive isocurvature
scalar, at unit sound speed, we compute the primordial bispectra generated by every cubic
interaction of dimension four or less in the unitary gauge, with the curvature--isocurvature
quadratic mixing $\rho\,\dot\pic\sigma$ resummed to all orders in the dimensionless mixing strength
$\la=\rho/H$. Placing the mixing in the free Hamiltonian and writing the exact linear solutions as
a single integral over one weight, every scale-invariant tree-level bispectrum reduces to a single
Schwinger-parameter integral over independent leg kernels, and what perturbation theory calls
single, double and triple exchange are contact diagrams of the resummed theory. At strong mixing
we unveil genuinely new, exact multifield bispectrum shapes with no correlation to the standard
shapes used in data analyses. The squeezed limit carries a cosmological collider signal with a
universal frequency set by the dressed mass parameter $\muf=\sqrt{\la^2+m^2/H^2-9/4}$ at every
exchange order. At weak mixing the clock coefficients are obtained in closed form and reduce, in
amplitude and phase, to the known perturbative single- and double-exchange results, exactly; the
triple-exchange leading coefficient is obtained in closed form at every mass, in terms of
hypergeometric functions, for the first time. At
strong mixing the collider amplitude scales as $e^{\pi(\la-2\muf)/2}$ over the whole $(\la,\muf)$
plane. This encompasses all previously studied regimes as particular cases: the perturbative
Boltzmann factor $e^{-\pi\muf}$ at weak mixing, half of it along every line of fixed bare mass, with
$\la\simeq\muf$, and $e^{\pi\la/2}$ at fixed effective mass. A first confrontation of the six exact
shapes with the Planck PR4 binned bispectrum shows a preference for strong mixing, where the
cosmological collider oscillations invade mildly squeezed configurations, although at no more than
$2.8\sigma$ before any look-elsewhere correction. The shapes of all interactions over the whole plane, whose computation was until
now the expensive step of any such analysis, are released with the code that produced them, so that
a template at any mixing and mass can be read off in milliseconds.}

\begin{document}
\maketitle

\section{Introduction}
\label{sec:intro}

Inflation explains the origin of the primordial fluctuations with an economy that nothing has
challenged in forty years: a phase of quasi-de Sitter expansion, during which the quantum
fluctuations of a clock field are stretched to cosmological scales and frozen into the adiabatic,
nearly scale-invariant, nearly Gaussian curvature perturbation that the cosmic microwave background
and the large-scale structure record~\cite{Planck:2018jri}. What inflation does not tell us is
what it is made of. The energy scale is unknown within many orders of magnitude, the field content
is unknown, and the two observables that are measured with precision, the amplitude and the tilt
of the scalar spectrum, are reproduced by any slowly rolling potential with two free parameters.

There are two reasons to expect that more than one field was active. From the top down, every
embedding of inflation in a theory of quantum gravity comes with many scalars: moduli and axions
in string compactifications, the partners imposed by supersymmetry, the fields that set the
couplings of the low-energy theory~\cite{Baumann:2014nda}; that exactly one of them was light
enough to matter and all the others decoupled is an assumption, not a prediction. From the bottom
up, the effective field theory of inflationary fluctuations~\cite{Cheung:2007st} describes a single
Goldstone boson of broken time translations, and its most general extension is the coupling of
that Goldstone to additional degrees of freedom with masses of order the Hubble scale, which
neither decouple nor dominate~\cite{Senatore:2010wk,Noumi:2012vr,Pinol:2024arz}. Heavy fields
coupled to the inflaton leave imprints even when they are too heavy to be excited on the
background trajectory, through the turns of that trajectory and the sound speed they
induce~\cite{Achucarro:2010jv,Achucarro:2010da,Cespedes:2012hu,Garcia-Saenz:2018ifx}.

The difficulty is that the power spectrum alone cannot disentangle these possibilities. A second
field that shares the expansion modifies the amplitude and the tilt in ways that a change of the
single-field potential mimics, and the tensor-to-scalar ratio, still undetected, constrains the
energy scale rather than the field content. The hope has therefore turned to primordial
non-Gaussianity~\cite{Maldacena:2002vr,Chen:2010xka,Meerburg:2019qqi,Achucarro:2022qrl}, whose
shape in momentum space encodes the interactions of the fields during inflation, and which carries
signatures that no single-field model can produce: a local bispectrum violating the single-field
consistency relation~\cite{Creminelli:2004yq}, isocurvature components, and, most specifically,
the oscillatory features imprinted by the propagation of massive fields.

The cosmological collider programme is the systematic study of the last of these. Its origin is
quasi-single-field inflation~\cite{Chen:2009we,Chen:2009zp,Chen:2012ge}, in which a field of mass
$m\sim H$ couples to the inflaton through a quadratic mixing and imprints, in the squeezed limit
of the bispectrum, an oscillation in the logarithm of the momentum ratio whose frequency is set by
the mass, and whose dependence on the angle between the soft and the hard momenta, a Legendre
polynomial, is set by the spin~\cite{Baumann:2011nk,Noumi:2012vr}. The physical
picture was made sharp in Ref.~\cite{Arkani-Hamed:2015bza}: the oscillation is the interference
between the two branches of a pair of massive particles produced during inflation, and its
Boltzmann suppression $e^{-\pi m/H}$ is the price of that production. The programme then grew in
two directions. On the theoretical side, the bootstrap reconstructed the correlators from
symmetries and singularities~\cite{Lee:2016vti,Arkani-Hamed:2018kmz,Baumann:2019oyu,Pimentel:2022fsc},
the Schwinger--Keldysh diagrammatics were systematised~\cite{Chen:2017ryl}, closed forms were
obtained for the exchange diagrams~\cite{Qin:2022fbv,Qin:2023ejc}, then for trees with any number
of massive lines~\cite{Liu:2024str}, with the non-analytic collider signal isolated by
dispersion relations~\cite{Liu:2024xyi} and by a cutting rule for in--in
correlators~\cite{Ema:2024hkj}, and the analysis was extended to several flavours of
massive fields~\cite{Pinol:2021aun,Aoki:2024jha}, to the exchange of two massive fields
from a single vertex, each converted into the curvature perturbation by one mixing
insertion~\cite{Aoki:2024uyi}, to reduced sound speeds and boost-breaking
exchanges~\cite{Jazayeri:2022kjy,Jazayeri:2023xcj,Qin:2025xct}, to light and spinning
fields~\cite{Bordin:2018pca}, to the scalaron of $R^2$ inflation~\cite{Wu:2024wti} and to
transient instabilities~\cite{McCulloch:2024hiz,Aoki:2026qea}. On the phenomenological side, the spectrum
of the Standard Model and of its extensions during inflation was mapped onto collider
signals~\cite{Chen:2016uwp,Kumar:2017ecc}, and the ways to make them large, chemical potentials
among them, were explored~\cite{Wang:2019gbi,Bodas:2020yho,Sou:2021juh}.

With a few exceptions, all of these works treat the quadratic mixing between the Goldstone and the
massive field perturbatively, for the technical reason that the mixed linear system had no known
analytical solution: the strongly mixed power spectrum was obtained numerically, from the exact
linear equations~\cite{An:2017hlx} or by resumming the local part of the massive
propagator~\cite{Iyer:2017qzw}, the passage to an effective single-field description at
strong coupling was examined in two-field models with curved field spaces~\cite{Cremonini:2010ua},
and the shift of the spectral index and of its running by a strongly mixed field was recently
used to reconcile plateau models with the ACT data~\cite{Aoki:2025ywt}, all of them at the
level of the two-point function. In that regime the expected signal is small: its size is
controlled by the mixing, and the mixing is what the expansion assumes to be small. The
perturbative templates have been searched for in the Planck data and in galaxy
surveys~\cite{Sefusatti:2012ye,Planck:2019kim,Sohn:2024xzd,Suman:2025vuf,Cabass:2024wob,Kumar:2026ogn},
without a detection and, more importantly, with little prospect of one: as argued in
Refs.~\cite{Pinol:2023oux,Philcox:2026njr}, the regime in which a collider signal could be seen is
the one in which the mixing is strong. In that regime the field that mixes with the Goldstone is
converted into curvature efficiently, the power spectrum is amplified, and the shapes themselves
change.

Until recently the strongly mixed regime had a single automated tool, the cosmological flow of
Refs.~\cite{Werth:2023pfl,Pinol:2023oux}, which evolves the correlators of any
multifield theory as a system of ordinary differential equations in time, mixing included, with no
expansion in any coupling. Its only implementation is numerical, the code
CosmoFlow~\cite{Werth:2024aui}, and the price is computational: each shape is an integration
over the whole history of every triangle, the squeezed limit is the most expensive corner, since
the modes must be followed over many decades of scale, and a scan over masses and mixings at the
resolution a data analysis needs is a substantial undertaking, feasible, as
Ref.~\cite{Philcox:2026tjj} showed by confronting tens of thousands of such templates with Planck,
but not something one can repeat at will or push into the deeply squeezed limit. The present
framework makes that cost negligible, and we release the six exact shapes over the whole plane of
mixings and masses, with the code and an interactive explorer, in a public
repository.\footnote{A short Python notebook with two modules, using only \texttt{numpy}, \texttt{scipy} and
\texttt{mpmath}: \url{https://github.com/lucaspinolCNRS/exact-collider}~\cite{Pinol:2026code}. The
same repository holds the raw tables of this paper and an explorer that evaluates the six shapes in
the browser, \url{https://lucaspinolcnrs.github.io/exact-collider/}.}

The opening came from the linear theory. Ref.~\cite{Huenupi:2026abj} solved the coupled system of the
Goldstone and the massive field exactly, at any mixing strength, in terms of confluent
hypergeometric functions. Ref.~\cite{Pinol:2026xnl} then showed that these exact solutions make the
bispectrum itself computable, exactly and to all orders in the mixing, as a single quadrature over
closed-form kernels, for the cubic self-interaction of the Goldstone, with closed forms in its limits,
and announced the present work. Two other papers appeared the
same day: Ref.~\cite{Huenupi:2026aqc}, by the authors of the exact solution, which applied it to bispectra
with the massive field at the vertex, and Ref.~\cite{Wang:2026lff}, which followed the same route; a
month later Ref.~\cite{Belrhali:2026uxn} obtained strongly mixed collider bispectra as convergent
series from a Laplace-space representation of the mode functions.

This paper formalises the construction of Ref.~\cite{Pinol:2026xnl} into a framework, and applies it
to every cubic interaction of the theory. The framework is general: the mixing is placed in the
free Hamiltonian, the exact linear solutions are written as one integral over a single weight,
and each mode of the theory becomes a superposition of plane waves with dressed frequencies. The
consequence is that the correlator of any set of fields, at any order in the interactions,
reduces to Schwinger-parameter integrals over independent leg kernels, one per external leg and
one per internal line, with the mixing entering only through those kernels. At tree level, for
the bispectrum, there is a single vertex and three legs, and what perturbation theory describes
as single, double or triple exchange of the massive field is, in the resummed theory, a contact
diagram with one, two or three legs of the massive field attached to a single vertex. The reduction is
the same for every interaction; only the vertex factor and the kernels change, and the
framework can be reused for the trispectrum, for loops, for other field contents, and for other
resummations.

We obtain the following results. The complete tree-level bispectrum of the effective theory with
one massive scalar, at unit sound speed and to dimension four in the unitary gauge, is given by
five master formulas, one per cubic operator, each a single one-dimensional integral over
closed-form kernels, exact in the mixing strength $\la=\rho/H$, from which closed forms follow
wherever a limit is taken, at weak mixing, at large mass and at strong mixing, and which one
quadrature evaluates everywhere else. Throughout, the two parameters of the theory are $\la$
and the effective mass parameter $\muf=\sqrt{\la^2+m^2/H^2-9/4}$, the mass that the massive field
acquires once the mixing is included, and we treat them as independent coordinates of a plane.
At strong mixing these are genuinely new shapes: their
correlation with the equilateral template vanishes along lines in the plane of mixing and mass,
where they are also seen to be neither orthogonal nor local, and the shapes of the different
interactions decorrelate from one another where the mixing is moderately strong, before
converging onto a common universal shape when it is very strong. The squeezed limit of every shape carries a
cosmological collider oscillation with one and the same frequency, $\muf$, at every exchange order
and at any mixing strength. At
weak mixing we obtain the coefficient of that oscillation, its amplitude and its phase, in closed
form for each interaction, and we recover the perturbative results of the literature exactly, in
amplitude and phase and at every mass, as particular cases of one formula: the single-exchange
signals of Refs.~\cite{Arkani-Hamed:2015bza,Qin:2023ejc,Pinol:2021aun} and the double-exchange
signal of Ref.~\cite{Aoki:2024uyi}; for triple exchange, for which no perturbative computation
exists, we obtain the leading coefficient of the signal in closed form at every mass, as a finite
combination of $\Gamma$ functions and ordinary hypergeometric functions, to our knowledge the
first. At strong mixing we treat the mixing strength and the
dressed mass as independent coordinates, with no hierarchy assumed between them, and we find that
the amplitude of the collider oscillation is governed over the whole plane by one exponential,
$e^{\pi(\la-2\muf)/2}$. This one exponent contains every regime studied before
as a particular case: the perturbative Boltzmann factor at weak mixing, half of it along any line
of fixed bare mass, which is the trajectory of a given model as its mixing grows and along which
the signal peaks at a mixing of order one before falling, and an exponential growth at fixed
dressed mass. Every analytic statement is checked numerically with two independent engines, and
every closed form is anchored, where an anchor exists, on the perturbative literature. Finally, a
first confrontation of the six exact shapes with the Planck PR4 binned bispectrum gives upper
bounds on the five cubic couplings, and shows a mild preference for strong mixing, where the
collider oscillations reach into mildly squeezed configurations.

The paper is organised as follows. \Cref{sec:eft} sets up the effective theory, its operator
basis and its regime of validity. \Cref{sec:linear} derives the exact linear solutions, the
resummation weight, the leg kernels and their soft, weak- and strong-mixing behaviours.
\Cref{sec:rules} gives the diagrammatic rules of the resummed theory and \cref{sec:bispectra} the
master formula of each cubic interaction, its implementation and the new shapes. \Cref{sec:squeezed}
is devoted to the squeezed limit: the weak-mixing closed forms, which recover the perturbative
literature, then the strong-mixing collider signal over the $(\la,\muf)$ plane. \Cref{sec:data}
confronts the six exact shapes with the Planck PR4 binned bispectrum, and \cref{sec:conclusion}
concludes. Every special function identity the paper uses is collected, with a line on what it
does, in \cref{app:formulae}; the remaining appendices collect the analytic continuation of the
weight at $u\to0$ (\cref{app:continuation}), the closed form of the weak-mixing $\sigma$ leg
(\cref{app:V2}), the double-exchange moment (\cref{app:double}), the triple-exchange moment in
closed form (\cref{app:triple}), the recovery of the perturbative
literature (\cref{app:dictionaries}), the numerical verification (\cref{app:numerics}) and a
statement on the use of AI assistance in this project (\cref{app:ai}).

\section{The effective theory and its regime of validity}
\label{sec:eft}

\subsection{Set-up}
\label{sec:eft:setup}

We work in the unitary-gauge effective field theory (EFT) of inflationary fluctuations
\cite{Cheung:2007st,Senatore:2010wk,Noumi:2012vr,Pinol:2024arz} extended by one additional scalar
$\sigma$, in the decoupling limit, with signature $(-,+,+,+)$ and $\dgz\equiv g^{00}+1$. Throughout
we assume exact de Sitter expansion ($H$ constant, slow-roll corrections neglected), constant Wilson
coefficients (hence exact scale invariance), and unit sound speed for both the curvature and the
isocurvature fluctuations. The Goldstone boson $\pi$ is introduced by the time diffeomorphism
$t\to t+\pi$, under which
\begin{equation}
  \dgz \;\longrightarrow\; -2\dot\pi + (\partial_\mu\pi)^2 ,
  \qquad (\partial_\mu\pi)^2 \equiv g^{\mu\nu}\partial_\mu\pi\partial_\nu\pi
  = -\dot\pi^2 + \frac{(\partial_i\pi)^2}{a^2} ,
\label{eq:eft:dg00}
\end{equation}
and $\zeta=-H\pi$ at linear order. The symmetry-breaking scale is $\fpi^4\equiv2\epsilon H^2\mpl^2$,
the canonically normalised Goldstone is $\pic=\fpi^2\pi$, and the undressed and dressed amplitudes
of the curvature power spectrum are related by
\begin{equation}
  \Dzo^2 = \frac{H^2}{8\pi^2\epsilon \mpl^2} = \frac{H^4}{4\pi^2\fpi^4},
  \qquad
  \Dz^2 = R\,\Dzo^2 ,
\label{eq:eft:Dz0}
\end{equation}
where $R(\la,\n)$, computed in \cref{sec:linear}, is the amplification of the power spectrum by
the mixing. Only $\Dz$ is observed, $\Dz^2=A_s=2.1\times10^{-9}$ \cite{Planck:2018vyg}, and we use
\cref{eq:eft:Dz0} repeatedly to trade $H/\fpi$ for it.

The action contains the mixing operators of dimension four or less in the unitary gauge that
involve no derivative of $\sigma$, together with the unique self-interaction of $\sigma$ of
dimension less than four:
\begin{align}
S \;=\; \int\! d^4x\,\sqrt{-g}\;\Big\{\;
  &\tfrac12 \mpl^2 R + \mpl^2\dot H g^{00} - \mpl^2(3H^2+\dot H)
   + \tfrac{1}{2!}M_2^4\,(\dgz)^2 + \tfrac{1}{3!}M_3^4\,(\dgz)^3
   \nonumber\\[2pt]
  &-\tfrac12(\partial_\mu\sigma)^2 - \tfrac12 m^2\sigma^2 - \gthree\,\sigma^3
   \nonumber\\[2pt]
  &+\;\tilde M^3\,\dgz\,\sigma
   \;+\;\tilde M_2^2\,\dgz\,\sigma^2
   \;+\;\tilde M_3^3\,(\dgz)^2\,\sigma \;\Big\} .
\label{eq:eft:action}
\end{align}
The first line is the single-field EFT of inflation, the second the $\sigma$ sector and the third
the mixing sector. Ref.~\cite{Pinol:2026xnl} retains $\tilde M^3\dgz\sigma$ alone; the operators
$\tilde M_2^2\dgz\sigma^2$, $\tilde M_3^3(\dgz)^2\sigma$ and $\gthree\sigma^3$ extend its basis, each
with an independent Wilson coefficient and its own strong-coupling scale. Three remarks fix the
logic of what follows.

First, each of the three new operators is at least cubic in the fluctuations, because $\dgz$ has
no background value: $\dgz\sigma^2\supset-2\dot\pi\sigma^2$, $(\dgz)^2\sigma\supset 4\dot\pi^2\sigma$,
and $\sigma^3$ is manifestly cubic. No shift of $m^2$ or of $\rho$, and no $\sigma$ tadpole, is
generated at tree level, so the free theory, and with it every mode function, the resummation
weight, the boundary values and the power spectrum of \cref{sec:linear}, is exactly that of
Ref.~\cite{Pinol:2026xnl}. Second, the exchange order $n_\sigma=0$ is the single-field operator
$(\dgz)^3$ of the first line, $\dgz\sigma$ generates $n_\sigma=1$, $\dgz\sigma^2$ is the unique
source of $n_\sigma=2$ at dimension four or less, and $\sigma^3$ the unique source of
$n_\sigma=3$; the operator $(\dgz)^2\sigma$ adds nothing new in order but frees one coefficient at
$n_\sigma=1$. Third, at dimension four the only self-interaction of $\sigma$ is $\sigma^3$, since
$\dot\sigma\sigma^2=\tfrac13\partial_t(\sigma^3)$ is redundant up to the measure. Because no retained
cubic operator contains $\dot\sigma$, the momentum $p_\sigma$ receives no non-linear correction,
which is what makes the Legendre transform of \cref{sec:eft:hamiltonian} close simply.

\paragraph{Symbols.} Throughout, $\gthree$ denotes the $\sigma^3$ coupling, of mass dimension one,
and the collider frequency of the resummed theory is always written $\muf$, with its subscript,
$\n=\ii\muf$ being the effective index of \cref{eq:eft:nueff}; the two never appear without their
distinguishing marks, with one declared exception: in the weak-mixing parts of the paper
(\cref{sec:lin:weak,sec:sq:weak} and \cref{app:V2,app:double,app:dictionaries}), where the bare and
the effective mass coincide at the order considered, the bare index and the bare mass parameter are
written $\nbare=\ii\mu$, $\mu\equiv\sqrt{m^2/H^2-9/4}$. The reader consulting
Ref.~\cite{Huenupi:2026abj} should note that its $\mu$ is the effective mass, our $\meff$
(\cref{sec:lin:system}).

\subsection{Quadratic and cubic Lagrangians}
\label{sec:eft:lagrangians}

Following Ref.~\cite{Pinol:2026xnl} we set $M_2=0$ throughout, so that the Goldstone propagates at the
speed of light, $c_s=1$. This is more than a simplification of the interaction basis: it makes the
Goldstone and the massive field share a single light cone, and the collapse of the two dressed
carriers of each channel onto a single plane wave (\cref{sec:lin:omega}) rests on both fields
propagating at the same speed. A non-zero $M_2$ would require the linear theory to be redone, not
supplemented. Expanding \cref{eq:eft:action} to quadratic order in the decoupling limit and
normalising canonically, one then finds
\begin{equation}
  \frac{\mathcal{L}^{(2)}}{a^3}
  = \frac{1}{2}\Big[\dot\pic^{\,2} - \frac{(\partial_i\pic)^2}{a^2}\Big]
  + \rho\,\dot\pic\,\sigma
  + \frac{1}{2}\Big[\dot\sigma^2 - \frac{(\partial_i\sigma)^2}{a^2} - m^2\sigma^2\Big],
\qquad
  \rho = -\frac{2\tilde M^3}{\fpi^2},\quad
  \la \equiv \frac{\rho}{H} ,
\label{eq:eft:L2}
\end{equation}
with $\pic=\fpi^2\pi$ the canonically normalised Goldstone, $\sigma$ canonically normalised
already, and the sign of the mixing term the convention of Refs.~\cite{Pinol:2026xnl,Werth:2024aui}.

At cubic order, with $c_s=1$, the Lagrangian reads
\begin{equation}
\frac{\mathcal{L}^{(3)}}{a^3}
= \underbrace{-\lambda_2\,\dot\pic^{\,3}}_{n_\sigma=0}
\;\underbrace{-\;\frac{1}{2\Lambda_2}\,\dot\pic^{\,2}\sigma\;-\;\frac{1}{2\Lambda_1}\,
   \frac{(\partial_i\pic)^2}{a^2}\,\sigma}_{n_\sigma=1}
\;\underbrace{-\;\frac{\alpha}{2}\,\dot\pic\,\sigma^2}_{n_\sigma=2}
\;\underbrace{-\;\gthree\,\sigma^3}_{n_\sigma=3}\,,
\label{eq:eft:L3}
\end{equation}
with
\begin{equation}
  \lambda_2 = \frac{4M_3^4}{3\fpi^6},\qquad
  \frac{1}{\Lambda_1} = \frac{\rho}{\fpi^2} = \frac{H\la}{\fpi^2},\qquad
  \frac{1}{\Lambda_2} = -\frac{8\tilde M_3^3}{\fpi^4}-\frac{\rho}{\fpi^2},\qquad
  \alpha = \frac{4\tilde M_2^2}{\fpi^2}.
\label{eq:eft:couplings}
\end{equation}
This follows the notation of Ref.~\cite{Pinol:2023oux}, at $c_s=1$. The four exchange orders are carried by operators of dimension six, five, four and
three respectively, matching $[\lambda_2]=-2$, $[\Lambda_{1,2}]=1$, $[\alpha]=0$ and $[\gthree]=1$.

Exactly one cubic coupling is fixed by the non-linearly realised boosts. The operator
$\tilde M^3\dgz\sigma$ produces the boost-invariant combination $(\partial_\mu\pi)^2\sigma$, which
contributes $-\rho/\fpi^2$ to the coefficient of $\dot\pic^{\,2}\sigma$ and $+\rho/\fpi^2$ to that
of $(\partial_i\pic)^2\sigma/a^2$. Since $(\dgz)^2\sigma$ contributes to $\dot\pic^{\,2}\sigma$ only,
the gradient coupling stays locked to the mixing, $\Lambda_1^{-1}=\rho/\fpi^2$, whereas the
velocity coupling $\Lambda_2$ is freed by $\tilde M_3^3$. The two lock together,
$\Lambda_2^{-1}=-\Lambda_1^{-1}$, only at the Lorentz-invariant point $\tilde M_3^3=0$. Retaining
$\tilde M_3^3$ therefore promotes single exchange from a one-shape to a two-shape channel. Neither
$\alpha$ nor $\gthree$ is constrained by the non-linearly realised symmetry.
\Cref{tab:eft:dictionary} collects the couplings.

\begin{table}[t]
\centering
\begin{tabular}{l|l|c|l}
\hline
coupling & vertex in $\mathcal{L}^{(3)}/a^3$ & $n_\sigma$ & status \\
\hline
$\rho$ & $+\rho\,\dot\pic\sigma$ \ (quadratic) & --- & resummed; $\la\equiv\rho/H$ \\
$\lambda_2$ & $-\lambda_2\dot\pic^{\,3}$ & $0$ & free \\
$1/\Lambda_2$ & $-\frac{1}{2\Lambda_2}\dot\pic^{\,2}\sigma$ & $1$ & free, through $\tilde M_3^3$ \\
$1/\Lambda_1$ & $-\frac{1}{2\Lambda_1}(\partial_i\pic)^2\sigma/a^2$ & $1$ & fixed: $\Lambda_1^{-1}=\rho/\fpi^2$ \\
$\alpha$ & $-\frac{\alpha}{2}\dot\pic\sigma^2$ & $2$ & free \\
$\gthree$ & $-\gthree\,\sigma^3$ & $3$ & free \\
\hline
\end{tabular}
\caption{The cubic couplings at $c_s=1$, in the notation of Ref.~\cite{Pinol:2023oux}.}
\label{tab:eft:dictionary}
\end{table}

\subsection{Hamiltonian and the resummed free theory}
\label{sec:eft:hamiltonian}

The conjugate momenta follow from $\mathcal{L}^{(2)}+\mathcal{L}^{(3)}$, every non-linear piece
included,
\begin{equation}
  p_\pi = a^3\Big[\dot\pic + \rho\sigma - 3\lambda_2\dot\pic^{\,2}
          - \frac{1}{\Lambda_2}\dot\pic\,\sigma - \frac{\alpha}{2}\,\sigma^2\Big],
  \qquad
  p_\sigma = a^3\dot\sigma .
\label{eq:eft:momenta}
\end{equation}
Inverting them perturbatively and performing the Legendre transform, one finds that the $\alpha$
terms cancel identically between $p_\pi\dot\pic$ and $\mathcal{L}$, that the $\lambda_2$ and
$\Lambda_2^{-1}$ terms flip sign, and that the entire quadratic theory, mixing included, can be
placed in the free Hamiltonian:
\begin{align}
  H_{\rm free} &= \int\! d^3x\;\Big[\frac{p_\pi^2}{2a^3} + \frac{p_\sigma^2}{2a^3}
     + \frac{a}{2}(\partial_i\pic)^2 + \frac{a}{2}(\partial_i\sigma)^2
     + \frac{a^3}{2}\meff^2\sigma^2 - \rho\,\sigma\,p_\pi\Big],
  \qquad \meff^2 = m^2+\rho^2,
  \label{eq:eft:Hfree}\\[4pt]
  H_{\rm int} &= \int\! d^3x\; a^3\Big[\lambda_2\,u_\pi^3
     + \frac{u_\pi^2\sigma}{2\Lambda_2}
     + \frac{1}{2\Lambda_1}\frac{(\partial_i\pic)^2}{a^2}\sigma
     + \frac{\alpha}{2}\,u_\pi\,\sigma^2
     + \gthree\sigma^3\Big],
  \qquad u_\pi \equiv \frac{p_\pi}{a^3}-\rho\sigma ,
  \label{eq:eft:Hint}
\end{align}
$u_\pi$ being the velocity operator of $\pic$. This split is exact: no expansion in $\rho$ has been
made, and $H_{\rm free}+H_{\rm int}$ is the Hamiltonian of \cref{eq:eft:action} to cubic order.

Perturbation theory is then set up in the usual way, with the difference that the free evolution
already contains the mixing. Interaction-picture fields $\pic^{I}$, $\sigma^{I}$ and their momenta
are defined by evolving with $H_{\rm free}$ alone, so that they obey the linear equations of
\cref{sec:linear}; in particular the free Hamilton equation for $\pic^I$ reads
$\dot\pic^{\,I}=p_\pi^I/a^3-\rho\sigma^I$, i.e.\ $u_\pi^I=\dot\pic^{\,I}$. Substituting the
interaction-picture fields into \cref{eq:eft:Hint} therefore gives back, term by term, minus the
cubic Lagrangian of \cref{eq:eft:L3} evaluated on those same fields,
\begin{equation}
  H_{\rm int}^{I}\big[\pic^I,\sigma^I\big] \;=\; -\,\mathcal{L}^{(3)}\big[\pic^I,\sigma^I\big] ,
\label{eq:eft:HintI}
\end{equation}
which is the object that enters the in--in formula. From here on we drop the superscript $I$: every field in
\cref{sec:rules,sec:bispectra} is an interaction-picture field, and every $\dot\pic$ is the
free-theory velocity.

\Cref{eq:eft:Hfree} is that of Ref.~\cite{Pinol:2026xnl}, so that
\begin{equation}
  \n \equiv \sqrt{\frac94-\frac{\meff^2}{H^2}}=\sqrt{\nbare^2-\la^2} ,\qquad
  \n=\ii\muf ,\qquad \muf = \sqrt{\la^2+\frac{m^2}{H^2}-\frac94}
\label{eq:eft:nueff}
\end{equation}
for an effectively heavy field, with $\nbare=\sqrt{9/4-m^2/H^2}$ the bare index, are unmodified.
The bare mass never appears in isolation: every
linear mode function depends on $(m,\rho)$ only through $(\n,\la)$. For $m^2\ge0$ the field is
necessarily effectively heavy once $\la\ge3/2$; conversely, at fixed $\muf$, $\la^2>9/4+\muf^2$
means a tachyonic bare mass.

\Cref{eq:eft:HintI} holds exactly at cubic order, with no extra vertex, for the structural reason
that the first deviation of the Legendre transform from the naive sign flip is quadratic in
$\partial\mathcal{L}^{(3)}/\partial\dot\phi$ and therefore quartic. That quartic non-Lagrangian
vertex is
\begin{equation}
  \frac{\Delta\mathcal{H}_4}{a^3}
  = \frac12\Big(3\lambda_2 \dot\pic^{\,2} + \frac{\dot\pic\,\sigma}{\Lambda_2}
    + \frac{\alpha}{2}\sigma^2\Big)^{\!2},
\label{eq:eft:H4}
\end{equation}
but they are needed at trispectrum order only, together with terms from $\mathcal{L}^{(4)}$ not
written here.

\subsection{Regime of validity}
\label{sec:eft:validity}

What is needed is a ceiling on each of the five dimensionless couplings
\begin{equation}
  c_n \;\in\; \Big\{\;\lambda_2H^2,\quad H/\Lambda_2,\quad H/\Lambda_1,\quad
  \alpha,\quad \gthree/H \;\Big\} ,
\label{eq:eft:cn}
\end{equation}
one of which, $H/\Lambda_1=2\pi\Dzo\la$, is not free. Every bound below is dimensional analysis, an
order-of-magnitude statement in which no factor of $4\pi$ is tracked.

At weak mixing, $\la\lesssim1$, both fields are relativistic and the criterion is that the cubic
Lagrangian not overtake the quadratic one at the frequency at which the correlator is generated,
$\mathcal{L}_3/\mathcal{L}_2\lesssim1$ at $\omega\sim H$; with every field scaling as $H$ this reads
$|c_n|\lesssim1$ for each entry of \cref{eq:eft:cn}:
\begin{equation}
  \lambda_2H^2\lesssim1,\qquad H/\Lambda_2\lesssim1,\qquad
  H/\Lambda_1\lesssim1\ \Leftrightarrow\ \la\lesssim\frac{1}{2\pi\Dzo},\qquad
  |\alpha|\lesssim1,\qquad \gthree/H\lesssim1 .
\label{eq:eft:weakbounds}
\end{equation}
For $\lambda_2\dot\pic^{\,3}$ the perturbative-unitarity analysis of Ref.~\cite{Cheung:2007st},
from $2\to2$ scattering in the decoupling limit, places the breakdown scale above this estimate by
loop factors of $4\pi$, and since nothing in the derivation uses the field content the same
relaxation is expected for each entry. The third entry is the cutoff of the
symmetry-fixed operator, $\Lambda_\la=2\fpi^2/(H\la)$, in dimensionless form; at weak mixing
$R\simeq1$ and it reads $\la\lesssim(2\pi\Dz)^{-1}$, which is not constraining since $\la\lesssim1$
here already.

\begin{figure}[t]
\centering
\includegraphics[width=\textwidth]{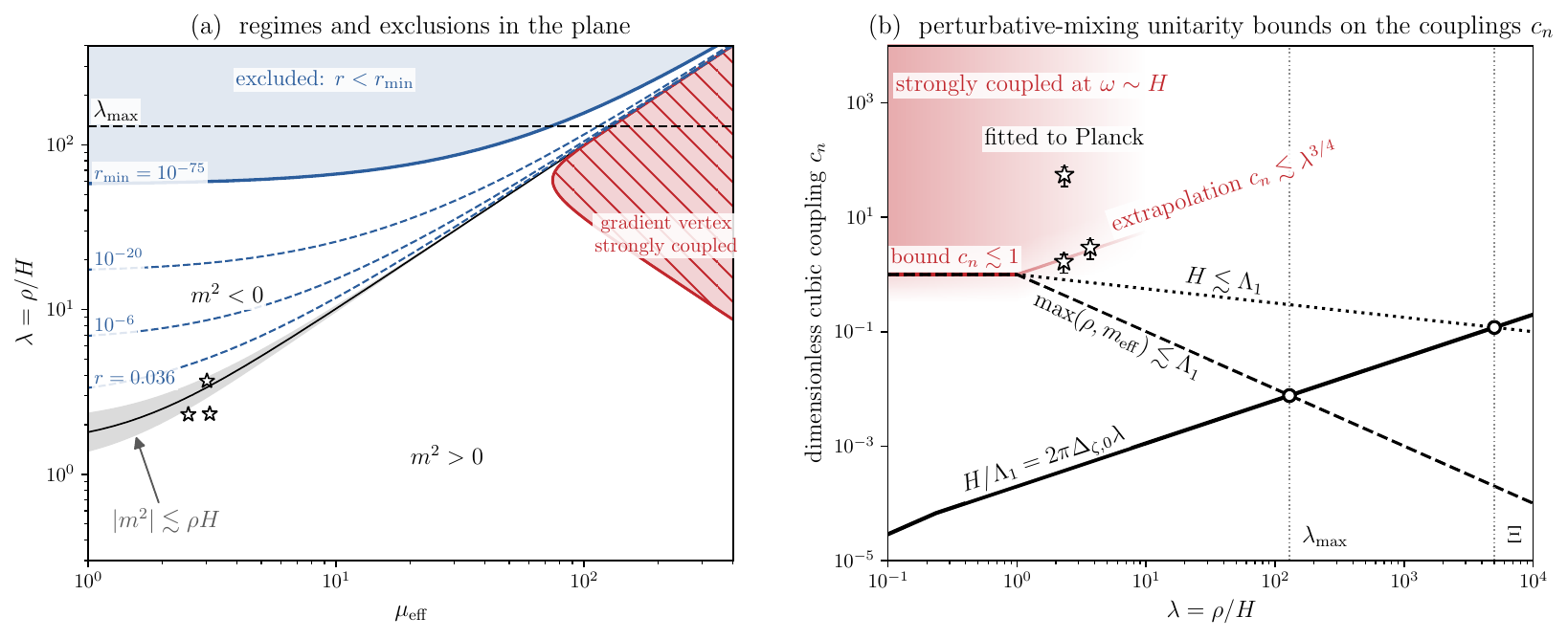}
\caption{Left: the $(\muf,\la)$ plane. Grey band: the strong-mixing regime
\cref{eq:eft:strongregime}, $|m^2|\lesssim\rho H$, straddling the bare-massless line $m^2=0$ (thin
solid); to its right the field is heavy on its own, above it the bare mass is tachyonic. Red
($\backslash\backslash$): the symmetry-fixed gradient vertex is strongly coupled, $\max(\rho,\meff)
\gtrsim\Lambda_1$, which is \cref{eq:eft:lammax} continued off the band. Black dashed: the
self-consistency bound $\la\le\la_{\max}$ of \cref{eq:eft:lammax}, its form on the band. Blue:
contours of constant tensor-to-scalar ratio, $R(\la,\muf)=16/r$ from \cref{eq:eft:Rbound}; a
measurement of $r$ would exclude everything above its own contour, and the solid one,
$r_{\rm min}\simeq10^{-75}$, is the unconditional exclusion set by reheating before
nucleosynthesis. Stars: the three best fits of \cref{sec:data}. Right: the dimensionless couplings
$c_n$ of \cref{eq:eft:cn} against $\la$. Red: the region excluded by the perturbative-mixing
unitarity bound at $\omega\sim H$, $c_n\lesssim1$ of \cref{eq:eft:weakbounds}, common to the four
free couplings, drawn with the confidence we have in it: it fades across the bound, which is an
order-of-magnitude statement, so that a coupling sitting on the line is neither excluded nor
safe, and it fades with the mixing, since the bound is established for $\la\lesssim1$ (solid
boundary) and only extrapolated beyond, with the $\la^{3/4}$ scaling of \cref{eq:eft:strongbounds}
(dashed boundary), until nothing is left of it at $\la\simeq10$. Black solid: the symmetry-fixed $H/\Lambda_1=2\pi\Dzo\la$, which is a
value and not a bound, since it has no freedom. Dotted and dashed: the two ceilings on that value,
$H\lesssim\Lambda_1$ and $\max(\rho,\meff)\lesssim\Lambda_1$ of \cref{eq:eft:lammax}, which it
meets at $\Xi$ and at $\la_{\max}$ (circles). Stars: the couplings
fitted in \cref{sec:data}, shown against the extrapolated bound for what it is worth.}
\label{fig:eft:validity}
\end{figure}

These are perturbative-mixing bounds, and they are the only ones we have. Now at strong mixing,
$\la\gg1$, no unitarity computation exists, and what we draw in \cref{fig:eft:validity} is an
extrapolation. It follows the estimates of Ref.~\cite{Pinol:2023oux}: the mixing dominates the
quadratic action, $\omega=k^2/\rho$, a single degree of freedom propagates, and rescaling
$\tilde x=\rho^{1/2}x$ makes a cubic vertex with $n_\partial$ pairs of spatial derivatives carry
$\rho^{-3/4+n_\partial}$. Asking again that the vertex be perturbative at $\omega\sim H$,
\begin{equation}
  \lambda_2H^2,\;\; H/\Lambda_2,\;\; |\alpha|,\;\; \gthree/H \;\lesssim\; \la^{3/4} ,
  \qquad
  H/\Lambda_1\lesssim\la^{-1/4} .
\label{eq:eft:strongbounds}
\end{equation}
The gradient vertex is the exception twice over: its ceiling \emph{falls} with the mixing,
gradients being enhanced when $\omega\ll k$, and its coupling grows with $\la$, so that the entry is
a bound on $\la$ itself. Even as an extrapolation, \cref{eq:eft:strongbounds} is derived for
$|m^2|/\rho\lesssim H$, that is, in the modified-dispersion regime
\begin{equation}
  \frac{|m^2|}{\rho}\lesssim H
  \quad\Longleftrightarrow\quad
  \Big|\la-\sqrt{\muf^2+\tfrac52}\,\Big|\lesssim\tfrac12 ,
\label{eq:eft:strongregime}
\end{equation}
of unit width in $\la$ straddling the bare-massless line $\la=\sqrt{\muf^2+\tfrac94}$; outside it
the dispersion relation is relativistic again and even that scaling argument fails. The same
restriction applies to every strong-mixing formula of this subsection, including the bound on
$\la$ derived next: they use the modified-dispersion form of the amplification,
$R^{1/2}=(\sqrt\pi/\Gamma(3/4))\la^{1/4}$ of \cref{sec:lin:R}, which holds inside the band
\cref{eq:eft:strongregime} only, so that they are statements about that narrow band of the plane,
drawn off it in \cref{fig:eft:validity} for orientation, and none is precise elsewhere. We
therefore treat the strong-mixing ceilings on the four free couplings as indicative at best, and
nothing below is conditioned on them.

The one bound that does not concern a free parameter is the one on $\la$, because $\Lambda_1$ is
fixed by the non-linearly realised boosts. Requiring only that the Hubble scale lie below the
cutoff, $H\lesssim\Lambda_1$, and converting $\Dzo=\Dz R^{-1/2}$ with the modified-dispersion
$R^{1/2}$ self-consistently, gives
$\la\lesssim\Xi\equiv[2\sqrt\pi\,\Gamma(3/4)\,\Dz]^{-1}\simeq5.0\times10^3$, the bound of
Ref.~\cite{Pinol:2023oux} at $c_s=1$. But $H$ is not the largest scale in the
problem once the mixing is strong: the spectrum contains a state of mass $\meff$, and the resummed
mixing carries frequencies up to $\rho$ within a single mode, so that the operator must
accommodate $E=\max(\rho,\meff)=H\max(\la,\sqrt{\muf^2+\tfrac94})$. Requiring $E\lesssim\Lambda_1$
and repeating the same self-consistent step, now on $\la^2$, gives
\begin{equation}
  \max(\rho,\meff)\lesssim\Lambda_1 \ \Longleftrightarrow\
  \la^{7/4}\lesssim\Xi \ \Longrightarrow\
  \la\lesssim\la_{\max}=\Xi^{4/7}\simeq1.3\times10^2 ,
\label{eq:eft:lammax}
\end{equation}
stronger by $\Xi^{3/7}\simeq40$ and equivalent to the transparent statement $\rho\lesssim\fpi$:
the mixing cannot exceed the scale at which time translations are broken. This is the bound
$\la\ll\Dzo^{-1/2}$ quoted in Ref.~\cite{Pinol:2026xnl}, and the one we use. For the four free
couplings we quote
the weaker $\omega\sim H$ ceiling, which is the one a measurement of a correlator at that frequency
directly tests; requiring them to survive up to $E$ instead would tighten each by $(H/E)^{d-4}$.

One more relation is not a statement about couplings at all. The mixing amplifies the curvature
spectrum, $\Dz=R^{1/2}\Dzo$ with the $R(\la,\muf)$ of \cref{sec:lin:R}, while the tensor spectrum
does not feel it,
\begin{equation}
  r \;=\; \frac{16\epsilon}{R(\la,\muf)} ,
  \qquad
  \frac{H}{\mpl}=\pi\Dz\sqrt{\frac{r}{2}} ,
\label{eq:eft:Rbound}
\end{equation}
so that a measurement of $r$ bounds the amplification by $R\le16/r$ through $\epsilon\le1$, and
everything above the contour $R(\la,\muf)=16/r$ is excluded. In the modified-dispersion regime,
inside the grey band of \cref{fig:eft:validity}, the amplification costs nothing,
$R\simeq2.1\sqrt\la$; above it $R$ grows like $e^{\pi\la}$, so that a large mixing
is bought with an exponentially amplified scalar spectrum and hence a small $r$. A value of $r$ at
its present upper limit, $0.036$, would exclude $\la\gtrsim4.1$ at $\muf=2$, each decade of
smaller $r$ pushes the boundary out by only $\Delta\la=\ln10/\pi\simeq0.7$, and the requirement
that reheating precede nucleosynthesis excludes $\la\gtrsim59$ there
unconditionally.\footnote{$T_{\rm reh}\gtrsim4\,{\rm MeV}$ gives $H\gtrsim3\times10^{-42}\mpl$,
hence $r\gtrsim r_{\min}=10^{-75}$.} \Cref{fig:eft:validity} collects the
band, the two ceilings and the tensor contours.

The theory \cref{eq:eft:action} with $M_2=0$ is thus under control provided
\begin{equation}
  \la \lesssim \la_{\max} \simeq 1.3\times10^2,
  \qquad
  |\lambda_2| H^2,\; \Big|\frac{H}{\Lambda_2}\Big|,\; |\alpha|,\; \frac{|\gthree|}{H} \;\lesssim\;
  \begin{cases}1 & \la\lesssim1\\ \la^{3/4} & \la\gg1\end{cases} ,
\label{eq:eft:validity}
\end{equation}
together with the amplification bound $R\le16/r$ of \cref{eq:eft:Rbound}, every entry being an
order-of-magnitude estimate, anchored on a unitarity computation in its weak-mixing half and an
extrapolation in the other. Three qualifications are open questions rather than details. The bounds are tree-level statements
evaluated on the dressed background; every leg of the resummed theory carries the Boltzmann weight
$e^{a\pi\la/2}$ of \cref{sec:linear}, and a naive count would enhance an $L$-loop diagram by
$e^{I\pi\la/2}$ with $I$ internal lines, but the channel sum is not sign-definite and the
tree-level squeezed limit exhibits exactly such a cancellation (\cref{sec:sq:strong}), so that the
loop expansion parameter of the resummed theory is not established by \cref{eq:eft:validity}. We
use $\zeta=-H\pi$, and the quadratic relation $\zeta=-H\pi+H\pi\dot\pi+\tfrac12\dot H\pi^2$
contributes an extra bispectrum term, exponentially suppressed for an effectively heavy field,
which we have not evaluated in the resummed theory~\cite{Maldacena:2002vr}. Finally, we truncate
the mixing and massive-field sectors at dimension four in the unitary gauge and, within that, to
operators with no derivative of $\sigma$ beyond the kinetic term: operators such as
$\sigma(\partial_\mu\sigma)^2$ or $\dgz g^{0\mu}\partial_\mu\sigma$ would correct $p_\sigma$ and
invalidate the simple form of \cref{eq:eft:Hint}.

\section{Exact linear theory}
\label{sec:linear}

\subsection{The coupled system}
\label{sec:lin:system}

The quadratic Lagrangian \cref{eq:eft:L2} at $c_s=1$ mixes the two fluctuations through the single
dimensionful coupling $\rho$. We do not treat it perturbatively: it sits in the free Hamiltonian
\cref{eq:eft:Hfree}, and everything in this section is exact in $\la$. Every mode function, kernel
and coefficient below carries the effective index $\n$ of \cref{eq:eft:nueff}; the bare
$\nbare$ appears only where a limit is taken at fixed bare mass, and we flag it there. In conformal
time, $a=-1/(H\tau)$, and in terms of $x\equiv-k\tau\in(0,\infty)$, with a prime denoting
$\partial_x$, the linear equations are
\begin{align}
  \label{eq:lin:P}
  \Big(x^{-2}\,\mathcal{D}\pic\Big)' + x^{-2}\pic &= 0 ,
  &&\mathcal{D}\pic \;\equiv\; \pic' - \frac{\la}{x}\,\sigma ,\\[2pt]
  \label{eq:lin:S}
  \big(x^{-2}\sigma'\big)' + x^{-2}\sigma + \frac{m^2}{H^2}\,x^{-4}\sigma
  + \la\,x^{-3}\,\pic' &= 0 .
\end{align}
The combination $\mathcal{D}\pic$ is the curvature momentum, $p_\pi=a^3(\dot\pic+\rho\sigma)
=-(k^3/H^2)\,x^{-2}\mathcal{D}\pic$, and \cref{eq:lin:P} states that it is conserved up to the
gradient source. \Cref{eq:lin:P,eq:lin:S} are the coupled system solved exactly in
Ref.~\cite{Huenupi:2026abj}, up to a relabelling of the two channels introduced below; we stress that the
index the exact solution is naturally written with is the effective one.

At $\la=0$, \cref{eq:lin:P} is the equation of a massless field in de Sitter, with Bunch--Davies
solutions
\begin{equation}
  \label{eq:lin:carriers}
  u^1_\pi(x) = \ii(1-\ii x)e^{\ii x},\qquad u^2_\pi = (u^1_\pi)^*,\qquad
  {u^1_\pi}' = \ii x\,e^{\ii x},\qquad
  u^1_\pi {u^2_\pi}' - u^2_\pi {u^1_\pi}' = -2\ii x^2 .
\end{equation}
These two carriers are used only as a basis on which to expand the interacting curvature mode: the
expansion coefficients are exact functions of $x$, and nothing assumes $\la$ small. No isocurvature
carrier is needed, because $\sigma$ is slaved to $\pic$ by \cref{eq:lin:P} and is obtained by
solving that constraint.

\subsection{Bogoliubov decomposition and the exact solution}
\label{sec:lin:bogo}

In this subsection we follow the strategy of Refs.~\cite{Huenupi:2026abj,Huenupi:2026aqc}, in the form given in the
first of them, and recast its result in the form we need. We expand each field on two oscillators
$\alpha=1,2$,
\begin{equation}
  \label{eq:lin:expansion}
  X_{\vec k}(\tau) = X^\alpha(\tau,k)\,\hat a^{\vec k}_\alpha + \text{h.c.},
  \qquad X\in\{\pic,\sigma\},
  \qquad [\hat a_\alpha^{\vec k},\hat a_\beta^{\vec k'\dagger}]=\delta_{\alpha\beta}(2\pi)^3\delta^{(3)}(\vec k-\vec k'),
\end{equation}
and write the curvature mode on the two carriers with time-dependent Bogoliubov coefficients,
\begin{equation}
  \label{eq:lin:split}
  \begin{gathered}
  \pic^\alpha(\tau,k) = \frac{H}{\sqrt{2k^3}}\Big[\pi^\alpha_1(x)\,u^1_\pi(x) + \pi^\alpha_2(x)\,u^2_\pi(x)\Big],\\[2pt]
  \mathcal{D}\pic^\alpha(\tau,k) = \frac{H}{\sqrt{2k^3}}\Big[\pi^\alpha_1(x)\,{u^1_\pi}'(x) + \pi^\alpha_2(x)\,{u^2_\pi}'(x)\Big].
  \end{gathered}
\end{equation}
The first relation writes one function as two and is underdetermined; the second removes the
redundancy. It is a choice, and the natural one: since $\mathcal{D}\pic\propto p_\pi$, the pair
$(\pi_1,\pi_2)$ diagonalises the field and its conjugate momentum simultaneously, and at $\la=0$
the split reduces to the textbook one. Applying $\mathcal{D}$ to the first relation and demanding
consistency with the second, then inserting the first into \cref{eq:lin:P} and using that the
carriers solve the free equation, one obtains two relations for the derivatives of the Bogoliubov
coefficients, which the Wronskian \cref{eq:lin:carriers} solves as
\begin{equation}
  \label{eq:lin:pidot}
  \frac{H}{\sqrt{2k^3}}\,{\pi^\alpha_1}' = \frac{\la\,\sigma^\alpha}{2x^2}\,e^{-\ii x},
  \qquad
  \frac{H}{\sqrt{2k^3}}\,{\pi^\alpha_2}' = \frac{\la\,\sigma^\alpha}{2x^2}\,e^{+\ii x} .
\end{equation}
These two relations drive the rest of the section. Their ratio,
${\pi^\alpha_2}'=e^{2\ii x}{\pi^\alpha_1}'$, gives the second coefficient from the first with no
reference to $\sigma$, and either of them returns $\sigma^\alpha$ algebraically once $\pi_1^\alpha$
is known.

Eliminating three of the four Bogoliubov coefficients produces a single fourth-order equation for
$\pi^\alpha_1$ \cite{Huenupi:2026abj}. Its four solutions are $\widehat{D}_{\n,a}F(\ii a\la/2,1,-2\ii x)$
with $F=U,M$ the Tricomi and Kummer functions, $a=\pm$, and
\begin{equation}
  \label{eq:lin:Dhat}
  \widehat{D}_{\n,a} \;\equiv\; \hyp\!\Big(\tfrac12-\n,\ \tfrac12+\n;\ 1+\ii a\la;\ \tfrac{\ii}{2}\partial_x\Big)
  \;=\;\sum_{n\ge0}\frac{(\tfrac12-\n)_n(\tfrac12+\n)_n}{(1+\ii a\la)_n\,n!}\Big(\frac{\ii}{2}\partial_x\Big)^{\! n}.
\end{equation}
The label $a=\pm$ is the \emph{channel} index. In the Bunch--Davies vacuum only the two $U$
solutions survive \cite{Huenupi:2026abj}.

\subsection{The weight \texorpdfstring{$\omega_a(u)$}{omega\_a(u)} and the collapse of the carriers}
\label{sec:lin:omega}

It is convenient to name the combination in which the channel index and the mixing strength always
appear,
\begin{equation}
  \label{eq:lin:za}
  z_a \;\equiv\; \frac{\ii a\la}{2},\qquad a=\pm ,\qquad z_{-a}=-z_a=z_a^{*} ,
\end{equation}
so that the third parameter of \cref{eq:lin:Dhat} is $1+2z_a$. The series \cref{eq:lin:Dhat} is an
infinite-order differential operator, and acting on $U(z_a,1,-2\ii x)$ it produces a doubly
infinite series. It resums in closed form once the Tricomi function is written in the integral
representation \cref{eq:uf:tricomi}, $U(c,1,-2\ii x)=\Gamma(c)^{-1}\int_0^\infty\dd u\,u^{c-1}(1+u)^{-c}e^{2\ii ux}$
over a dimensionless variable $u\ge0$, in which the $x$-dependence sits in a single plane wave and
everything else is a weight. Our case $c=z_a$ is purely imaginary, i.e.\ exactly on the boundary
of convergence of that representation: near $u=0$ the integrand has modulus $u^{-1}$ while its
phase oscillates ever faster. The left-hand side is nevertheless entire in $c$, and the right-hand
side is to be read as its analytic continuation, made explicit in \cref{app:continuation}:
subtracting the Taylor terms of the regular factor at $u=0$ and integrating them analytically
defines the integral, and this is at the same time the proof of the continuation and the algorithm
a quadrature must implement. The variable $u$ has a direct reading: the mode is a superposition of
plane waves of dressed frequency $k(1+2u)$, and at $\la\to0$ the weight collapses onto $u=0$ and
returns the undressed massless mode (\cref{sec:lin:weak}).

With the Bunch--Davies prescription $x\to x(1-\ii\epsilon)$ each derivative in \cref{eq:lin:Dhat}
acts on a plane wave, $\tfrac{\ii}{2}\partial_x e^{2\ii ux}=-u\,e^{2\ii ux}$, and the operator
collapses to multiplication by a ${}_2F_1$ function of $-u$:
\begin{equation}
  \label{eq:lin:f1}
  \pi^\alpha_1 = A^\alpha_a f^a_1 ,\qquad
  f^a_1(x) = \int_0^\infty\!\dd u\;\omega_a(u)\,e^{2\ii ux},
\end{equation}
\begin{equation}
  \label{eq:lin:omega}
  \boxed{\ \
  \omega_a(u) = \frac{u^{z_a-1}(1+u)^{-z_a}}{\Gamma(z_a)}\;
  \hyp\Big(\tfrac12-\n,\ \tfrac12+\n;\ 1+2z_a;\ -u\Big),
  \qquad \omega_{-a}=\omega_a^{*}, \ \ }
\end{equation}
with summation over the repeated channel index and the constants $A^\alpha_a$ fixed in
\cref{sec:lin:bogcoeff}. The weight $\omega_a$ encodes the whole linear theory. Its large-$u$
behaviour follows from the connection formula \cref{eq:uf:connection},
\begin{equation}
  \label{eq:lin:omegatail}
  \omega_a(u) \;\simeq_{u\gg1}\; \frac{1}{\Gamma(z_a)}\sum_{b=\pm}
  \frac{\Gamma(1+2z_a)\,\Gamma(2b\n)}{\Gamma(\tfrac12+b\n)\,\Gamma(\tfrac12+b\n+2z_a)}\;
  u^{-3/2+b\n}\Big[1+\mathcal{O}(u^{-1})\Big],
\end{equation}
and the two powers $u^{-3/2\pm\n}$ are the clock: they are the only source of the non-analytic
$\kappa^{1/2\mp\n}$ of the squeezed bispectrum, and of the $x^{3/2\mp\n}$ late-time behaviour of
$\sigma$.

The second coefficient follows from the ratio of \cref{eq:lin:pidot} by one integration,
\begin{equation}
  \label{eq:lin:f2}
  \pi^\alpha_2 = A^\alpha_a f^a_2 ,\qquad
  f^a_2(x)=\int_0^\infty\!\dd u\;\omega_a(u)\,\frac{u}{1+u}\,e^{2\ii(1+u)x},
\end{equation}
the integration constant being zero by the Bunch--Davies condition that the negative-frequency
carrier be unpopulated deep inside the horizon. Assembling \cref{eq:lin:split}, the two carriers combine into a single dressed plane
wave:
\begin{equation}
  \label{eq:lin:Mpi}
  \boxed{\ \ \pic^\alpha(\tau,k) = \frac{H}{\sqrt{2k^3}}\,A^\alpha_a\,\Ma^\pi_a(x),
  \qquad
  \Ma^\pi_a(x) = \ii\int_0^\infty\!\dd u\;\omega_a(u)\,\frac{1-\ii x(1+2u)}{1+u}\;e^{\ii x(1+2u)} . \ \ }
\end{equation}
This is the structural fact that makes the resummed theory tractable: each channel mode is a
single integral over $u$ of one plane wave, because $u^1_\pi(x)e^{2\ii ux}$ and
$u^2_\pi(x)e^{2\ii(1+u)x}$ carry the same phase $e^{\ii x(1+2u)}$. Its derivative is equally simple,
\begin{equation}
  \label{eq:lin:Mpiprime}
  \partial_x \Ma^\pi_a(x) = \ii x\int_0^\infty\!\dd u\;\omega_a(u)\,\frac{(1+2u)^2}{1+u}\;e^{\ii x(1+2u)} ,
\end{equation}
and the isocurvature mode follows algebraically from the first of \cref{eq:lin:pidot},
\begin{equation}
  \label{eq:lin:Msig}
  \boxed{\ \sigma^\alpha(\tau,k) = \frac{H}{\sqrt{2k^3}}\,A^\alpha_a\,\Ma^\sigma_a(x),
  \quad
  \Ma^\sigma_a(x) = \frac{2x^2}{\la}e^{\ii x}\,{f^a_1}'(x)
  = \frac{4\ii x^2}{\la}\int_0^\infty\!\dd u\;\omega_a(u)\,u\;e^{\ii x(1+2u)} . \ }
\end{equation}
Both channel modes are integrals of the same weight against the same plane wave, and
differ only in the rational function of $u$ they carry. Representations of this kind, a
continuous superposition of simpler objects against a weight, have been used for massive
correlators before: the spectral representation of Ref.~\cite{Werth:2024mjg} writes a massive
propagator as an integral over conformal ones, the partial Mellin--Barnes representation of
Ref.~\cite{Qin:2022fbv} writes it as a contour integral of powers, and the Laplace-space form of
Ref.~\cite{Belrhali:2026uxn} is the closest to ours, being also taken at the level of the mixed mode
functions; what is specific here is that the weight is one function of one variable, exact in the
mixing, and that the same weight carries both fields. The explicit $1/\la$ in \cref{eq:lin:Msig} is
not a singularity: as $\la\to0$ the measure collapses onto $u=0$ where the weight $u$ vanishes, so
the $u$-integral is itself $\mathcal{O}(\la)$; the cancellation is made quantitative in
\cref{sec:lin:weak}.

\subsection{Channel diagonalisation, boundary values and the power spectrum}
\label{sec:lin:bogcoeff}

Matching, as in Ref.~\cite{Huenupi:2026abj}, at a fiducial sub-horizon time $x_0\gg1$ to the decoupled Bunch--Davies solutions, mode
$\alpha=1$ purely curvature and mode $\alpha=2$ purely isocurvature, using
$f^a_1(x)\to e^{-a\pi\la/4}(2x)^{-\ii a\la/2}$ and
$\Ma^\sigma_a(x)\to-\ii a\,x\,e^{\ii x}e^{-a\pi\la/4}(2x)^{-\ii a\la/2}$, one finds
\begin{equation}
  \label{eq:lin:A}
  A^1_\pm = \tfrac12 e^{\pm\pi\la/4}\,(2x_0)^{\pm\ii\la/2},
  \qquad
  A^2_\pm = \pm\tfrac{\ii}{2} e^{\pm\pi\la/4}\,(2x_0)^{\pm\ii\la/2} .
\end{equation}
The matching phases are pure phases and cancel identically in every observable, since
\begin{equation}
  \label{eq:lin:diag}
  \boxed{\ \ \sum_{\alpha=1,2}A^\alpha_a\big(A^\alpha_b\big)^{*}
  = \tfrac12\,e^{a\pi\la/2}\,\delta_{ab} \ \ }
\end{equation}
is diagonal in the channel index. The residue $\tfrac12e^{a\pi\la/2}$ is a Boltzmann-like weight,
appearing exactly once per propagator, and it is where the exponential enters the theory. It plays
the role that a chemical potential plays in the collider literature, where a coupling to a
background rotation lifts the Boltzmann suppression of a heavy
particle~\cite{Bodas:2020yho,Sou:2021juh}; here no such coupling is introduced and the weight is
generated by the mixing itself. Note that it is not by itself the size of any observable: the
boundary values and the kernels carry their own exponentials and partly cancel it, and the leading
behaviour of a correlator is fixed only after the channel sum, where cancellations between $\vec a$
and $-\vec a$ are the rule (\cref{sec:sq:strong}).

\paragraph{Boundary values.} At $x\to0$ the bracket in \cref{eq:lin:Mpi} tends to $1/(1+u)$, so
$\Ma^\pi_a(0)=\ii r_a$ with $r_a=\int_0^\infty\dd u\,\omega_a(u)/(1+u)$. Substituting $u=t/(1-t)$
and applying Pfaff's transformation \cref{eq:uf:pfaff} turns the measure into
\begin{equation}
  \label{eq:lin:pfaff}
  \omega_a(u)\,\dd u
  = \frac{t^{z_a-1}(1-t)^{-1/2-\n}}{\Gamma(z_a)}\,
    \hyp\Big(\tfrac12-\n,\ \tfrac12-\n+2z_a;\ 1+2z_a;\ t\Big)\,\dd t ,
  \qquad t\in(0,1),
\end{equation}
a form in which both endpoint powers are explicit, and which is also the better one for numerics.
Euler's integral \cref{eq:uf:euler} then gives a ${}_3F_2$ at unit argument of a special type
that does collapse to $\Gamma$'s, and after one use of the duplication formula
\cref{eq:uf:gamma}
\begin{equation}
  \label{eq:lin:raclosed}
  \boxed{\ \ r_a = \frac{\Gamma\!\big(\tfrac12+\tfrac{\ii a\la}{2}\big)\,
  \Gamma\!\big(\tfrac34-\tfrac{\n}{2}\big)\Gamma\!\big(\tfrac34+\tfrac{\n}{2}\big)}
  {\sqrt{\pi}\;\Gamma\!\big(\tfrac34-\tfrac{\n}{2}+\tfrac{\ii a\la}{2}\big)
  \Gamma\!\big(\tfrac34+\tfrac{\n}{2}+\tfrac{\ii a\la}{2}\big)} ,
  \qquad r_{-}=r_{+}^{*} , \ \ }
\end{equation}
even in $\n$ and finite for all $\la$ and $\Real\,\n<3/2$. The isocurvature mode instead vanishes
at the boundary: its $u$-integral is dominated by the tail \cref{eq:lin:omegatail}, and
\begin{equation}
  \label{eq:lin:Msigbdry}
  \Ma^\sigma_a(x) \;\simeq_{\,x\to0}\;
  \frac{4\ii}{\la}\sum_{b=\pm}\Wcal^a_b\,(-2\ii)^{-1/2-b\n}\;x^{3/2-b\n} ,
  \qquad
  \boxed{\ \Wcal^a_b \equiv \frac{\Gamma(1+2z_a)\,\Gamma(2b\n)}{\Gamma(z_a)\,\Gamma(\tfrac12+b\n+2z_a)} .\ }
\end{equation}
The powers are exactly $x^{3/2\mp\n}$, the standard super-horizon behaviour of a field of effective
mass $\meff$: the resummed mixing enters the late-time exponents only through $m\to\meff$, and
enters their amplitudes in full. The
coefficients $\Wcal^a_b$ are the same ones that govern the soft limit of the leg kernels below.
Since $\Ma^\sigma_a(0)=0$ while
$\Ma^\pi_a(0)\neq0$ for $\Real\,\n<3/2$, \emph{a propagator end at the boundary is necessarily a
$\pic$ end}, which is what allows the diagrammatic rules of \cref{sec:rules} to be so compact. The
exactly massless case $\n=3/2$, where $\sigma$ does not decay and the power spectrum grows
secularly \cite{Huenupi:2026abj}, is excluded throughout.

\paragraph{The power spectrum.}
\label{sec:lin:R}
With $\zeta=-H\pic/\fpi^2$ and \cref{eq:lin:diag},
$\Dz^2=(k^3/2\pi^2)\sum_\alpha|\zeta^\alpha|^2=R\,\Dzo^2$ with
$R=\tfrac12\sum_ae^{a\pi\la/2}|r_a|^2$. Since $|\Gamma(\tfrac12+\ii y)|^2=\pi/\cosh\pi y$, the
$\Gamma(\tfrac12+\ii a\la/2)$ of \cref{eq:lin:raclosed} cancels the $\cosh(\pi\la/2)$ of the
channel sum exactly, and
\begin{equation}
  \label{eq:lin:R}
  \boxed{\ \ R(\la,\n) = \left|\,
  \frac{\Gamma\!\big(\tfrac34-\tfrac{\n}{2}\big)\Gamma\!\big(\tfrac34+\tfrac{\n}{2}\big)}
       {\Gamma\!\big(\tfrac34-\tfrac{\n}{2}+\tfrac{\ii\la}{2}\big)\Gamma\!\big(\tfrac34+\tfrac{\n}{2}+\tfrac{\ii\la}{2}\big)}
  \right|^{2} , \ \ }
\end{equation}
with the exact special values $R(0,\n)=1$ and $R(\la,\tfrac12)=\sinh(\pi\la)/(\pi\la)$. For an
effectively heavy field, $\n=\ii\muf$, the two denominators carry the sum and the difference of the
two variables of the problem,
\begin{equation}
  \label{eq:lin:Rplane}
  R(\la,\ii\muf)=\frac{\big|\Gamma\big(\tfrac34+\tfrac{\ii\muf}{2}\big)\big|^{4}}
  {\big|\Gamma\big(\tfrac34+\tfrac{\ii(\la-\muf)}{2}\big)\big|^{2}\,
   \big|\Gamma\big(\tfrac34+\tfrac{\ii(\la+\muf)}{2}\big)\big|^{2}} ,
\end{equation}
which is \cref{eq:lin:R} rewritten and is the form we use at strong mixing (\cref{sec:lin:strong}).
With $\muf>0$ the denominator factor $|\Gamma(\tfrac34+\tfrac{\ii}{2}(\la+\muf))|^{-2}$ is large as
soon as $\la\gg1$, whatever the ratio $\muf/\la$. Since $\muf+|\la-\muf|\ge\la$, at most one of the
two other $\Gamma$ functions can have an argument of order one. Three
limits are useful. At weak mixing and fixed bare mass,
\begin{equation}
  \label{eq:lin:Rweak}
  R = 1 + \frac{\la^2}{4}\Big[\psi^{(1)}\big(\tfrac34-\tfrac{\nbare}{2}\big)
                             +\psi^{(1)}\big(\tfrac34+\tfrac{\nbare}{2}\big)\Big] + \mathcal{O}(\la^4),
\end{equation}
where the $\mathcal{O}(\la)$ term vanishes because it is $\Real[\ii\la\,\cdots]$ and the
$\mathcal{O}(\la^2)$ coefficient may be evaluated at the bare index. At strong mixing and fixed
bare mass, $\muf=\sqrt{\la^2-\nbare^2}\simeq\la-\nbare^2/2\la$, so one $\Gamma$ in the denominator
stays at finite argument while the other runs to infinity, and
\begin{equation}
  \label{eq:lin:Rstrongbare}
  R = \frac{\pi\sqrt{\la}}{\Gamma(3/4)^2}\left[1+\frac{\pi\nbare^2}{4\la}
  +\mathcal{O}(\la^{-2})\right].
\end{equation}
At strong mixing and fixed effective mass both denominators run to infinity and
\begin{equation}
  \label{eq:lin:Rstrongeff}
  R \;\simeq\; \frac{e^{\pi\la}}{2\pi^2\la}\,
  \Big|\Gamma\big(\tfrac34-\tfrac{\n}{2}\big)\Gamma\big(\tfrac34+\tfrac{\n}{2}\big)\Big|^2 ,
\end{equation}
i.e.\ $e^{\pi\la/2}$ per external leg: the Boltzmann weight of \cref{eq:lin:diag} surviving into
the observable. The coefficient of \cref{eq:lin:Rstrongbare} and the two $\Gamma$-limits were
verified numerically at the level quoted in \cref{app:numerics}. As noted in Ref.~\cite{Pinol:2026xnl}, the strong-mixing amplification
$\Dz^2\to[\pi/(4\pi^2\Gamma(3/4)^2)](H/\fpi)^4\sqrt\la$ coincides exactly with the amplification found in
Ref.~\cite{Pinol:2023oux}, since $\pi/(4\pi^2\Gamma(3/4)^2)=2\Gamma(5/4)^2/\pi^3$ by
$\Gamma(1/4)\Gamma(3/4)=\pi\sqrt2$; this is an independent check of the resummed linear theory.

\subsection{Leg kernels: one seed function}
\label{sec:lin:kernels}

Every vertex integral of \cref{sec:rules} reduces, through the Schwinger identity, to integrals
of $\omega_a(u)e^{-\beta u}$ against the rational weights appearing in
\cref{eq:lin:Mpi,eq:lin:Mpiprime,eq:lin:Msig}. We define, for $\Real\,\beta>0$,
\begin{equation}
  \label{eq:lin:kernels}
  \boxed{\ \
  \Wn^a_n(\beta) = \int_0^\infty\!\dd u\;\omega_a(u)\,\frac{(1+2u)^n}{1+u}\,e^{-\beta u},
  \qquad
  \Vv^a(\beta) = \int_0^\infty\!\dd u\;\omega_a(u)\,u\,e^{-\beta u} ,\ \ }
\end{equation}
with $[\Wn^a_n]^*=\Wn^{-a}_n$ and $[\Vv^a]^*=\Vv^{-a}$ by reality of the fields. The entire toolkit
is one function of one variable. Indeed, for $\Real\,\beta>0$ differentiation under the integral
sign is legitimate, $\partial_\beta$ inserts $-u$, and $e^{-\beta u}$ is an eigenfunction of
$1-2\partial_\beta$ with eigenvalue $1+2u$, so that
\begin{equation}
  \label{eq:lin:master}
  \boxed{\ \ \Wn^a_n = \big(1-2\partial_\beta\big)^n\,\Wn^a_0 ,\qquad
  \Vv^a = -{\Wn^a_0}'+{\Wn^a_0}'' . \ \ }
\end{equation}
These are exact identities, not approximations. Polynomial division gives moreover
\begin{equation}
  \label{eq:lin:division}
  \frac{(1+2u)^2}{1+u} = 4u + \frac{1}{1+u}
  \qquad\Longrightarrow\qquad
  \boxed{\ \ \Wn^a_2(\beta) = \Wn^a_0(\beta) + 4\,\Vv^a(\beta)\qquad\text{for all } \la,\ \n,\ \beta , \ \ }
\end{equation}
an exact algebraic relation whose consequences for the collider signal are drawn in
\cref{sec:lin:tails}. Every kernel inherits the normalisation $\Wn^a_0(0)=r_a$ of
\cref{eq:lin:raclosed}; $\Wn^a_1(0)$ converges for $\Real\,\n<\tfrac12$; $\Wn^a_2(0)$ and $\Vv^a(0)$
diverge, and those divergences are the soft tails discussed below. For $\Real\,\beta>0$ all kernels
are analytic for every real $\la$ and every $\n$ with $|\Real\,\n|<\tfrac32$, i.e.\ for every mass
$m^2>0$: the threshold $\Real\,\n<\tfrac12$ met elsewhere in this paper concerns the boundary values
and the soft tails at $\beta\to0$, where a light field's power $\beta^{-1/2-\n}$ ceases to be
integrable, not the kernels at finite argument. For an implementation, the consequence is the
following: one adaptive quadrature over $u$ for $\Wn^a_0(\beta)$, tabulated in $\beta$, for each point of the two-parameter family $(\la,\n)$; everything
downstream, at every exchange order, is analytic differentiation of that one tabulated function. No
closed form is known for $\Wn^a_0(\beta)$ itself at general $\la$ and $\n$; its boundary value, its
small- and large-$\beta$ asymptotics and its weak-mixing expansion are in closed form, and finding
one for the function would make the whole construction analytic (\cref{sec:conclusion}).

The kernels never appear alone in a correlator: each external leg carries its channel index through
the weight $e^{a\pi\la/2}r_a$ of the boundary end and the Boltzmann factor, and the channel sum
factorises leg by leg at fixed Schwinger parameter (\cref{sec:rules}). It is therefore convenient
to define once and for all the \emph{dressed legs},
\begin{equation}
  \label{eq:lin:dressed}
  \boxed{\ \
  \begin{gathered}
  \bW_n(\beta) \equiv \sum_{a=\pm}e^{a\pi\la/2}r_a\,\Wn^{-a}_n(\beta),\qquad
  \bV(\beta) \equiv \frac{4}{\la}\sum_{a=\pm}e^{a\pi\la/2}r_a\,\Vv^{-a}(\beta),\\[2pt]
  \bP(\beta) \equiv \bW_0(\beta)+\tfrac{\beta}{2}\,\bW_1(\beta),
  \end{gathered}\ \ }
\end{equation}
which are the Schwinger-space bulk-to-boundary propagators of a velocity leg, of a $\sigma$ leg
and of an undifferentiated leg respectively; the constant $4/\la$ in $\bV$ is the one carried by
\cref{eq:lin:Msig}. Two exact consequences of \cref{eq:lin:division,eq:lin:raclosed} are
\begin{equation}
  \label{eq:lin:dressedid}
  \bV(\beta) = \frac{\bW_2(\beta)-\bW_0(\beta)}{\la},\qquad \bW_0(0) = 2R .
\end{equation}

\subsection{Soft behaviour and the exact \texorpdfstring{$\tfrac14$}{1/4} theorem}
\label{sec:lin:tails}

After the reduction of \cref{sec:rules} a leg is evaluated at $\beta_j=2\xi e_j$ with $e_j=k_j/k_t$
and $\xi=\mathcal{O}(1)$; a leg is soft when its momentum is small compared with the total, and
then $\beta_j\ll1$. A small $\beta$ probes the weight at large $u$. Indeed the exponential in
$\int\dd u\,\omega_a(u)\chi(u)e^{-\beta u}$, with $\chi(u)$ the rational function a kernel carries
($(1+2u)^n/(1+u)$ for $\Wn_n$ and $u$ for $\Vv$), cuts the integral off at $u\sim1/\beta$, so that a tail
$u^{-p}$ with $\Real\,p<1$ is not integrable on its own and controls the small-$\beta$ behaviour
alone, \cref{eq:uf:watson}:
\begin{equation}
  \label{eq:lin:softmech}
  \int_0^\infty\!\dd u\;u^{-p}\,e^{-\beta u} \;=\; \Gamma(1-p)\,\beta^{\,p-1} .
\end{equation}
The soft limit of a leg is therefore the large-$u$ tail of $\omega_a$: a mode that leaves the
horizon long before the others is sensitive only to the very high dressed frequencies, and those
are precisely the ones carrying the oscillating powers $u^{-3/2\pm\n}$ of \cref{eq:lin:omegatail}.
This is the mechanism behind the collider signal in this formalism. Applying
\cref{eq:lin:softmech} to \cref{eq:lin:omegatail}, with $p=\tfrac12-b\n$ after the weight
$(1+2u)^2/(1+u)\to4u$ has been used, the $\Gamma(\tfrac12+b\n)$ it produces cancels the one in
\cref{eq:lin:omegatail}, and
\begin{equation}
  \label{eq:lin:tails}
  \boxed{\ \
  \Wn^a_2(\beta) \;\simeq_{\,\beta\to0}\; 4\sum_{b=\pm}\Wcal^a_b\,\beta^{-1/2-b\n},
  \qquad
  \Vv^a(\beta) \;\simeq_{\,\beta\to0}\; \sum_{b=\pm}\Wcal^a_b\,\beta^{-1/2-b\n},
  \ \ }
\end{equation}
with $\Wcal^a_b$ of \cref{eq:lin:Msigbdry}. For a heavy field, $\n=\ii\muf$, both powers have
modulus $\beta^{-1/2}$ and differ only by the phase $\mp\muf\ln\beta$; they are equally leading
and their sum is an oscillation of fixed amplitude in $\ln\beta$, the clock. The complete
small-$\beta$ expansion of a kernel runs over the three families $\beta^{-1/2-\n+n}$,
$\beta^{-1/2+\n+n}$ and $\beta^{n}$ with $n\ge0$, the third being the analytic background.

\paragraph{The $\tfrac14$ theorem.} By \cref{eq:lin:division}, $\Vv^a=\tfrac14(\Wn^a_2-\Wn^a_0)$
identically, and the small-$\beta$ expansion of $\Wn^a_0$ contains the non-analytic powers
$\beta^{3/2-b\n+n}$ only (its integrand tail is $u^{-5/2+b\n}$), i.e.\ exactly the powers of
$\Wn^a_2$ shifted by two units. Therefore \emph{the coefficients of $\beta^{-1/2-b\n}$ and
$\beta^{1/2-b\n}$ in $\Vv^a$ are $\tfrac14$ of those in $\Wn^a_2$, exactly, for every $\la$ and
$\n$}; the two families first differ at relative order $\beta^2$. No expansion in $\la$ and no
asymptotic argument enters. Its physical consequence is the one advertised in
Ref.~\cite{Pinol:2026xnl}: because a
$\sigma$ leg and a velocity leg have proportional soft tails, every squeezed clock, at every
exchange order, sits at the same $\kappa^{1/2-b\n}$ with the single frequency $\muf$, never
$2\muf$ or $3\muf$, since only one external leg can be soft. In terms of the dressed legs,
\begin{equation}
  \label{eq:lin:dressedtails}
  \bW_2(\beta)\;\simeq_{\beta\to0}\;4\sum_{b=\pm}\Wcal_b\,\beta^{-1/2-b\n},\qquad
  \bV(\beta)\;\simeq_{\beta\to0}\;\frac4\la\sum_{b=\pm}\Wcal_b\,\beta^{-1/2-b\n},
\end{equation}
with
\begin{equation}
  \label{eq:lin:Wcaldef}
  \boxed{\ \ \Wcal_b\equiv\sum_{a=\pm}e^{a\pi\la/2}\,r_a\,\Wcal^{-a}_b\ \ ,\ \ }
\end{equation}
where the channel-summed tail coefficient $\Wcal_b$ is a two-term sum of ratios of $\Gamma$
functions: the entire soft-leg content of every squeezed limit in this paper is in closed form, at
every $\la$.

\paragraph{Undifferentiated legs.} The remaining kernels are less singular than $\Wn_2$ by two
units of $\beta$, $\Wn_0\sim\beta^{3/2-b\n}$ and $\Wn_1\sim\beta^{1/2-b\n}$, and $\Wn^a_0(0)=r_a$ is
finite: an undifferentiated leg carries no soft enhancement. It never occurs alone, however, since
$\pic$ enters the cubic Lagrangian only through derivatives, and the gradient vertex delivers its
two curvature legs together with explicit powers of their momenta, collected in the combination
$\Pp_a(\beta)=\Wn^a_0+\tfrac\beta2\Wn^a_1=(1+\tfrac\beta2)\Wn^a_0-\beta\,{\Wn^a_0}'$. The explicit
$\beta/2$ exactly compensates the one-unit gap between the clock tails of $\Wn_0$ and $\Wn_1$, so
that both pieces contribute at order $\beta^{3/2-b\n}$ and partially cancel:
\begin{equation}
  \label{eq:lin:Ptail}
  \Pp_a(\beta)\;=\;r_a+\mathcal{O}(\beta)
  \;+\;\sum_{b=\pm}\frac{\Wcal^a_b}{b\n-\tfrac32}\;\beta^{3/2-b\n}
  \;+\;\mathcal{O}\big(\beta^{5/2-b\n}\big) .
\end{equation}
An undifferentiated pair thus carries the same two clock powers as a velocity leg, weighted by
$(b\n-\tfrac32)^{-1}$ and suppressed by $\beta^{2}$, which is why in the gradient channel the
permutation with the soft $\sigma$ dominates the one with a soft gradient leg
(\cref{sec:squeezed}). Defining the soft index $d$ by $|\mathcal{K}(\beta)|\lesssim\beta^{-d}$ as
$\beta\to0$, a heavy field has $d=\tfrac12$ for $\Wn_2$ and $\Vv$ and $d=0$ for $\Wn_1$, $\Wn_0$
and $\Pp$ (for real $\n$, $d=\tfrac12+\n$, $\max(0,\n-\tfrac12)$ and $\max(0,\n-\tfrac32)$
respectively); the soft indices govern the convergence of the Schwinger integral
(\cref{sec:rules:contour}).

\subsection{Weak mixing}
\label{sec:lin:weak}

Weak mixing means $\la\to0$. Whether the bare or the effective mass is held fixed does not matter
for what follows: since $\n^2=\nbare^2-\la^2$, the two indices differ at relative
$\mathcal{O}(\la^2)$, which is beyond the first non-vanishing order of every quantity below, the
$\mathcal{O}(\la^0)$ curvature legs, the $\mathcal{O}(\la)$ isocurvature leg and the
$\mathcal{O}(\la^2)$ tail coefficients alike. We therefore write the bare index $\nbare$ throughout this
subsection and, for a heavy field, the bare mass parameter $\mu\equiv-\ii\nbare$, the
convention of the other weak-mixing parts of the paper (\cref{sec:sq:weak}, \cref{app:V2,app:double,app:dictionaries}).
The order in which $\la\to0$ is taken relative to a soft momentum
requires some care: we shall see that at weak mixing it does not affect the coefficient of a
non-analytic power, only the window in $\beta$ over which the soft form is accurate, and we treat
the legs at a generic argument and at a soft one in turn.

\paragraph{General statements.} Two facts organise everything. The first is that the channel index
and the mixing strength enter every kernel only through $z_a=\ii a\la/2$, so that each one-channel
kernel of \cref{eq:lin:kernels}, $\mathcal{K}^a\in\{\Wn^a_n,\Vv^a\}$, is one analytic function of a
single variable evaluated at $z_a$,
\begin{equation}
  \label{eq:lin:Kexpand}
  \mathcal{K}^a(\beta)=\mathcal{K}_0(\beta)+z_a\mathcal{K}_1(\beta)+z_a^2\mathcal{K}_2(\beta)
  +\mathcal{O}(z_a^3) ,
\end{equation}
$\mathcal{K}_n$ denoting the coefficient of $z_a^n$, real and independent of the channel, so that the
parity of each term under $a\to-a$ is the parity of its order. The second is that the
weight collapses onto the origin. Let $\varphi$ be analytic at $u=0$ and decaying at infinity;
splitting the integral at $u=1$ and subtracting $\varphi(0)$ on the left piece,
\begin{equation}
  \label{eq:lin:distrib}
  \begin{gathered}
  \frac{1}{\Gamma(z)}\int_0^\infty\!\dd u\;u^{z-1}\varphi(u)
  = \frac{1}{\Gamma(z)}\Big[\frac{\varphi(0)}{z}
    + \underbrace{\int_0^1\!\dd u\,u^{z-1}\big(\varphi(u)-\varphi(0)\big)
    + \int_1^\infty\!\dd u\,u^{z-1}\varphi(u)}_{\text{finite at }z=0}\Big]\\[2pt]
  \xrightarrow[\;z\to0\;]{}\; \varphi(0) ,
  \end{gathered}
\end{equation}
because $1/\Gamma(z)=z+\gamma z^2+\mathcal{O}(z^3)$ turns the pole into $\varphi(0)$ and kills the
two finite pieces. In other words, on functions analytic at the origin,
\begin{equation}
  \label{eq:lin:deltalimit}
  \frac{u^{z_a-1}}{\Gamma(z_a)}\;\xrightarrow[\ \la\to0\ ]{}\;\delta(u) ,
\end{equation}
and since $\omega_a$ of \cref{eq:lin:omega} is $u^{z_a-1}/\Gamma(z_a)$ times such a
function, every $u$-integral of \cref{sec:lin:kernels} tends at $\la\to0$ to its integrand at
$u=0$: $\mathcal{K}_0(\beta)=\chi(0)$, with $\chi$ the rational weight of the kernel,
$(1+2u)^n/(1+u)$ for $\Wn_n$ and $u$ for $\Vv$. Finally, the external leg factor
$e^{a\pi\la/2}r_a$, with $r_a$ of \cref{eq:lin:raclosed}, is also a function of $z_a$ alone, and
expands as $e^{a\pi\la/2}r_a=1+a\la\varrho_1+\mathcal{O}(\la^2)$ with
\begin{equation}
  \label{eq:lin:rfrak}
  \varrho_1 = \frac{\pi}{2}+\frac{\ii}{2}\Big[\psi(\tfrac12)-\psi(\tfrac34-\tfrac\nbare2)-\psi(\tfrac34+\tfrac\nbare2)\Big],
\end{equation}
whose real part is $\pi/2$ for every $\nbare$. Multiplying the two expansions, with $\mathcal{K}^{-a}$
evaluated at $z_{-a}=-z_a$, and summing over the channels with $\sum_a1=2$, $\sum_aa=0$,
$\sum_aa^2=2$, the dressed combination of \cref{eq:lin:dressed} is
\begin{equation}
  \label{eq:lin:universalleg}
  \sum_{a=\pm}e^{a\pi\la/2}r_a\,\mathcal{K}^{-a}(\beta)
  \;=\; 2\,\mathcal{K}_0(\beta)\big[1+\mathcal{O}(\la^2)\big]
  \;+\;\la^2\Big[-\ii\varrho_1\,\mathcal{K}_1(\beta)
  -\tfrac12\mathcal{K}_2(\beta)\Big] \;+\;\mathcal{O}(\la^4) .
\end{equation}
The two written terms of order $\la^2$ are products of pieces of equal parity: $\mathcal{K}_2$
against the leading $1$ of the external leg, and the two odd pieces, $a\la\varrho_1$ from the
external leg and $-z_a\mathcal{K}_1$ from the kernel, whose product survives the sum because
$a^2=1$; the odd pieces alone are annihilated, so there is no term linear in $\la$, and since the
summand is invariant under $(a,\la)\to(-a,-\la)$ the sum is even in $\la$, which is why
weak-mixing coefficients in this problem come out even in $\la$. Two cases follow, and they are the
two kinds of leg. If $\mathcal{K}_0\ne0$ the answer is $2\mathcal{K}_0$, and everything else,
including the $\mathcal{O}(\la^2)$ term of $e^{a\pi\la/2}r_a$ that multiplies $\mathcal{K}_0$ and that we
have not written, is a relative $\mathcal{O}(\la^2)$ correction we never need. If $\mathcal{K}_0=0$
that term is absent and the leading term is the bracket, in which the even part $\mathcal{K}_2$ and
the odd part $\mathcal{K}_1$ contribute at the same order: this is the case of an isocurvature leg,
whose weight $\chi(u)=u$ vanishes at the origin.

\paragraph{The legs at a generic argument.} We now take $\la\to0$ with $\beta$ held fixed, which is
the relevant order when no momentum is soft: $\la$ is smaller than every dimensionless ratio in the
kinematics. For a curvature kernel the weight is $1$ at the origin, $\Wn^a_n(\beta)=1+\mathcal{O}(z_a)$
by \cref{eq:lin:distrib}, and the two channels of \cref{eq:lin:universalleg} add, so that
\begin{equation}
  \label{eq:lin:W2chan}
  \bW_n(\beta) = 2 + \mathcal{O}(\la^2),\qquad \bP(\beta) = 2\big(1+\tfrac\beta2\big)+\mathcal{O}(\la^2) ,
\end{equation}
uniformly in $\beta$: at leading order a curvature leg does not feel the mixing at all, and the
factor $2$ is the channel sum.

An isocurvature leg is the second case. Since its weight $u$ vanishes at the origin, $\Vv_0=0$
and the expansion \cref{eq:lin:Kexpand} of $\Vv^a$ starts at order $z_a$, with the free rotated
mode as its first coefficient,
\begin{equation}
  \label{eq:lin:V0}
  \boxed{\ \ \Vv_1(\beta) = \int_0^\infty\!\dd u\;\hyp\big(\tfrac12-\nbare,\tfrac12+\nbare;1;-u\big)\,e^{-\beta u}
  = \frac{e^{\beta/2}}{\sqrt{\pi\beta}}\,K_{\nbare}\!\Big(\frac{\beta}{2}\Big) , \ \ }
\end{equation}
the Bessel form being the solution of the second-order equation $\mathcal{L}\Vv_1=0$ of
\cref{app:V2} that decays at large $\beta$, normalised by its soft tail $G_b$ of
\cref{eq:lin:Wcalweak}; it is the Hankel function of the free massive mode
$u_\sigma(x)\propto x^{3/2}H^{(1)}_{\nbare}(x)$ at the imaginary argument of the rotated contour of
\cref{sec:rules:contour}, $K_{\nbare}$ being $H^{(1)}_{\nbare}$ on the imaginary axis. At this order the only trace of the mixing is the factor $z_a$ left by
$u^{z_a}/\Gamma(z_a)$ in the weight, and $\Vv_1$ is the $\sigma$ mode with no insertion. The second coefficient collects the $z_a$-dependence of
$(1+u)^{-z_a}$, of the third parameter of the $\hyp$ and of $1/\Gamma(z_a)$, and it is real and in
closed form,
\begin{equation}
  \label{eq:lin:V2closed}
  \boxed{\ \
  \begin{gathered}
  \Vv_2(\beta)=-\frac{\pi}{\sin\pi\nbare}\,\frac{e^{\beta/2}}{\sqrt{\pi\beta}}
  \big[\Psi_+\,I_{-\nbare}\big(\tfrac\beta2\big)-\Psi_-\,I_{\nbare}\big(\tfrac\beta2\big)\big]
  +\frac{4}{1-4\nbare^2}\,{}_2F_2\big(1,1;\tfrac32-\nbare,\tfrac32+\nbare;\beta\big) ,\\[4pt]
  \Psi_b\equiv\psi\big(\tfrac12+b\nbare\big)+\tfrac\gamma2 ,
  \end{gathered}\ \ }
\end{equation}
derived in \cref{app:V2} as the solution of the second-order equation that $\Vv_1$ obeys, with a
constant source: the first term is $\Vv_1$ with its two Bessel halves re-weighted by $-2\Psi_+$
and $-2\Psi_-$ instead of equally, and the second is entire. With $\mathcal{K}_0=0$, $\mathcal{K}_1=\Vv_1$
and $\mathcal{K}_2=\Vv_2$ in \cref{eq:lin:universalleg}, and the leg constant $4/\la$ of
\cref{eq:lin:dressed}, the dressed isocurvature leg is the sum of its two parities,
\begin{equation}
  \label{eq:lin:sigmapower}
  \boxed{\ \
  \bV(\beta) \;=\; 4\la\Big[\underbrace{-\,\tfrac12\Vv_2(\beta)}_{\text{even}\times\text{even}}
   \;\underbrace{-\,\ii\varrho_1\Vv_1(\beta)}_{\text{odd}\times\text{odd}}\Big]
   \;+\;\mathcal{O}(\la^3) .\ \ }
\end{equation}
Together with \cref{eq:lin:W2chan},
\begin{equation}
  \label{eq:lin:legcount}
  \text{(dressed leg)} \;=\;
  \begin{cases}
    \mathcal{O}(\la^0), & \dot\pic \text{ or undifferentiated } \pic,\\
    \mathcal{O}(\la^1), & \sigma ,
  \end{cases}
  \qquad \text{as } \la\to0 ,
\end{equation}
the perturbative counting of one mixing insertion per $\sigma$ leg, reproduced with no diagrammatic
input.

\paragraph{The legs at a soft argument.} The squeezed limit takes $\beta\to0$ first, at fixed $\la$,
and only then expands in the mixing. That order is available in closed form, because the soft
behaviour \cref{eq:lin:tails} is exact in $\la$: it is enough to expand its coefficient. With
$1/\Gamma(z_a)=z_a(1+\gamma z_a)$ and
$1/\Gamma(\tfrac12+b\nbare+2z_a)=[1-2z_a\psi(\tfrac12+b\nbare)]/\Gamma(\tfrac12+b\nbare)$ in
\cref{eq:lin:Msigbdry},
\begin{equation}
  \label{eq:lin:Wcalweak}
  \Wcal^a_b = G_b\Big[z_a-2z_a^2\,\Psi_b\Big]+\mathcal{O}(z_a^3),
  \qquad
  G_b\equiv\frac{\Gamma(2b\nbare)}{\Gamma(\tfrac12+b\nbare)}=\frac{2^{2b\nbare-1}}{\sqrt\pi}\,\Gamma(b\nbare) ,
\end{equation}
$G_b$ being the coefficient with which the power $\beta^{-1/2-b\nbare}$ appears in $\Vv_1$, and the two
terms having, again, opposite parities in $a$. The two orders agree where they overlap: the channel
sum \cref{eq:lin:universalleg} turns \cref{eq:lin:Wcalweak} into
$\Wcal_b=\sum_ae^{a\pi\la/2}r_a\Wcal^{-a}_b=\la^2G_b[\Psi_b-\ii\varrho_1]+\mathcal{O}(\la^4)$, and
the
soft tail of \cref{eq:lin:sigmapower}, read off \cref{eq:lin:V0,eq:lin:V2closed}, gives the same
coefficient (\cref{app:V2}). The combination $\Psi_b-\ii\varrho_1$ is the scalar that organises
every weak-mixing clock coefficient of \cref{sec:squeezed}. At weak
mixing the order of limits thus does not matter for the coefficient of a non-analytic power; what
depends on it is how small $\beta$ must be for the soft form to be accurate, since the analytic
background of a leg is $\mathcal{O}(\la^0)$ for a curvature leg and $\mathcal{O}(\la)$ for an
isocurvature one.

\subsection{Strong mixing}
\label{sec:lin:strong}

Strong mixing means $\la\gg1$, and we start by assuming no hierarchy between $\la$ and $\muf$: the
two are independent coordinates of the quarter plane, drawn throughout this paper with $\muf$
horizontal and $\la$ vertical, and a given Lagrangian with fixed bare mass $m$ moves along
$\muf=\sqrt{\la^2+m^2/H^2-9/4}$, which approaches the diagonal $\la=\muf$ from the right as $\la$
grows; the fixed-effective-mass line of Ref.~\cite{Pinol:2026xnl} is a vertical in the same plane. We write
\begin{equation}
  \label{eq:lin:delta}
  \delta\equiv\la-\muf ,
\end{equation}
so that the diagonal is $\delta=0$ and fixed effective mass is $\delta\to\infty$. Three quantities
must be sized before the bispectra of \cref{sec:bisp:strong,sec:sq:strong} can be taken to strong
mixing: the amplification $R$, the soft-leg coefficient $\Wcal_b$, and the argument at which a hard
kernel is sampled. The first is the exact \cref{eq:lin:Rplane}.

\begin{figure}[t]
\centering
\includegraphics[width=\textwidth]{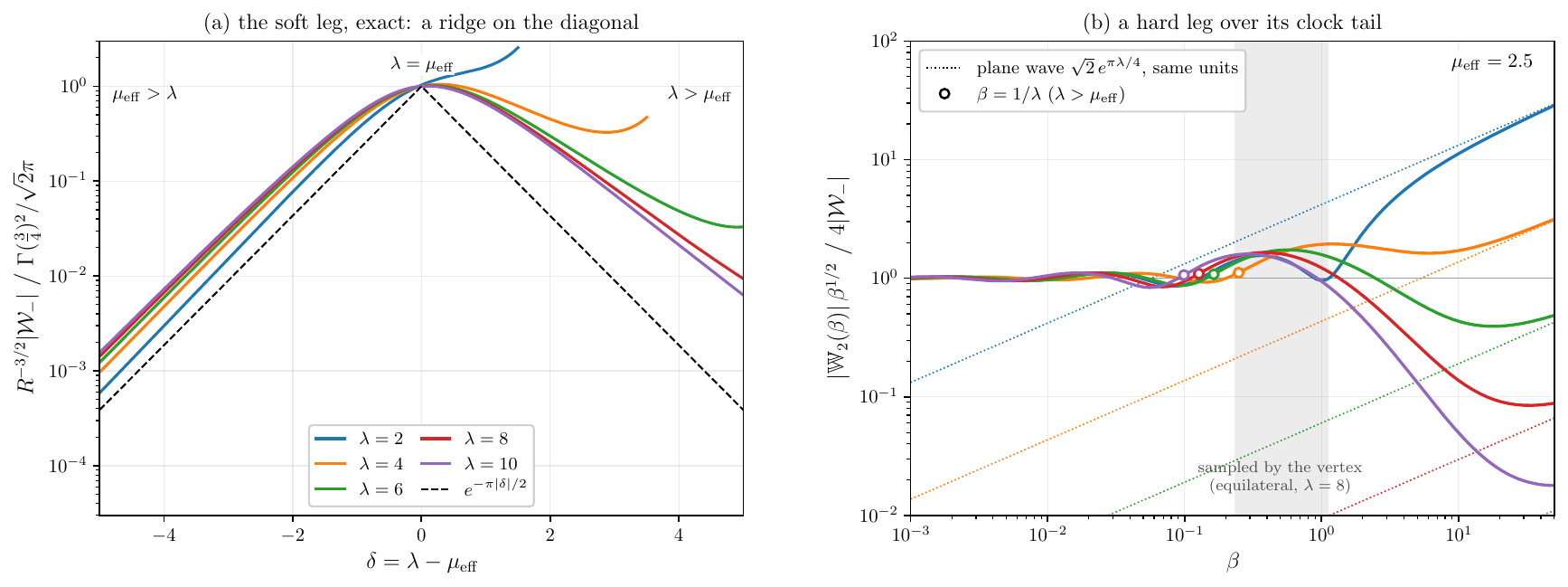}
\caption{The two ingredients of the strong-mixing bispectrum, from the linear theory. (a) The soft
leg normalised as a bispectrum, $R^{-3/2}|\Wcal_-|$, from the exact \cref{eq:lin:Wplane,eq:lin:Rplane},
against $\delta=\la-\muf$ for $\la=2,4,6,8,10$ (each curve ends where $\muf=1/2$): the curves collapse onto
a ridge of height $\Gamma(\tfrac34)^2/\sqrt2\pi$ on the diagonal, with flanks $e^{-\pi|\delta|/2}$,
\cref{eq:lin:ridge}, and rise again only where $\muf$ is no longer large. (b) A hard leg over its
clock tail: the dressed velocity leg divided by the dominant branch of its tail,
$|\bW_2(\beta)|\beta^{1/2}/4|\Wcal_-|$, against its argument at $\muf=2.5$, for the same five mixings
and in the same colours, from the double-precision engine. The ratio is unity, within the subleading
terms of the tail, over the range a hard leg is sampled on (shaded: the central four fifths of the
equilateral no-exchange integrand at $\la=8$); the leg leaves its tail at $\beta\sim1/\la$ (circles)
and tends at large argument towards the sub-horizon plane wave $\sqrt2\,e^{\pi\la/4}$, drawn dotted
in the same units for each mixing.}
\label{fig:lin:ingredients}
\end{figure}

\paragraph{The soft leg.} Each per-channel tail coefficient has an elementary modulus,
\begin{equation}
  \label{eq:lin:Wcalmod}
  \big|\Wcal^a_b\big|^2 = \frac{\la^2}{4\muf}\;
  \frac{\sinh(\pi\la/2)\,\cosh\pi(a\la+b\muf)}{\sinh(\pi\la)\,\sinh(2\pi\muf)}
  \qquad(\n=\ii\muf),
\end{equation}
by \cref{eq:uf:gammamod}, since every $\Gamma$ in $\Wcal^a_b$ has an argument of the form $\ii y$,
$1+\ii y$ or $\tfrac12+\ii y$; the ratio $|\Wcal^+_b/\Wcal^-_b|=[\cosh\pi(\la+b\muf)/\cosh\pi(\la-b\muf)]^{1/2}$
is a free per-point test of an implementation, insensitive to every normalisation and off by order
one if the channel index sits on the wrong factor. What the bispectra need is the channel sum
\cref{eq:lin:dressedtails}, and it too is elementary:
\begin{equation}
  \label{eq:lin:Wplane}
  \boxed{\ \
  \big|\Wcal_b\big| = \la\,\sinh\!\Big(\frac{\pi\la}{2}\Big)\,e^{-b\pi\muf/2}
  \sqrt{\frac{R}{\muf\,\sinh(2\pi\muf)}}\ ,
  \qquad
  \Big|\frac{\Wcal_+}{\Wcal_-}\Big|=e^{-\pi\muf} ,\ \ }
\end{equation}
an identity at every $(\la,\muf)$ (\cref{app:numerics}). It says three things. The two branches of the soft leg are separated by the full Boltzmann factor
$e^{-\pi\muf}$ everywhere, not asymptotically and not only on one line. The whole dependence on the
mixing sits in the prefactor $\la\sinh(\pi\la/2)$ and in $R$, so that the soft leg normalised to the
spectrum is free of $R$,
\begin{equation}
  \label{eq:lin:softfree}
  R^{-1/2}\big|\Wcal_-\big|=\frac{\la\,\sinh(\pi\la/2)\,e^{\pi\muf/2}}{\sqrt{\muf\,\sinh2\pi\muf}}
  \;\xrightarrow[\ \la,\muf\gg1\ ]{}\;\frac{\la}{\sqrt{2\muf}}\,e^{\pi\delta/2} ,
\end{equation}
which carries the exponential $e^{\pi(\la-\muf)/2}$ exactly, and whose limit requires only that
$\la$ and $\muf$ be large, whatever their ratio: the soft leg grows with the mixing and pays the
mass, and the two cancel on the diagonal. And, divided by $R^{3/2}$ instead, which is the
normalisation of a bispectrum, the soft leg is a ridge along the diagonal,
\begin{equation}
  \label{eq:lin:ridge}
  \begin{gathered}
  R^{-3/2}\big|\Wcal_-\big|\;\underset{\la,\,\muf\gg1,\ |\delta|\gtrsim1}{\simeq}\;
  \frac{\la}{\sqrt2}\,\frac{\sqrt{|\la^2-\muf^2|}}{\muf^{3/2}}\,e^{-\pi|\delta|/2} ,\\[3pt]
  R^{-3/2}\big|\Wcal_-\big|\Big|_{\delta=0}\;\xrightarrow[\ \la\to\infty\ ]{}\;\frac{\Gamma(\tfrac34)^2}{\sqrt2\,\pi}=0.337989\ldots ,
  \end{gathered}
\end{equation}
of $\la$-independent height and with flanks $e^{-\pi|\delta|/2}$ on both sides
(\cref{fig:lin:ingredients}a). The flank form follows from \cref{eq:lin:Rplane} with all three
$\Gamma$'s expanded, which requires $|\delta|\gtrsim1$; on the layer $|\delta|\lesssim1$ the factor
$\Gamma(\tfrac34+\tfrac{\ii\delta}{2})$ must be kept exact, which replaces the vanishing
$\sqrt{|\la^2-\muf^2|}$ by the finite height of the ridge. The form fails once more at the right
end of the curves of \cref{fig:lin:ingredients}a, where $\muf=\la-\delta$ becomes of order one and
the $\Gamma$ functions of $\muf$ take over: the ridge is a statement about $\la,\muf\gg1$, not
about large $\delta$ at fixed $\la$.

\paragraph{A hard leg, and where it is sampled.} The two other legs of a bispectrum are hard,
$e_j=\mathcal{O}(1)$: after the reduction of \cref{sec:rules} they are evaluated at $\beta_j=2\xi e_j$
and integrated against the vertex measure $\xi^Ne^{-\xi}$, so that what matters is the
$\beta$-dependence of a dressed kernel over the range the vertex samples. That range is not
$\mathcal{O}(1)$ at strong mixing: because the kernels grow as $\beta^{-1/2}$ towards small
argument, a hard leg is sampled at $\beta\lesssim1$ and, for $\la\gg\muf$, at $\beta\propto1/\la$.

Over that range a hard leg is its own clock tail. \Cref{fig:lin:ingredients}b shows the dressed
velocity leg divided by the dominant branch of its tail, $|\bW_2(\beta)|\beta^{1/2}/4|\Wcal_-|$, at
$\muf=2.5$ for five mixings: over the range the vertex samples the ratio stays within a factor
$1.6$ of unity for every $\la$, the excursions being the subleading terms of the tail expansion; at
larger argument, on a scale of order $1/\la$ times a number that grows with the mass (circles at
$\beta=1/\la$), the leg leaves its tail and decays towards the sub-horizon plane wave
$\Wn_n^a(\beta)\simeq\beta^{-\ii a\la/2}$, $R^{-1/2}|\bW_2|\to\sqrt2\,e^{\pi\la/4}$, where half the
Boltzmann weight of the boundary end survives the normalisation and no mass dependence remains.
The estimate of a hard leg over the range the vertex samples is
therefore the tail itself, with the exact prefactor of \cref{eq:lin:softfree},
\begin{equation}
  \label{eq:lin:hardleg}
  R^{-1/2}\big|\bW_2(\beta)\big|\;\simeq\;4R^{-1/2}|\Wcal_-|\,\beta^{-1/2}
  \;\xrightarrow[\ \la,\muf\gg1\ ]{}\;\frac{4\la}{\sqrt{2\muf}}\,e^{\pi\delta/2}\,\beta^{-1/2}
  \qquad(\beta\lesssim1) ,
\end{equation}
so that a hard leg carries the same $e^{\pi(\la-\muf)/2}$ as the soft leg and one power of $\la$, and
nothing of the plane-wave regime, which sets in far beyond the sampled range. Only the tail
carries the clock, and it is read through the vertex integral, which sits at the time at which the
heavy quanta are produced, $k/a\sim\meff$, and pays their Boltzmann factor there
(\cref{sec:sq:strong}). The two panels of \cref{fig:lin:ingredients} are thus the two ingredients of
the strong-mixing collider signal: the exact soft leg of panel (a), and the two hard legs of panel
(b), each on its tail and each carrying $e^{\pi\delta/2}$, integrated against the vertex, where those
two exponentials cancel against each other between the two clock branches and leave a power of
$\la$ that \cref{sec:sq:strong} measures.

\paragraph{An isocurvature leg is a velocity leg over the mixing.} For $\la\gg\muf$ the vertex
samples $\beta_j\lesssim1$ on \emph{all three} legs at \emph{any} configuration, the equilateral one
included, as just shown. In that region the exact division identity \cref{eq:lin:division} may be
read backwards,
\begin{equation}
  \label{eq:lin:quarter}
  \Vv^a(\beta)=\tfrac14\Wn^a_2(\beta)\Big[1-\frac{\Wn_0^a(\beta)}{\Wn_2^a(\beta)}\Big],
  \qquad
  \Big|\frac{\Wn^a_0}{\Wn^a_2}\Big|\;\simeq\;\frac{\muf}{\sqrt2}\sqrt{\frac{\beta}{\la^3}}
  \quad\text{for }\beta\lesssim1 .
\end{equation}
For $\la\gg\muf$, then, an isocurvature leg is a velocity leg divided by the mixing,
\begin{equation}
  \label{eq:lin:VoverW}
  \bV\;\longrightarrow\;\bW_2/\la
  \qquad\text{up to a relative }\mathcal{O}\big(\la^{-3/2}\beta^{1/2}\big),
\end{equation}
which, since the vertex samples $\beta\propto1/\la$, is a relative $\mathcal{O}(\la^{-2})$ on every
integrated quantity, as \cref{sec:bisp:strong} measures; this is all that section needs to relate
the exchange channels to the no-exchange one in that region.

\paragraph{Soft and strong do not commute.} A leg is its clock tail only for $\beta\lesssim1/\la$,
a window that shrinks with the mixing; a leg is soft in the sense of \cref{sec:lin:tails} only when
its argument lies inside it. Which of the two limits, $\beta\to0$
or $\la\to\infty$, is taken first therefore matters, and we state the order wherever we use them.

\section{Diagrammatic rules for the resummed theory}
\label{sec:rules}

\tikzset{
  bpt/.style  = {draw, fill=black, rectangle, inner sep=2.0pt},
  vp/.style   = {draw, fill=black, circle,    inner sep=2.3pt},
  vm/.style   = {draw, fill=white, circle,    inner sep=2.3pt},
  vs/.style   = {draw, fill=white, circle,    inner sep=2.3pt, dash pattern=on 1.3pt off 1.0pt},
  pil/.style  = {thick},
  sigl/.style = {thick, dash pattern=on 3pt off 2.2pt},
  jct/.style  = {fill=black, circle, inner sep=0.55pt},
}
\newcommand{\mixedline}[2]{%
  \draw[pil]  (#1) -- ($(#1)!0.5!(#2)$);
  \draw[sigl] ($(#1)!0.5!(#2)$) -- (#2);
  \node[jct] at ($(#1)!0.5!(#2)$) {};}

The rules of this section are the Schwinger--Keldysh (SK) rules of Ref.~\cite{Chen:2017ryl} with
the free propagators of $\pic$ and $\sigma$ replaced by the dressed lines of \cref{sec:linear}.
Three structural features distinguish the resummed formulation. The propagators are mixed: because
the quadratic mixing sits in the free Hamiltonian, $\pic$ and $\sigma$ are expanded on the same pair
of oscillators, \cref{eq:lin:expansion}, so that $[\hat\pic,\hat\sigma]\neq0$ even for
interaction-picture fields and a single Wightman function contracts any pair of fields; there are
three independent pairings and no undressed $\sigma$ line. The boundary is a $\pic$ boundary: every
propagator end at the final slice contributes the number $\ii r_a$ (\cref{sec:lin:bogcoeff}). And
there is no quadratic vertex left to insert: what is conventionally called an exchange diagram is
here a contact diagram, and at tree level a bispectrum has exactly one vertex, the exchange order
being the number of $\sigma$ legs attached to it. Bulk-to-bulk lines appear from quartic order on
and in loops.

\subsection{Propagators}
\label{sec:rules:props}

For $X,Y\in\{\pic,\sigma\}$, the mode expansion \cref{eq:lin:expansion} with the channel
diagonalisation \cref{eq:lin:diag} gives the dressed Wightman function
\begin{equation}
  \label{eq:rules:wightman}
  \begin{gathered}
  \Dw^{XY}_k(\tau_1,\tau_2)\;\equiv\;\langle X_{\vec k}(\tau_1)\,Y_{-\vec k}(\tau_2)\rangle'
  \;=\;\frac{H^2}{2k^3}\sum_{a=\pm}\frac{e^{a\pi\la/2}}{2}\;
  \Ma^X_a(x_1)\,\big[\Ma^Y_a(x_2)\big]^{*},\\[2pt]
  \big[\Dw^{XY}_k(\tau_1,\tau_2)\big]^{*} = \Dw^{YX}_k(\tau_2,\tau_1),
  \end{gathered}
\end{equation}
with $x_i=-k\tau_i$ and the prime denoting the stripped momentum delta. The four pairings are built
from the same two channel modes, and the sum over $a$ is a sum over two decoupled species with
Boltzmann weights $e^{\pm\pi\la/2}$. Two immediate checks: $\Dw^{\pi\pi}_k(0,0)=(H^2/2k^3)R$
reproduces the power spectrum \cref{eq:lin:R}, and $\Dw^{\sigma Y}_k(0,\tau_2)=0$ is the statement
that the boundary is a $\pic$ boundary. Because $\Dw^{XY}$ is not symmetric under
$X\leftrightarrow Y$ at fixed times, a line must record which field sits at which end, and we
decorate the two half-lines separately, solid for $\pic$ and dashed for $\sigma$. Bulk points
carry the SK branch of their vertex, $+$ time-ordered and $-$ anti-time-ordered, and we draw a
filled circle for a $+$ vertex, an open circle for a $-$ vertex, and a dashed circle for a vertex
summed over both branches:
\begin{center}
\begin{tikzpicture}[baseline=-2pt]
  \node[vp](p) at (0,0){}; \node at (0.55,0){\footnotesize $+$};
  \node[vm](m) at (1.6,0){}; \node at (2.15,0){\footnotesize $-$};
  \node[vs](s) at (3.2,0){}; \node[anchor=west] at (3.45,0){\footnotesize $\textstyle\sum_{\pm}$};
\end{tikzpicture}
\end{center}
Labelling by the branches of the two vertices, the four branch propagators are
\begin{equation}
  \label{eq:rules:SK}
  \begin{gathered}
  \Gsk^{XY}_{-+} = \Dw^{XY}_k(\tau_1,\tau_2), \qquad
  \Gsk^{XY}_{+-} = \Dw^{YX}_k(\tau_2,\tau_1), \\[2pt]
  \Gsk^{XY}_{++} = \theta(\tau_1-\tau_2)\,\Dw^{XY}_k(\tau_1,\tau_2)
                 + \theta(\tau_2-\tau_1)\,\Dw^{YX}_k(\tau_2,\tau_1),
  \end{gathered}
\end{equation}
with $\Gsk_{--}=[\Gsk_{++}]^{*}$ and $\Gsk_{+-}=[\Gsk_{-+}]^{*}$; the three line types, drawn
here with the branch assignment of $\Gsk_{-+}$, i.e.\ with the Wightman function
\cref{eq:rules:wightman} read from left to right, are
\begin{center}
\begin{tikzpicture}[baseline=-2pt]
  \node[vm](a) at (0,0){}; \node[vp](b) at (2.1,0){};
  \draw[pil](a)--(b); \node at (1.05,-0.42){\footnotesize $\Gsk^{\pi\pi}_{-+}=\Dw^{\pi\pi}$};
  \node[vm](c) at (4.3,0){}; \node[vp](d) at (6.4,0){};
  \draw[sigl](c)--(d); \node at (5.35,-0.42){\footnotesize $\Gsk^{\sigma\sigma}_{-+}=\Dw^{\sigma\sigma}$};
  \node[vm](e) at (8.6,0){}; \node[vp](f) at (10.7,0){};
  \mixedline{e}{f} \node at (9.65,-0.42){\footnotesize $\Gsk^{\pi\sigma}_{-+}=\Dw^{\pi\sigma}$};
  \node at (8.6,0.38){\footnotesize $\pic$}; \node at (10.7,0.38){\footnotesize $\sigma$};
\end{tikzpicture}
\end{center}
There is no fourth line type, since $\Dw^{\sigma\pi}$ is the mixed line read from the other end.

\paragraph{Bulk-to-boundary propagators.} An external operator sits on the final slice
$\tau=0$, where the isocurvature mode vanishes, $\Ma^\sigma_a(x)\propto x^{3/2\mp\n}$ by
\cref{eq:lin:Msigbdry}: a $\sigma$ end at the boundary is suppressed by powers of $(-k\tau)$
relative to a $\pic$ end, whose mode tends to the finite $\ii r_a$, so that at $\tau\to0$ every
external line is a $\pic$ line. We draw the boundary end as a square, which is therefore always
solid, and label the line by the branch of the bulk vertex only, since a boundary point carries no
branch label; this leaves two bulk-to-boundary propagators per bulk field $Y$:
\begin{center}
\begin{tikzpicture}[baseline=-2pt]
  \node[bpt](a) at (0,0){}; \node[vp](b) at (2.1,0){};
  \draw[pil](a)--(b); \node at (1.05,-0.42){\footnotesize $\Kb^{\pi}_{+}$};
  \node[bpt](c) at (3.4,0){}; \node[vm](d) at (5.5,0){};
  \draw[pil](c)--(d); \node at (4.45,-0.42){\footnotesize $\Kb^{\pi}_{-}$};
  \node[bpt](e) at (6.8,0){}; \node[vp](f) at (8.9,0){};
  \mixedline{e}{f} \node at (7.85,-0.42){\footnotesize $\Kb^{\sigma}_{+}$};
  \node[bpt](g) at (10.2,0){}; \node[vm](h) at (12.3,0){};
  \mixedline{g}{h} \node at (11.25,-0.42){\footnotesize $\Kb^{\sigma}_{-}$};
\end{tikzpicture}
\end{center}
\begin{equation}
  \label{eq:rules:b2b}
  \Kb^{Y}_{+}(k;\tau') = \Dw^{\pi Y}_k(0,\tau')
  = \frac{H^2}{4k^3}\sum_{a=\pm}e^{a\pi\la/2}\,(\ii r_a)\,\big[\Ma^Y_a(x')\big]^{*},
  \qquad
  \Kb^{Y}_{-} = \big[\Kb^{Y}_{+}\big]^{*} ,
\end{equation}
$Y$ being the field at the bulk vertex, so that $\Kb^\sigma_\pm$ is the mixed line with its dashed
half at the vertex. The dressed legs of \cref{eq:lin:dressed} are these propagators after the
Schwinger reduction below. The would-be $\Kb$ with a $\sigma$ at the boundary, $\Dw^{\sigma Y}_k(0,\tau')$,
vanishes at $\tau=0$, and at a late but finite time it is down by $(-k\tau)^{3/2-|\Real\n|}$ with
the coefficients $\Wcal^a_b$ of \cref{eq:lin:Msigbdry}; it would enter only a correlator of
$\sigma$ itself, which is not observed.

\subsection{Vertices}
\label{sec:rules:vertices}

With \cref{eq:eft:HintI}, a term $c\,\mathcal{O}$ in
$H_{\rm int}/a^3$ gives the pair of vertices
\begin{equation}
  \label{eq:rules:vertexrule}
  V_{\pm} = \mp\,\ii\,c\;n_{\rm W}\int_{-\infty}^{0}\!\dd\tau'\;a(\tau')^{q}\;
  \big[\text{momentum factors}\big]\;\big[\text{attached propagators}\big] ,
\end{equation}
the filled and open circles of \cref{sec:rules:props}, where $n_{\rm W}$ is the multiplicity of the Wick
contraction ($3!$ for three distinct external momenta attached to $\dot\pic^{\,3}$, $2!$ for the two
identical $\dot\pic$ of $\dot\pic^{\,2}\sigma$, and so on) and $a^{q}$ is the scale factor left after
reducing $\int\dd t\,a^3\mathcal{O}$ to conformal time, $\int\dd t\,a^3=\int\dd\tau\,a^4$, and
converting each time derivative to $a^{-1}\partial_\tau$ and each gradient pair to
$a^{-2}\partial_i\partial_i$:
\begin{equation}
  \label{eq:rules:q}
  q = 4 - n_{\partial_t} - 2\,n_{\partial_i} .
\end{equation}
Thus $q=1$ for $\dot\pic^{\,3}$, $q=2$ for $\dot\pic^{\,2}\sigma$ and for
$(\partial_i\pic)^2\sigma/a^2$, $q=3$ for $\dot\pic\sigma^2$ and $q=4$ for $\sigma^3$. Each spatial
derivative acting on a leg of momentum $\vec k_j$ becomes $\ii\vec k_j$, so a gradient pair gives
$-\vec k_b\cdot\vec k_c$; each $\partial_\tau$ acts on the attached propagator, including, for
$\Gsk_{\pm\pm}$, on the step functions, whose equal-time terms are the usual local contributions and
must be kept \cite{Chen:2017ryl}; they are absent from every single-vertex diagram. A time
derivative and a gradient pair therefore do different things: the first lowers $q$ by one and
promotes the kernel of the attached leg, the second lowers $q$ by two, leaves the propagator
untouched and delivers a momentum factor. The $\dot\pic^{\,2}\sigma$ and
$(\partial_i\pic)^2\sigma/a^2$ vertices are drawn identically and differ precisely here.

\subsection{Reduction to leg kernels and the master formula}
\label{sec:rules:reduction}

Writing $x'=-k_j\tau'$ and using \cref{eq:lin:Mpi,eq:lin:Mpiprime,eq:lin:Msig}, each bulk end of a
$+$ line takes the universal form
\begin{equation}
  \label{eq:rules:legform}
  \big[D\,\Ma^{Y}_{a}(x')\big]^{*}
  \;=\;\ell\;k_j^{\,m}\;(-\tau')^{\,p}\int_0^\infty\!\dd u_j\;\omega_{-a}(u_j)\;g(u_j)\;
       e^{\,\ii k_j\tau'(1+2u_j)} ,
\end{equation}
with the dictionary
\begin{equation}
  \label{eq:rules:legtable}
  \begin{array}{lccclc}
   \text{bulk end} & \ell & m & p & g(u_j) & \text{kernel}\\[3pt]
   \hline\\[-8pt]
   \pic \ \ (\text{undifferentiated}) & -\ii & 0 & 0 & 1/(1+u_j) & \Wn_0\\[4pt]
        & +1 & 1 & 1 & (1+2u_j)/(1+u_j) & \Wn_1\\[4pt]
   \partial_\tau\pic & +\ii & 2 & 1 & (1+2u_j)^2/(1+u_j) & \Wn_2\\[4pt]
   \sigma & -4\ii/\la & 2 & 2 & u_j & \Vv
  \end{array}
\end{equation}
one integration variable $u_j$ per leg (the undifferentiated $\pic$ has two rows, to be added);
for a $-$ line one conjugates,
$a\to-a$, $\ell\to\ell^*$, $e^{\ii k\tau'(\cdots)}\to e^{-\ii k\tau'(\cdots)}$. Collecting
$a^q\propto(-H\tau')^{-q}$ with the $(-\tau')^{p_j}$ of the legs, a single vertex with all legs on
the boundary gives $\int\dd\tau'(-\tau')^{N}e^{\ii k_t\tau'\mathcal{E}}$ with
\begin{equation}
  \label{eq:rules:N}
  N = \sum_j p_j - q ,\qquad
  \mathcal{E} = 1+2\sum_j e_j u_j,\qquad e_j=\frac{k_j}{k_t},\ \ k_t=\sum_j k_j .
\end{equation}
For a vertex $\dot\pic^{\,n}\sigma^{V-n}$ one has $q=4-n$ and $\sum_jp_j=n+2(V-n)$, hence
\begin{equation}
  \label{eq:rules:Nuniv}
  \boxed{\ \ N = 2V-4 \quad\text{independently of } n , \ \ }
\end{equation}
so that $N=2$ for all four cubic exchange orders $\dot\pic^{\,3}$, $\dot\pic^{\,2}\sigma$,
$\dot\pic\sigma^2$ and $\sigma^3$: each replacement of a velocity leg by a $\sigma$ leg raises $q$ by
one and raises that leg's time power from $1$ to $2$, and the two increments cancel. The measure
$\xi^2e^{-\xi}$ below is thus fixed by the vertex weight and the leg time powers alone, and
forgetting the extra time power of a $\sigma$ leg would shift $N$ by one unit per $\sigma$ leg,
multiplying every squeezed coefficient by a mass-dependent ratio of $\Gamma$'s. The one cubic
vertex that departs from this rule is $(\partial_i\pic)^2\sigma/a^2$: it has $q=2$, but each
undifferentiated leg contributes $p=0$ or $p=1$, so it is a sum of three terms with $N=0,1,2$,
which we collect by writing the vertex at $N=0$ with the kernel $\Pp_a(\beta)=\Wn^a_0+\tfrac\beta2\Wn^a_1$
on each undifferentiated leg.

With the Bunch--Davies contour understood (\cref{sec:rules:contour}), the Schwinger identity
\begin{equation}
  \label{eq:rules:schwinger}
  \int_{-\infty}^{0}\!\dd\tau'\,(-\tau')^{N}e^{\ii k_t\tau'\mathcal{E}}
  = \frac{N!}{(\ii k_t\mathcal{E})^{N+1}}
  = \frac{1}{(\ii k_t)^{N+1}}\int_0^\infty\!\dd\xi\;\xi^{N}e^{-\xi}\prod_j e^{-2\xi e_j u_j}
\end{equation}
disentangles the frequencies: the $u_j$ integrals factorise into one integral per leg at
$\beta_j=2\xi e_j$, i.e.\ into the kernels \cref{eq:lin:kernels}. Since the channel indices of different legs never meet,
the sum over the $2^V$ channel assignments factorises at fixed $\xi$, and each leg appears in
exactly the dressed combinations \cref{eq:lin:dressed}. Assembling, a single vertex with all legs
on the boundary gives
\begin{equation}
  \label{eq:rules:master}
  \boxed{\ \
  \langle Q\rangle' = 2\,\Imag\Big[\ \Ncal_{\rm v}\;
  \int_0^\infty\!\dd\xi\;\xi^{N}e^{-\xi}\prod_{j=1}^{V}\mathbb{K}_j(2\xi e_j)\ \Big],
  \quad \mathbb{K}_j\in\{\bW_2,\ \bP,\ \bV\},\ \ }
\end{equation}
where $\Ncal_{\rm v}$ collects the coupling, the multiplicity, the momentum factors, the
$H^2/4k_j^3$ of each propagator, the leg constants $\ell_j$ with the $4/\la$'s already absorbed in
$\bV$, and $(\ii k_t)^{-(N+1)}$. The projection $2\,\Imag$ is the first-order in--in formula,
$\langle\hat O\rangle=2\,\Imag\int\dd t\,\langle\hat O H_{\rm int}(t)\rangle$, which is the only
convention-free statement; it becomes $\pm2\,\Real$ of the $\xi$-integral after the time integral
has been done, the sign being that of $\ii^{-(N+1)}$ times the product of leg constants, and it is
fixed channel by channel in \cref{sec:bispectra}. Two rules of thumb summarise the reduction: a
velocity leg is $\bW_2$, a $\sigma$ leg is $\bV$, an undifferentiated leg is $\bP$; and replacing a
velocity leg by a $\sigma$ leg costs a real factor, $-4/(\la aH)$ times the swap $\Wn_2\to\Vv$,
whose $1/a=-H\tau'$ is exactly the extra time power that keeps $N=2$.

Complex conjugation maps a diagram to the same diagram with all branch labels flipped, so that for
any correlator with all external points on the boundary the sum over branch assignments is
$2\,\Real$ of the sum over assignments with one reference vertex on the $+$ branch: the result is
manifestly real and only half the assignments must be computed. For a single vertex this is the
$2\,\Imag$ of \cref{eq:rules:master}. A separate reality property holds channel by channel,
$[\Wn^a_n]^*=\Wn^{-a}_n$, $[\Vv^a]^*=\Vv^{-a}$ and $r_{-a}=r_a^*$, so that the channel sum pairs
$\vec a$ with $-\vec a$; it is the origin of the branch cancellations of \cref{sec:sq:strong}.

The rules can be summarised as follows. Draw a square for each external operator and all
topologies connecting squares to vertices; decorate each half-line with the field of the operator
it touches, a square end being always $\pic$; make every vertex branch-summed (a dashed circle),
or fix its branch (filled or open) when a single term is wanted; assign
$\Gsk^{XY}_{ab}$ to internal lines and $\Kb^Y_\pm$ to external ones; assign the vertex factor
\cref{eq:rules:vertexrule} with $q$ from \cref{eq:rules:q}; reduce each bulk end to its dressed
leg, and each vertex time to a Schwinger parameter with $N$ from \cref{eq:rules:N}; sum over
channels through the dressed legs. Nothing in this list assumes a particular number of legs,
vertices or loops, and in particular the reduction covers vertices containing $\hat\sigma$,
precisely because $[\hat\pic,\hat\sigma]\ne0$ in the resummed interaction picture.

\subsection{Convergence and the Bunch--Davies contour}
\label{sec:rules:contour}

Both \cref{eq:rules:schwinger} and the interchange of the $\tau'$ and $u_j$ integrations rest on a
contour statement. Write each leg as its reduced kernel at the complex argument
$\beta_j=-2\ii k_j\tau'$; the vertex integrand is
$(-\tau')^N e^{\ii k_t\tau'}\prod_j\mathcal{K}_j(-2\ii k_j\tau')$, integrated along the ray
$\arg(-\tau')=-\theta$, the physical contour being $\theta\to0^+$ and the fully rotated one
$\theta=\pi/2$, on which the integral \emph{is} the Schwinger representation with
$\xi=-k_t\tau'$. The kernels are integrals of $e^{-\beta u}$ over the half-line, hence analytic for
$\Real\beta>0$, which contains the sector; at large $|\tau'|$ they are bounded and the decay is
carried by $e^{\ii k_t\tau'}$, exponential for any $\theta>0$ and only algebraic at $\theta=0$, which
is what the Bunch--Davies $\ii\epsilon$ supplies; and the small arc contributes
$\mathcal{O}(r^{N+1-\sum_jd_j})$ with the soft indices of \cref{sec:lin:tails}, so that the
rotation is legitimate iff
\begin{equation}
  \label{eq:rules:convergence}
  \boxed{\ \ N+1 \;>\; \sum_j d_j ,\ \ }
\end{equation}
which is also the convergence condition of the $\xi$-integral at $\xi\to0$. For a heavy field all
four cubic channels have $N=2$ and $\sum_jd_j=\tfrac32$, and the gradient vertex has
$\sum_jd_j=\tfrac12$ at $N=0$: every channel of this paper is safe.

\paragraph{Light fields.} For a light isocurvature field, $\n$ real, the soft indices grow with
$\n$ and \cref{eq:rules:convergence} with three singular legs reads $\n<\tfrac12$: every formula of
this paper holds as it stands in that window, with the pure power law $\kappa^{1/2-\n}$ of the
$b=+$ branch in place of the clock (the $b=-$ term then lies below the analytic background), but
the interval $\tfrac12<\n<\tfrac32$, which contains most of the phenomenologically
interesting light range, is not directly computable from the reduced form. The divergence is an
artefact of the split and not of the correlator, which is analytic in $m^2$, hence in $\n^2$, so
that the reduced expression defines it by analytic continuation in $\n$: one subtracts from the
product of the legs the finitely many small-$\xi$ powers $\xi^{N+s}$ with $\Real(N+s)\le-1$, taken
from the complete power set of each leg including the analytic family, which for $\n>\tfrac12$ is
larger than the $b=-$ clock term, integrates them in closed form through the continued
$\Gamma(N+s+1)$, and integrates the remainder numerically. We have not carried this out channel by
channel, the rotation has not been tested at real $\n$, and everything below is stated for an
effectively heavy field.

\subsection{Examples}
\label{sec:rules:examples}

We give five examples, chosen to span the structural cases rather than to enumerate the channels:
no vertex; a time-derivative vertex; a spatial-derivative vertex; a bulk-to-bulk mixed line; a loop.

\paragraph{(a) Power spectrum.} No vertex at all: one line with both ends on the boundary.
\begin{center}
\begin{tikzpicture}[baseline=-2pt]
  \node[bpt](a) at (0,0){}; \node[bpt](b) at (2.0,0){}; \draw[pil](a)--(b);
\end{tikzpicture}
\end{center}
Each square gives $\ii r_a$, the line gives one channel sum with weight $\tfrac12e^{a\pi\la/2}$ and
one $H^2/2k^3$,
\begin{equation}
  \label{eq:rules:ex-a}
  \langle\zeta_{\vec k}\zeta_{-\vec k}\rangle' = \frac{H^2}{\fpi^4}\,\Dw^{\pi\pi}_k(0,0)
  = \frac{2\pi^2}{k^3}\,\Dzo^2\,R ,
  \qquad R=\tfrac12\sum_a e^{a\pi\la/2}|r_a|^2 ,
\end{equation}
which is the only place where the rules are checked against something known in closed form.

\paragraph{(b) Contact bispectrum from $\lambda_2\dot\pic^{\,3}$.} The vertex is branch-summed
(dashed circle), so the single diagram is the whole branch sum, $V_++V_-$ with the three external
lines $\Kb^\pi_+$ or $\Kb^\pi_-$ accordingly.
\begin{center}
\begin{tikzpicture}[baseline=-2pt]
  \node[vs](v) at (0,0){};
  \node[bpt](a) at (1.5,0.9){}; \node[bpt](b) at (1.8,0){}; \node[bpt](c) at (1.5,-0.9){};
  \draw[pil](v)--(a); \draw[pil](v)--(b); \draw[pil](v)--(c);
  \node at (0.95,0.78){\footnotesize $k_1$}; \node at (1.00,0.24){\footnotesize $k_2$};
  \node at (0.95,-0.78){\footnotesize $k_3$};
\end{tikzpicture}
\end{center}
Reading off: Wick multiplicity $3!$, $c=\lambda_2$, $q=1$ by \cref{eq:rules:q}, three
$\partial_\tau\pic$ legs with $p_j=1$ by \cref{eq:rules:legtable}, hence $N=3-1=2$ and each leg
calls $\Wn_2$. Then \cref{eq:rules:master} gives
\begin{equation}
  \label{eq:rules:ex-b}
  B_\zeta = \frac{3\lambda_2 H^8}{16\,\fpi^6\,k_1k_2k_3\,k_t^3}\;
  \Real\Big[\sum_{\vec a}\Big(\prod_{j=1}^{3}e^{a_j\pi\la/2}r_{a_j}\Big)
  \int_0^\infty\!\dd\xi\,\xi^2e^{-\xi}\prod_{j=1}^{3}\Wn^{-a_j}_2(2\xi e_j)\Big] .
\end{equation}
We do the conversion step by step, since it fixes the sign of everything downstream. Each leg is
$\ii$-free: its boundary end gives $\ii r_{a_j}$ and its velocity row gives $+\ii$ against
$(-\tau')$, so that $(+\ii)^3\prod_j(\ii r_{a_j})=-\prod_j r_{a_j}$, in the leg-constant convention of
\cref{sec:rules:reduction}. The conversion $\zeta=-H\pic/\fpi^2$ supplies a further $(-1)^3$, and the
one explicit $\ii$ left is the $(\ii k_t)^{-3}=+\ii k_t^{-3}$ of \cref{eq:rules:schwinger}, so that
$2\,\Imag[\ii Z]=+2\,\Real[Z]$; the three signs multiply to $+$. With
$(k_1k_2k_3)^2/(k_1k_2k_3k_t^3)=e_1e_2e_3$ and $\fpi^2=R^{1/2}H^2/2\pi\Dz$ this is
the shape function $S$ of \cref{sec:bispectra}, $S=\Ncal_0\,e_1e_2e_3\Real[\cdots]$, with
\begin{equation}
  \label{eq:rules:ex-bshape}
  \Ncal_0=\frac{3}{32\pi R^{3/2}}\lambda_2H^2\,\frac{1}{\Dz},
\end{equation}
the no-exchange prefactor of Ref.~\cite{Pinol:2026xnl} and the first rung of the ladder of
\cref{sec:bispectra}.

\paragraph{Other vertices without spatial derivatives.} Nothing above used the field content, only
$q$ and the leg time powers, so the $\Lambda_2$, $\alpha$ and $\gthree$ vertices are read off the
same diagram with one, two or three solid legs replaced by mixed ones:
\begin{center}
\begin{tikzpicture}[baseline=-2pt]
  \node[vs](v) at (0,0){};
  \node[bpt](a) at (1.7,0.95){}; \node[bpt](b) at (2.0,0){}; \node[bpt](c) at (1.7,-0.95){};
  \draw[pil](v)--(a); \mixedline{b}{v} \mixedline{c}{v}
  \node at (0.55,1.18){\footnotesize $-\tfrac{\alpha}{2}\dot\pic\sigma^2$};
  \begin{scope}[xshift=6cm]
  \node[vs](w) at (0,0){};
  \node[bpt](d) at (1.7,0.95){}; \node[bpt](e) at (2.0,0){}; \node[bpt](f) at (1.7,-0.95){};
  \mixedline{d}{w} \mixedline{e}{w} \mixedline{f}{w}
  \node at (0.55,1.18){\footnotesize $-\gthree\sigma^3$};
  \end{scope}
\end{tikzpicture}
\end{center}
Each replacement does three things and only three. It removes one time derivative and so raises $q$
by one; it raises that leg's $p$ from $1$ to $2$, and the two increments cancel in $N=\sum_jp_j-q$,
so that $N=2$ is unchanged, \cref{eq:rules:Nuniv}; and it swaps $\Wn^{-a_j}_2\to\Vv^{-a_j}$
together with the leg constant, a real factor $-4/\la aH$ per mixed leg, whose sign is the
alternating $(-1)^{n_{\dot\pi}}$ of the ladder of \cref{sec:bispectra}. Hence \cref{eq:rules:ex-b}
carries over verbatim,
\begin{equation}
  \begin{gathered}
  \label{eq:rules:ex-c}
  B_\zeta = \Big[\text{prefactor of \cref{eq:rules:ex-b}},\ \lambda_2\to c,\ 3!\to S_{\rm W}\Big]
  \Big(\!-\frac{4}{\la H}\Big)^{n_\sigma} \\[2pt]
  \times\;
  \Real\Big[\sum_{\vec a}\Big(\prod_{j}e^{a_j\pi\la/2}r_{a_j}\Big)
  \int_0^\infty\!\!\dd\xi\,\xi^2e^{-\xi}
  \prod_{\text{solid}}\Wn^{-a_j}_2\prod_{\text{mixed}}\Vv^{-a_j}\Big],
  \end{gathered}
\end{equation}
summed over which external legs carry the $\sigma$. The four exchange orders are indeed a single
calculation.

\paragraph{(c) Spatial derivatives: the $\Lambda_1$ vertex.} This is the qualitatively different
case, and the one where reading $N$ off the diagram matters.
\begin{center}
\begin{tikzpicture}[baseline=-2pt]
  \node[vs](v) at (0,0){};
  \node[bpt](a) at (1.7,0.95){}; \node[bpt](b) at (2.0,0){}; \node[bpt](c) at (1.7,-0.95){};
  \draw[pil](v)--(a); \draw[pil](v)--(b); \mixedline{c}{v}
  \node at (1.05,-0.98){\footnotesize $k_3$};
\end{tikzpicture}
\end{center}
Here $n_{\partial_t}=0$ and $n_{\partial_i}=1$, so $q=4-0-2=2$; the two solid legs are
\emph{undifferentiated}, $p=0$ in the first row of \cref{eq:rules:legtable}, and the mixed leg has
$p=2$. Hence
\begin{equation}
  \label{eq:rules:ex-N0}
  N = (0+0+2)-2 = 0 ,
\end{equation}
the measure is $e^{-\xi}$ with no power of $\xi$, and the two solid legs call $\Pp$ rather than
$\Wn_2$: the missing powers $\xi^1$ and $\xi^2$ are not absent but inside
$\Pp_a=\Wn^a_0+\tfrac{\beta}{2}\Wn^a_1$, one per undifferentiated leg. The gradients supply
$(\ii\vec k_1)\!\cdot\!(\ii\vec k_2)=\tfrac12(k_1^2+k_2^2-k_3^2)$. Two sign sources differ from (b)
and must be tracked together: an undifferentiated leg carries no velocity constant, so the product
of leg constants is $+(4/\la)\prod_jr_{a_j}$ rather than alternating, while
$(\ii k_t)^{-1}=-\ii k_t^{-1}$ turns $2\,\Imag$ into $-2\,\Real$. The two compensate, and the
prefactor of the gradient channel is positive, $\Ncal_{1\perp}>0$, opposite in sign to its velocity
partner; this is precisely the relative sign that makes the boost-invariant combination
$\Lambda_2^{-1}=-\Lambda_1^{-1}$ of \cref{sec:bispectra} a difference of the two reduced shapes.
Assembling as in (b) and summing over which external leg carries the $\sigma$,
\begin{equation}
  \label{eq:rules:ex-lam1}
  B_\zeta^{(\Lambda_1)} = \frac{H^7}{8\,\la\,\Lambda_1\,\fpi^6\,k_t}
  \sum_{c=1}^{3}\frac{\tfrac12\big(k_a^2+k_b^2-k_c^2\big)}{k_a^3k_b^3k_c}
  \Real\Big[\sum_{\vec a}\Big(\prod_{j}e^{a_j\pi\la/2}r_{a_j}\Big)
  \!\int_0^\infty\!\!\dd\xi\,e^{-\xi}\Pp_{-a_a}\Pp_{-a_b}\Vv^{-a_c}\Big],
\end{equation}
the kernels at $\beta_j=2\xi e_j$, $c$ labelling the leg that carries the $\sigma$ and $(a,b)$ the two
others.

\paragraph{(d) Beyond a single vertex.} The reduction survives at any number of vertices and any
loop order, because the bulk-to-bulk Wightman function \cref{eq:rules:wightman} is again a single
frequency in each of its arguments: for opposite branches the two vertex integrals reduce
separately, the internal line's two $u$-integrals being ordinary kernels, while for equal branches
the step function nests the times, and with $k_t^{(i)}$ the sum of the momenta entering vertex $i$
(the internal momentum counted at both), $\mathcal{E}_i=1+2\sum_{j\in i}e^{(i)}_ju_j$ and
$\tau_1<\tau_2$,
\begin{equation}
  \begin{gathered}
  \label{eq:rules:nested}
  \int_{-\infty}^{0}\!\!\dd\tau_2(-\tau_2)^{N_2}e^{\ii k_t^{(2)}\mathcal{E}_2\tau_2}
  \int_{-\infty}^{\tau_2}\!\!\dd\tau_1(-\tau_1)^{N_1}e^{\ii k_t^{(1)}\mathcal{E}_1\tau_1}
  \\[2pt]
  =\sum_{m=0}^{N_1}\binom{N_1}{m}
  \frac{(N_2+m)!\,(N_1-m)!}
       {\big[\ii\big(k_t^{(1)}\mathcal{E}_1+k_t^{(2)}\mathcal{E}_2\big)\big]^{N_2+m+1}
        \big[\ii\,k_t^{(1)}\mathcal{E}_1\big]^{N_1-m+1}}
  \\[2pt]
  =\sum_{m=0}^{N_1}\binom{N_1}{m}\frac{1}{\ii^{N_1+N_2+2}}
  \int_0^\infty\!\dd\xi_1\,\xi_1^{N_1-m}e^{-\xi_1\,k_t^{(1)}\mathcal{E}_1}
  \int_0^\infty\!\dd\xi_2\,\xi_2^{N_2+m}e^{-\xi_2\,(k_t^{(1)}\mathcal{E}_1+k_t^{(2)}\mathcal{E}_2)} ,
  \end{gathered}
\end{equation}
obtained by expanding $(-\tau_1)^{N_1}=[(-\tau_2)+(\tau_2-\tau_1)]^{N_1}$ and Schwinger-parametrising
each factor as in \cref{eq:rules:schwinger}, $\xi_1$ conjugate to $k_t^{(1)}\mathcal{E}_1$ and $\xi_2$
to the sum; since the $\mathcal{E}_i$ are linear in the $u_j$
the exponentials still combine into one per leg, and the kernels are evaluated at shifted
arguments, $\beta_j=2\xi_1k_j/k_t^{(1)}+2\xi_2k_j/K$ for a leg at vertex $1$ and $\beta_j=2\xi_2k_j/K$
at vertex $2$, with $K=k_t^{(1)}+k_t^{(2)}$. Only the $\xi$-integrals become multi-dimensional, one
per vertex, and a loop adds the usual momentum integral.
\section{The four bispectra}
\label{sec:bispectra}

We now apply the rules of \cref{sec:rules} to the five cubic vertices of \cref{eq:eft:L3}.

\subsection{Master formulas}
\label{sec:bisp:master}

For the four vertices without spatial derivatives, assembling \cref{eq:rules:master} with the leg
constants of \cref{sec:rules:reduction}, $(-1)^{n_{\dot\pi}}(4/\la)^{n_\sigma}$, the three propagator factors $H^2/4k_j^3$, the momentum powers
$k_j^2$ of each leg and $a^q=(-H\tau')^{-q}$, one finds
\begin{equation}
  \label{eq:bisp:pic3}
  \langle\pic^3\rangle' = -\,(-1)^{n_\sigma}\,\frac{g_{n_\sigma}\,H^{5-n_\sigma}}{32\,k_1k_2k_3\,k_t^3}\;
  \Real\!\int_0^\infty\!\dd\xi\;\xi^2e^{-\xi}\sum_{\rm placements}\prod_{j=1}^3\mathbb{K}_j(2\xi e_j),
\end{equation}
where $g_{n_\sigma}$ is the coupling of the vertex times the number of ways of attaching its
identical fields to the three external legs, $g_{n_\sigma}=\{6\lambda_2,\ \Lambda_2^{-1},\ \alpha,\
6\gthree\}$ for $n_\sigma=0,1,2,3$, and where the $(4/\la)^{n_\sigma}$ of
the leg constants has been absorbed into the $\sigma$ legs $\bV$.
With $\zeta=-H\pic/\fpi^2$, $(k_1k_2k_3)^2/(k_1k_2k_3k_t^3)=e_1e_2e_3$ and
$\fpi^6=R^{3/2}H^6/[(2\pi)^3\Dz^3]$ from \cref{eq:eft:Dz0},
\begin{equation}
  \label{eq:bisp:master}
  S^{(n_\sigma)} = \Ncal_{n_\sigma}\;e_1e_2e_3\;
  \Real\!\int_0^\infty\!\dd\xi\;\xi^2e^{-\xi}\sum_{\rm placements}\prod_{j=1}^3\mathbb{K}_j(2\xi e_j),
  \qquad
  \Ncal_{n_\sigma} = (-1)^{n_\sigma}\,\frac{g_{n_\sigma}\,H^{2-n_\sigma}}{64\pi\,R^{3/2}\,\Dz} ,
\end{equation}
that is,
\begin{equation}
  \label{eq:bisp:ladder}
  \begin{aligned}
  \Ncal_0 &= \frac{3}{32\pi R^{3/2}}\lambda_2H^2\,\frac{1}{\Dz},&
  \Ncal_{1\parallel} &= -\frac{1}{64\pi R^{3/2}}\frac{H}{\Lambda_2}\frac{1}{\Dz},\\
  \Ncal_{2} &= \frac{\alpha}{64\pi R^{3/2}}\frac{1}{\Dz},&
  \Ncal_{3} &= -\frac{3}{32\pi R^{3/2}}\frac{\gthree}{H}\frac{1}{\Dz} .
  \end{aligned}
\end{equation}
The first is the published no-exchange prefactor \cite{Pinol:2026xnl}; the four channels are governed
by the dimensionless combinations $\lambda_2H^2$, $H/\Lambda_2$, $\alpha$ and $\gthree/H$. The
explicit shapes are
\begin{equation}
  \label{eq:bisp:shapes}
  \boxed{\ \
  \begin{aligned}
  S^{(0)} &= \Ncal_0\,e_1e_2e_3\,\Real\!\int_0^\infty\!\dd\xi\,\xi^2e^{-\xi}\,
     \bW_2(2\xi e_1)\bW_2(2\xi e_2)\bW_2(2\xi e_3),\\
  S^{(1\parallel)} &= \Ncal_{1\parallel}\,e_1e_2e_3\sum_{c=1}^3\Real\!\int_0^\infty\!\dd\xi\,\xi^2e^{-\xi}\,
     \bV(2\xi e_c)\prod_{j\ne c}\bW_2(2\xi e_j),\\
  S^{(2)} &= \Ncal_{2}\,e_1e_2e_3\sum_{c=1}^3\Real\!\int_0^\infty\!\dd\xi\,\xi^2e^{-\xi}\,
     \bW_2(2\xi e_c)\prod_{j\ne c}\bV(2\xi e_j),\\
  S^{(3)} &= \Ncal_{3}\,e_1e_2e_3\,\Real\!\int_0^\infty\!\dd\xi\,\xi^2e^{-\xi}\,
     \bV(2\xi e_1)\bV(2\xi e_2)\bV(2\xi e_3),
  \end{aligned}\ \ }
\end{equation}
where $c$ labels the leg that carries the odd field of the vertex, as in \cref{eq:rules:ex-lam1}.
The $R^{-3/2}$ descends from the measured $\Dz^4\propto R^2$ in the shape denominator together with
one $\Dzo^{-1}=R^{1/2}\Dz^{-1}$ from the couplings. The relative signs are physical: they follow
from the alternation of the leg constants, i.e.\ from the relative sign between the isocurvature
and curvature channel modes in \cref{eq:lin:Msig}, and every cubic term of \cref{eq:eft:L3} enters
with the same sign.

The gradient vertex has $N=0$, two undifferentiated legs and the momentum factor
$-\vec k_a\cdot\vec k_b=\tfrac12(k_a^2+k_b^2-k_c^2)$, and its projection carries the opposite sign.
Summing over the leg $c$ that carries the $\sigma$,
\begin{equation}
  \label{eq:bisp:perp}
  \boxed{\ \
  \begin{gathered}
  S^{(1\perp)} = \Ncal_{1\perp}\sum_{c=1}^{3}F_c\;\Real\!\int_0^\infty\!\dd\xi\;e^{-\xi}\,
  \bP(2\xi e_a)\,\bP(2\xi e_b)\,\bV(2\xi e_c),\\[2pt]
  \Ncal_{1\perp}=+\frac{1}{64\pi R^{3/2}}\frac{H}{\Lambda_1}\frac{1}{\Dz},
  \qquad
  F_c\equiv\frac{e_c(e_a^2+e_b^2-e_c^2)}{2\,e_ae_b},
  \end{gathered}\ \ }
\end{equation}
$(abc)$ being a permutation of $(123)$. Relative to the velocity channel the leg constants keep
their sign while the projection flips, so the two single-exchange prefactors are opposite:
$\Ncal_{1\perp}/\Ncal_{1\parallel}=-\Lambda_2/\Lambda_1$. \Cref{eq:bisp:master,eq:bisp:perp} are the complete tree-level bispectrum
of the theory, exact in $\la$; the total is the sum of the five shapes with their couplings.

\subsection{Two immediate consequences}
\label{sec:bisp:consequences}

\paragraph{The single-field limit, as a check on the whole machinery.} As $\la\to0$ the weight
collapses onto $u=0$ by \cref{eq:lin:distrib}, so $\bW_2\to2$, $r_a\to1$, $R\to1$, and the
no-exchange bracket becomes $2^3\int\xi^2e^{-\xi}\dd\xi=16$:
\begin{equation}
  \label{eq:bisp:lam0}
  S^{(0)}\;\longrightarrow\;3\lambda_2 \fpi^2\;e_1e_2e_3,
  \qquad
  B_\zeta\;\longrightarrow\;\frac{3\lambda_2 H^8}{\fpi^6}\,\frac{1}{k_1k_2k_3\,k_t^3},
\end{equation}
the single-field equilateral result at $c_s=1$, exactly proportional to $e_1e_2e_3$: the textbook
computation with the undressed mode function $\pic\propto(1+\ii k\tau)e^{-\ii k\tau}$ gives
$\langle\pic^3\rangle'=-3\lambda_2H^5/(k_1k_2k_3k_t^3)$. Nothing was tuned to make them agree, and
they agree exactly, not up to a constant, so that the in--in sign, the vertex rule, the multiplicity
of a symmetric vertex, the measure $N=2$, the Schwinger identity and every normalisation in
\cref{eq:bisp:master} are confirmed at once. What this check does not reach is the isocurvature row
of the leg dictionary, which every $n_\sigma\ge1$ channel needs; that row is tested against the
perturbative literature in \cref{sec:sq:weak}.

The same limit on the other channels gives the perturbative shapes, the contact vertex with one
mixing insertion on each $\sigma$ leg, in our variables: by \cref{eq:lin:W2chan,eq:lin:sigmapower},
\begin{equation}
  \label{eq:bisp:lam0n}
  S^{(n_\sigma)}\;\simeq_{\la\ll1}\;\Ncal_{n_\sigma}\big|_{R=1}\;2^{3-n_\sigma}(4\la)^{n_\sigma}\;e_1e_2e_3
  \!\!\sum_{\rm placements}\!\!\Real\!\int_0^\infty\!\dd\xi\,\xi^2e^{-\xi}
  \!\!\prod_{\sigma\ \rm legs}\!\!\big[-\tfrac12\Vv_2-\ii\varrho_1\Vv_1\big](2\xi e_j),
\end{equation}
uniformly over the triangle of momentum ratios with an $\mathcal{O}(\la^2)$ relative remainder. These are not
proportional to $e_1e_2e_3$, because the one-insertion isocurvature leg is a non-trivial function
of $\beta$: by \cref{eq:lin:V2closed} it is a pair of Bessel branches with unequal weights plus one
entire ${}_2F_2$, and it is that ${}_2F_2$ which prevents the perturbative exchange shapes from
reducing to moments of Bessel functions, one for single exchange and products of two and three
for double and triple exchange. The power $\la^{n_\sigma}$ in front, one per mixing insertion, is
the counting of \cref{eq:lin:legcount}: the $\la$ dependence is entirely in the legs, the
prefactors \cref{eq:bisp:ladder} carry none.

\subsection{The universal shape at strong mixing}
\label{sec:bisp:strong}

At $\la\gg1$ and fixed effective mass the four channels stop being four shapes. The reason is one
line of \cref{sec:lin:strong}: every Schwinger integral is concentrated at $\xi\propto1/\la$, in
the tail regime of all three legs, and there $\Vv^a=\tfrac14\Wn^a_2[1-\Wn^a_0/\Wn^a_2]$ with
$|\Wn^a_0/\Wn^a_2|\ll1$, \cref{eq:lin:quarter}. An isocurvature leg is therefore a velocity leg divided by the mixing,
kernel by kernel and channel assignment by channel assignment, and the shapes of
\cref{eq:bisp:shapes} collapse onto the no-exchange one,
\begin{equation}
  \label{eq:bisp:prop}
  \boxed{\ \
  \begin{gathered}
  S^{(1\parallel)}\;\simeq_{\la\gg1}\;-\frac{H/\Lambda_2}{2\,\la\,\lambda_2H^2}\;S^{(0)},\qquad
  S^{(2)}\;\simeq_{\la\gg1}\;\frac{\alpha}{2\,\la^2\,\lambda_2H^2}\;S^{(0)},\\[3pt]
  S^{(3)}\;\simeq_{\la\gg1}\;-\frac{\gthree/H}{\la^3\,\lambda_2H^2}\;S^{(0)} ,
  \end{gathered}\ \ }
\end{equation}
each with a relative correction $\mathcal{O}(\muf^2n_\sigma/\la^2)$. The localisation holds at
every configuration of the triangle, so \cref{eq:bisp:prop} is a statement about the whole shape,
pointwise in $(k_1,k_2,k_3)$, and in particular about its squeezed limit, where it says that the
clock amplitudes of \cref{sec:squeezed} obey the same ratios, $\la^{n_\sigma}\Acal^{(n_\sigma)}/\Acal^{(0)}$
tending to the corresponding constant (\cref{sec:sq:pheno}). The universal shape is thus the
no-exchange shape $S^{(0)}(k_1,k_2,k_3;\la,\muf)$ itself, at the same point of the plane: the
statement is one of proportionality between channels, with constants that carry no dynamics, the
only $\la$-dependence being the explicit $1/\la$ of the isocurvature leg constant, one per $\sigma$
leg, the rational factors being placement counts, and nothing depending on the mass. $S^{(0)}$ is
not proportional to $e_1e_2e_3$ at strong mixing and we have no closed form for it; its equilateral
amplitude and its squeezed clock at large $\la$ are the subject of \cref{sec:bisp:pheno,sec:sq:strong},
and its appearance over the triangle is the first panel of \cref{fig:bisp:shapes}. The gradient
channel joins the same asymptotics one power further down, $S^{(1\perp)}/S^{(0)}=\mathcal{O}(\la^{-3})$,
since its two undifferentiated legs give $(\Pp/\Wn_2)^2=\mathcal{O}(\la^{-4})$, its measure $\xi^0$
in place of $\xi^2$ at $\xi\sim1/\la$ returns $\la^{+2}$, and its single $\sigma$ leg costs
$\la^{-1}$; its constant is the only one that is not universal, because $\Pp/\Wn_2$ is a boundary
value over a kernel and carries the mass.

\subsection{Implementation}
\label{sec:implementation}

The formulas of \cref{sec:bispectra} are exact but not closed: each shape is a one-dimensional
integral over $\xi$ of products of leg kernels, each kernel an integral over the weight
$\omega_a(u)$. They are nevertheless cheap, because the whole $(\la,\muf)$ dependence sits in the
kernels and the whole configuration dependence in the final quadrature, and the two are computed
separately. For a given point of the plane the seed kernel $\Wn^a_0$ and its three companions are
tabulated once on a logarithmic grid in $\beta$, the $u$-integral being done in the Pfaff variable
$t=u/(1+u)$ of \cref{eq:lin:pfaff}, in which the oscillating power at $t\to0$ is defined by the
continuation \cref{eq:app:continuation}; since the ${}_2F_1$ function in the weight does not depend
on $\beta$, it is evaluated once and all values of $\beta$ follow from one matrix product with
$e^{-\beta t/(1-t)}$. A full set of kernels at a few thousand values of $\beta$ costs a fraction of
a second in double precision; below the smallest tabulated $\beta$ the kernels are continued with
their soft tails \cref{eq:lin:tails}. Everything downstream is assembly: the dressed legs of
\cref{eq:lin:dressed} are formed once and interpolated in $\ln\beta$, and a shape at one
configuration is the $\xi$-quadrature of \cref{eq:bisp:shapes} or \cref{eq:bisp:perp} of three
interpolated legs, a few tens of Gauss--Legendre nodes sufficing.

The gain in cost is the point. In the cosmological flow, each triangle is an integration of the
correlators over the whole history of its three modes, mixing included, the squeezed corner is the
most expensive one since the modes must be followed over many decades, and the scan over masses and
mixings that a data analysis needs was, in Ref.~\cite{Philcox:2026tjj}, a computation of tens of
thousands of such templates on graphics processors. Here all five channels at one configuration take four milliseconds on
one core, the $171$ configurations of the Planck measurement of \cref{sec:data} a fraction of a
second, and the two-parameter plane of \cref{sec:bisp:pheno} a few minutes; the squeezed limit costs
nothing more than any other configuration, and its collider oscillation is moreover available in
closed form (\cref{sec:squeezed}). The shapes of all six channels over the plane are released with
this paper, together with the code that produced them,~\cite{Pinol:2026code} so that a template at any point of the
plane can be read off rather than recomputed.

Two independent engines certify the numbers. A second implementation, in arbitrary precision with
adaptive quadrature and library hypergeometric functions, shares no code with the first and agrees
with it to $10^{-13}$ or better wherever compared, the identity $\Wn^a_2=\Wn^a_0+4\Vv^a$ and the
boundary value $\Wn^a_0(0)=r_a$ being tested at every grid point; the pipeline reproduces the
single-field limit \cref{eq:bisp:lam0}, the closed-form squeezed clocks of all six channels and the
reduction to Ref.~\cite{Arkani-Hamed:2015bza} at the level quoted in \cref{app:numerics}. One
limitation matters for what follows: the observable shape is the real part of a complex bracket
whose modulus exceeds its real part by the Boltzmann cancellation of \cref{sec:sq:strong}, so that
double precision carries nine significant digits at $\la=6$ and one or two at $\la=10$; beyond
$\la=8$ the strong-mixing results of \cref{sec:sq:strong} were produced by a third engine, the
same quadrature in ball arithmetic at $40$ to $64$ digits (\cref{app:numerics}).

\subsection{New exact bispectrum shapes}
\label{sec:bisp:pheno}

We define the shape function and its equilateral amplitude as in Ref.~\cite{Pinol:2026xnl},
\begin{equation}
  S(k_1,k_2,k_3) \equiv \frac{(k_1k_2k_3)^2 B_\zeta}{(2\pi)^4\Dz^4},
  \qquad f_{\rm NL} \equiv \tfrac{10}{9}\,S(k,k,k),
\label{eq:bisp:shape}
\end{equation}
and since each channel is linear in one of the dimensionless couplings \cref{eq:eft:cn}, we quote
throughout the amplitude per unit coupling,
\begin{equation}
  \frac{f^{(n)}_{\rm NL}}{c_n}\;=\;\frac{10}{9}\,\frac{S^{(n)}(k,k,k)}{c_n} ,
  \qquad
  c_n\in\Big\{\lambda_2H^2,\ \frac{H}{\Lambda_2},\ \frac{H}{\Lambda_1},\ \alpha,\ \frac{\gthree}{H}\Big\} .
\label{eq:bisp:reduced}
\end{equation}
Whenever the dependence on the triangle is displayed, the shape is normalised to unity at the
equilateral configuration. No other rescaling is performed: \cref{eq:bisp:shape} is the convention
of the Planck measurement of \cref{sec:data}, so that the amplitudes quoted here and the couplings
constrained there can be compared directly. The size of \cref{eq:bisp:reduced} is not of order
unity: a cubic vertex of strength $c_n$ gives $B_\zeta\sim c_nP_\zeta^{3/2}$ while
\cref{eq:bisp:shape} divides by $\Dz^4\sim P_\zeta^2$, so that
\begin{equation}
  \frac{f^{(n)}_{\rm NL}}{c_n}\;\sim\;\Dz^{-1}\;=\;2.2\times10^{4} ,
\label{eq:bisp:size}
\end{equation}
which is why a Planck-level measurement constrains these couplings at the per-cent to order-one
level. For the no-exchange channel the weak-mixing limit \cref{eq:bisp:lam0} makes this exact,
$f^{(0)}_{\rm NL}/c_0\to\tfrac{5}{81\pi}\Dz^{-1}\simeq4.3\times10^{2}$ as $\la\to0$, which is where
the first column of \cref{tab:bisp:amplitudes} starts; the numbers below range from $10^{-1}$ to
$10^{5}$ around the estimate \cref{eq:bisp:size}, the spread being the exponential physics of the
mixing.

\begin{figure}[t]
\centering
\begin{minipage}[t]{0.485\textwidth}\centering
\includegraphics[width=\linewidth]{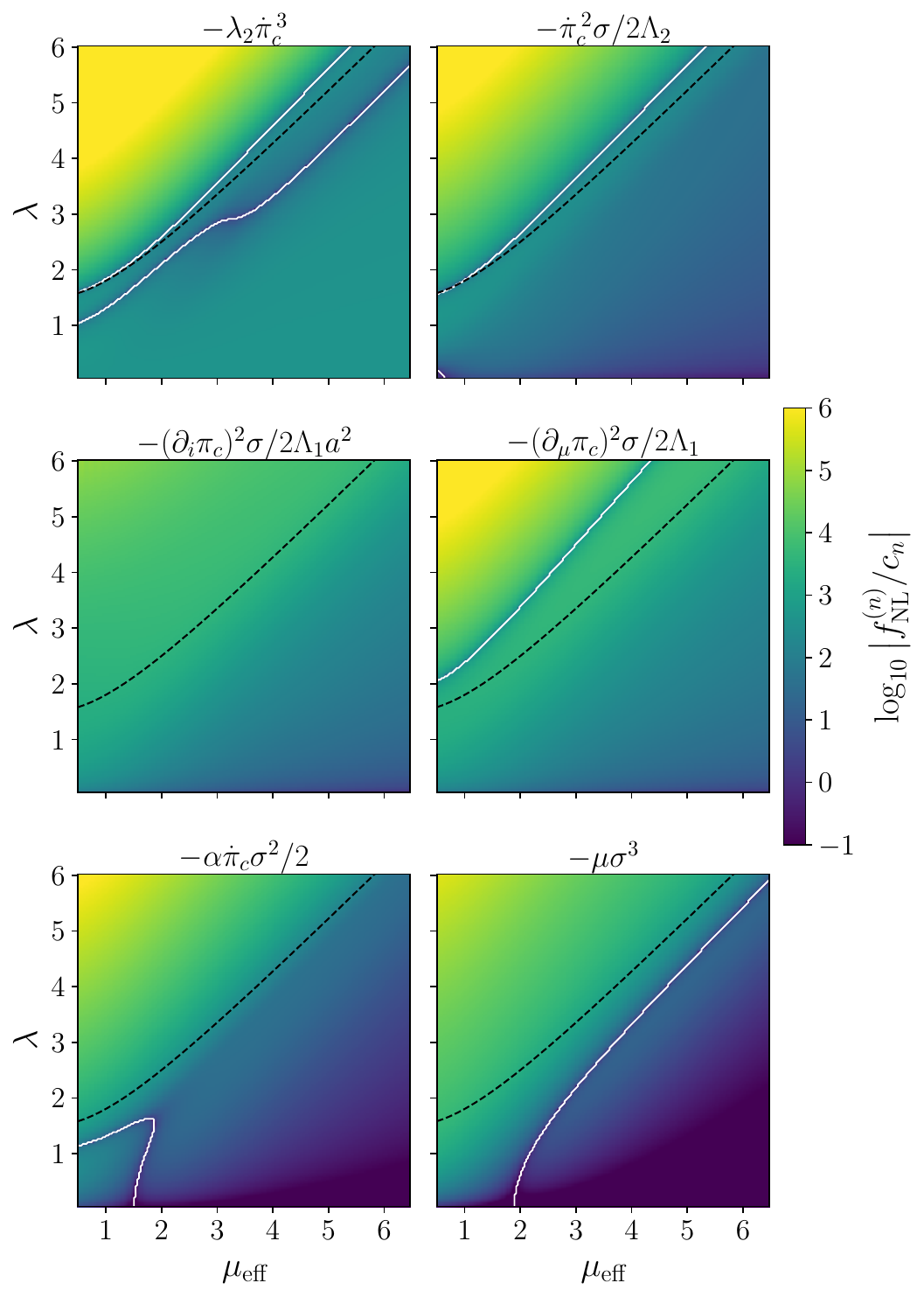}
\end{minipage}\hfill
\begin{minipage}[t]{0.485\textwidth}\centering
\includegraphics[width=\linewidth]{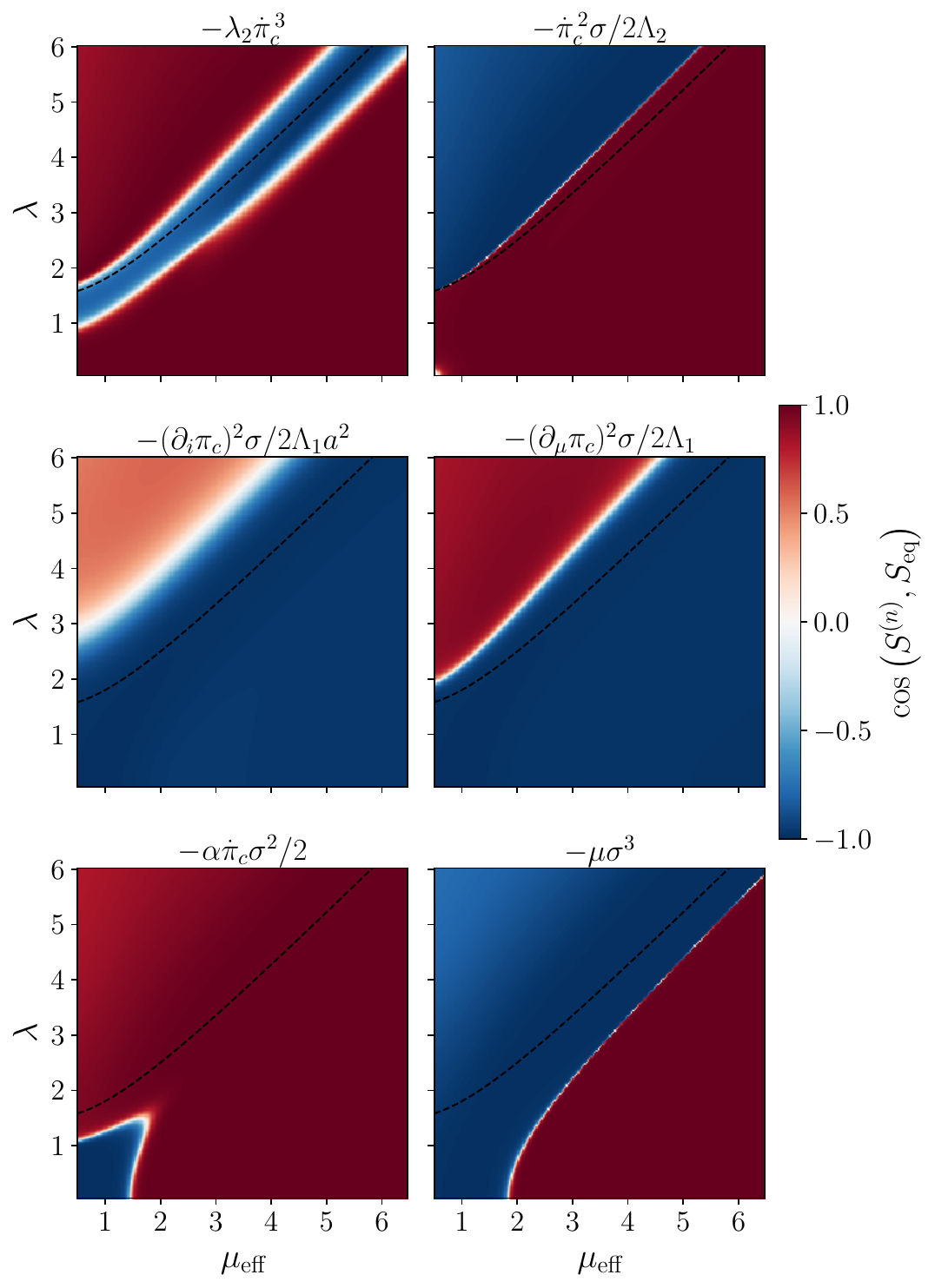}
\end{minipage}
\caption{Left: equilateral amplitude per unit dimensionless coupling, $f_{\rm NL}^{(n)}/c_n$ of
\cref{eq:bisp:reduced}, over the $(\la,\muf)$ plane, one panel per cubic interaction, on a
logarithmic colour scale of the modulus; white: the sign change. Right: cosine of each channel's
shape with the equilateral template, $\cos(S^{(n)},S_{\rm eq})$ of \cref{eq:bisp:cos}, over the same
plane; the white bands are the loci $\cos=0$, where the shape is a genuinely new function of the
triangle. Dashed: the bare-massless line $\la=\sqrt{9/4+\muf^2}$, along which the decorrelation
bands run.}
\label{fig:bisp:planes}
\end{figure}

\paragraph{Amplitudes.} The left panel of \cref{fig:bisp:planes} maps the equilateral amplitude per
unit coupling over
the plane, and \cref{tab:bisp:amplitudes} gives it at three mixings. The ordering between channels
is not the naive one. At weak mixing the no-exchange channel dominates and triple exchange is
negligible, each isocurvature leg costing one power of $\la$ by \cref{eq:lin:legcount}; the exchange
channels overtake it only in a window near $\la\simeq2$--$3$, where the no-exchange amplitude
crosses its own sign change; beyond it the counting of \cref{sec:bisp:strong}, one factor $1/\la$
per $\sigma$ leg, restores the original ordering, and beyond $\la\simeq3$ the no-exchange channel is
the largest of all. The behaviour of that reference channel is what the linear theory explains. At
fixed effective mass its bracket in \cref{eq:bisp:master} is a product of three dressed legs each
carrying $e^{\pi\la}$, the $e^{3\pi\la/4}$ of its boundary factor \cref{eq:lin:Rstrongeff} times the
$e^{\pi\la/4}$ of its kernel on its tail, so that its modulus grows as $e^{3\pi\la}$ against a prefactor
$\Ncal_0\propto R^{-3/2}\propto\la^{3/2}e^{-3\pi\la/2}$; the observable is the real projection, an
increasingly fine cancellation,
$|\Real\mathcal{B}^{(0)}|/|\mathcal{B}^{(0)}|\simeq3\times10^{-1},3\times10^{-2},10^{-4},3\times10^{-7}$ at $\la=2,4,6,8$,
which grows algebraically more slowly than the squeezed clock of \cref{sec:sq:strong} and changes
sign on the way, along the white contours of \cref{fig:bisp:planes}. At fixed bare mass, along
which a given Lagrangian moves when its mixing is dialled up, $R\propto\sqrt\la$ and neither the
prefactor nor the bracket carries an exponential: the amplitude saturates at a value of order $10^2$.
The gradient coupling is not free,
$H/\Lambda_1=2\pi\Dzo\la$, and once that is imposed its amplitude is a prediction of order
$0.1$--$1$, far below the sensitivity of any current dataset. On the sign-change contours an
equilateral $f_{\rm NL}$ is a meaningless label for the channel, a point \cref{sec:data} meets again
in the data.

\begin{table}[t]
\centering\small\setlength{\tabcolsep}{4pt}
\begin{tabular}{l|cccccc}
\hline
$\muf=2$ & $-\lambda_2\dot\pic^{\,3}$ & $-\frac{\dot\pic^{\,2}\sigma}{2\Lambda_2}$
 & $-\frac{(\partial_i\pic)^2\sigma}{2\Lambda_1a^2}$ & $-\frac{(\partial_\mu\pic)^2\sigma}{2\Lambda_1}$
 & $-\frac{\alpha}{2}\dot\pic\sigma^2$ & $-\gthree\sigma^3$ \\
\hline
$f_{\rm NL}/c_n$, $\la=0.5$ & $392$ & $38$ & $-164$ & $-202$ & $4.3$ & $0.17$ \\
$f_{\rm NL}/c_n$, $\la=2$   & $68$ & $271$ & $-2.0\times10^{3}$ & $-2.3\times10^{3}$ & $47$ & $-731$ \\
$f_{\rm NL}/c_n$, $\la=4$   & $1.1\times10^{5}$ & $-2.2\times10^{4}$ & $-6.4\times10^{3}$
 & $1.6\times10^{4}$ & $1.1\times10^{4}$ & $-1.2\times10^{4}$ \\
\hline
\end{tabular}
\caption{Equilateral amplitude per unit dimensionless coupling, \cref{eq:bisp:reduced}, with the
couplings $c_n$ of \cref{eq:eft:cn} in the order of \cref{tab:eft:dictionary}. These numbers include
the multiplicity of each vertex and the common normalisation $\Ncal_n$ of \cref{eq:bisp:master},
which no power counting tracks. For the symmetry-fixed channel the coupling is not free, and the
amplitude is a prediction: $f_{\rm NL}=-0.12$, $-0.69$, $-0.42$ at $\la=1$, $2$, $4$.}
\label{tab:bisp:amplitudes}
\end{table}

\paragraph{Correlations.} Throughout the paper the overlap of two shapes is measured by the cosine
\begin{equation}
  \cos(S,S')\;\equiv\;\frac{\sum S\,S'}{\sqrt{\sum S^2\;\sum S'^2}} ,
\label{eq:bisp:cos}
\end{equation}
the sum running over a uniform grid of triangles in $(x_1,x_2)=(k_1/k_3,k_2/k_3)$ with
$|1-x_2|\le x_1\le x_2\le1$ and $x_1\ge10^{-3}$, spaced by $\Delta x=0.01$ as in
Ref.~\cite{Pinol:2026xnl}; it is
the shape correlator of Ref.~\cite{Babich:2004gb} with the Fisher weight replaced by unity, the
appropriate choice when comparing theoretical shapes, the Fisher-weighted version appearing only in
\cref{sec:data}. The cosines between channels show how the universal shape of \cref{sec:bisp:strong} is
approached: at $\muf=2$ the three time-derivative exchange channels have $|\cos(S^{(n)},S^{(0)})|$
above $0.92$ at $\la\le1$, falling to $0.47$--$0.70$ at $\la=2$, where the reference shape itself
crosses a sign change, and returning to $0.96$--$0.998$ for $\la\ge4$; the convergence is not
monotonic, and the gradient channel never joins them, as it cannot, since its shape carries
$\sum_cF_c$ where the others carry $e_1e_2e_3$ (its cosine with $S^{(0)}$ stays between $0.07$ and
$0.98$ over the same range).

\begin{figure}[p]
\centering
\includegraphics[width=0.92\textwidth]{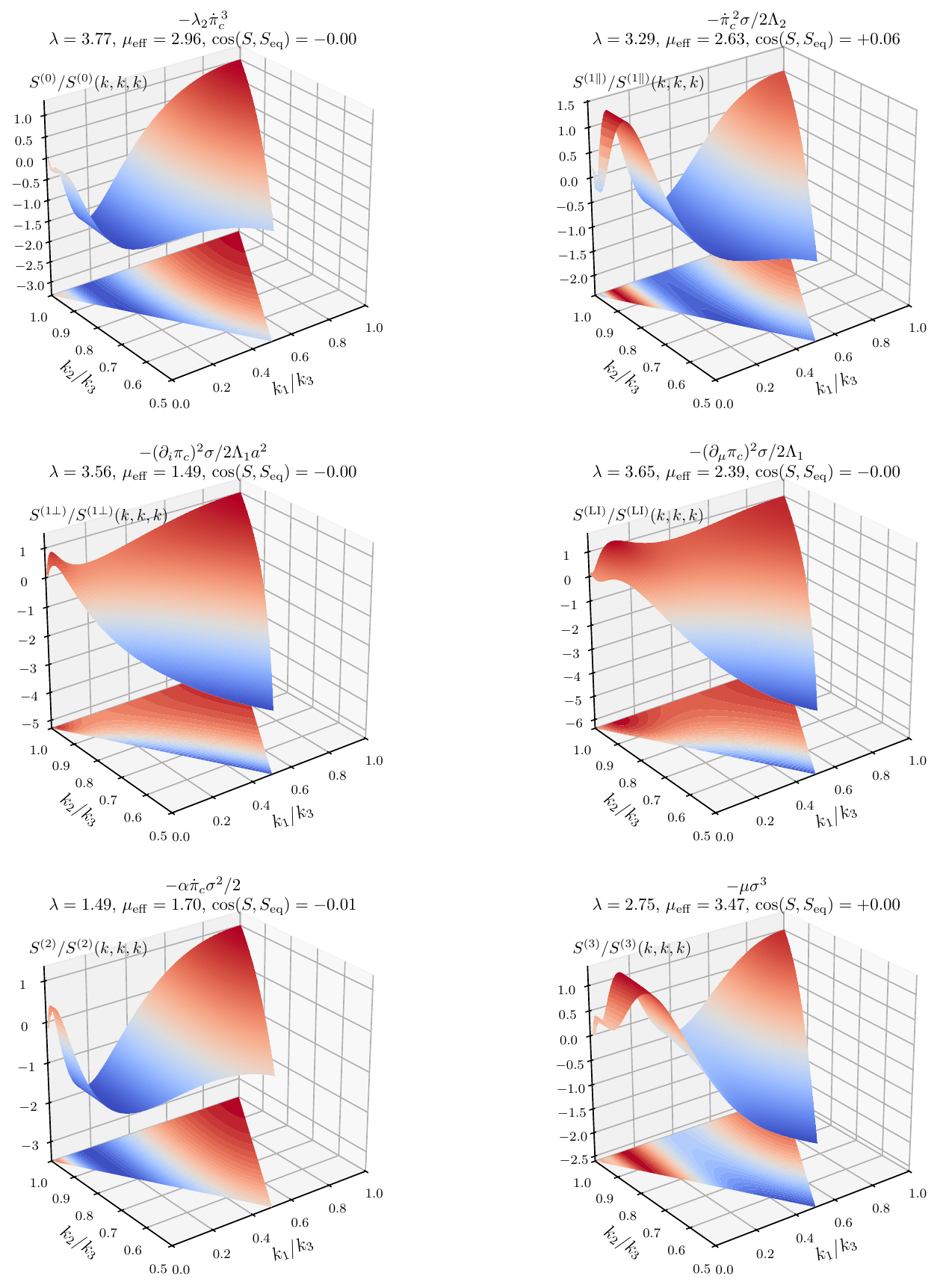}
\caption{The exact bispectrum shape of each cubic interaction over triangle configurations
$x=k_1/k_3$, $y=k_2/k_3$ with $|1-y|\le x\le y\le1$, normalised to its own equilateral value, at a
point of the plane where that channel decorrelates from the equilateral template. Each panel gives
the parameters and the cosine \cref{eq:bisp:cos}. The same field is projected below the surface,
where the altitude carries no meaning and only the colour does. The shapes peak in the folded
corner, change sign in the interior, and vanish as $\sqrt\kappa$ in the squeezed corner $k_1\to0$,
with the cosmological collider oscillations already visible at mildly squeezed configurations,
$k_1/k_3\simeq0.1$--$0.3$, and resolved in \cref{sec:sq:pheno}.}
\label{fig:bisp:shapes}
\end{figure}

The right panel of \cref{fig:bisp:planes} displays what makes a shape genuinely new: the cosine
with the equilateral
template $S_{\rm eq}=e_1e_2e_3$. At weak mixing, the only region of the plane that was accessible
without heavy numerical methods such as CosmoFlow~\cite{Werth:2024aui} before Ref.~\cite{Pinol:2026xnl}
and the present work, every channel is close to that template, the no-exchange one exactly so at
$\la\to0$ by \cref{eq:bisp:lam0}, and away from each channel's own sign-change locus every channel
has $|\cos(S,S_{\rm eq})|>0.92$ at $\la\lesssim1$: the six shapes are then indistinguishable from one
another and from the equilateral template, and only their amplitudes carry information. On the zeros of the cosine the shape is not the template, and each
channel has its own locus, because the sign change is controlled by the interference between its
legs and not by the linear theory alone. The decorrelation bands of the right panel sit on the
sign-change contours of the left one for a simple reason: the template $e_1e_2e_3$ peaks at the
equilateral point, so a shape whose equilateral value crosses zero changes sign inside the triangle
and its overlap with a template concentrated where it vanishes goes through zero with it. Those
of the no-exchange and single-exchange channels run
along the bare-massless line at different distances from it; double and triple exchange have in
addition a weak-mixing branch at nearly fixed effective mass, $\muf\simeq1.7$ and $\simeq2$, which
survives down to $\la\lesssim1$. \Cref{fig:bisp:shapes} draws the six shapes, each at a point of its
own $\cos=0$ locus. All six are strongly non-equilateral: the amplitude changes sign inside the
triangle, is largest in the folded corner $k_1+k_2\simeq k_3$, and vanishes in the squeezed corner
as $\sqrt\kappa$, the collider scaling of \cref{sec:squeezed}, with the oscillations of the
cosmological collider signal already visible at mildly squeezed configurations, $k_1/k_3\simeq0.1$--$0.3$.
Nothing of this structure survives the weak-mixing expansion.

\section{The squeezed limit and the collider clock}
\label{sec:squeezed}

\subsection{One soft leg}
\label{sec:sq:softleg}

Take $\kappa=k_1/k_3\ll1$ with $k_1$ soft, so that $e_1\to\kappa/2$, $e_{2,3}\to\tfrac12$,
$e_1e_2e_3\to\kappa/8$, $\beta_1\to\kappa\xi$ and $\beta_{2,3}\to\xi$. Because the measure
$e^{-\xi}$ confines $\xi$ to $\mathcal{O}(1)$, only the leg carrying $k_1$ is ever evaluated at
small argument, and only that leg is in the tail regime of \cref{sec:lin:tails}; the two hard legs
are analytic functions of $\kappa$. Inserting the dressed tails \cref{eq:lin:dressedtails} into
\cref{eq:bisp:shapes,eq:bisp:perp}, every channel takes the form
\begin{equation}
  \label{eq:sq:C}
  \boxed{\ \
  S^{(X)}\;\simeq_{\kappa\ll1}\;\Ncal_X\;\Real\Big[\sum_{b=\pm}\Ccal^{(X)}_b\,\kappa^{\frac12-b\n}\Big],
  \qquad
  \Ccal^{(X)}_b = c_{\rm soft}\;\Wcal_b\;J^{(N)}_b\big[\mathbb{K}_2,\mathbb{K}_3\big],\ \ }
\end{equation}
with the channel-summed tail coefficient $\Wcal_b$ of \cref{eq:lin:dressedtails}, the hard vertex
integral
\begin{equation}
  \label{eq:sq:J}
  J^{(N)}_b\big[f,g\big]\;\equiv\;\int_0^\infty\!\dd\xi\;\xi^{N-\frac12-b\n}\,e^{-\xi}\,
  f(\xi)\,g(\xi),
\end{equation}
at the channel's own measure, $N=2$ for the four vertices built from $\dot\pic$ and $\sigma$ and
$N=0$ for the gradient vertex, and a rational coefficient recording which field sits on the soft
leg:
$c_{\rm soft}=\tfrac12$ for a velocity leg, whose tail carries the $4$ of \cref{eq:lin:dressedtails},
and $c_{\rm soft}=1/(2\la)$ for a $\sigma$ leg, whose tail carries $4/\la$. Summing over the
placements of \cref{eq:bisp:shapes},
\begin{equation}
  \label{eq:sq:family}
  \begin{aligned}
  \Ccal^{(0)}_b &= \tfrac12\,\Wcal_b\,J^{(2)}_b[\bW_2,\bW_2],\\
  \Ccal^{(1\parallel)}_b &= \tfrac{1}{2\la}\,\Wcal_b\,J^{(2)}_b[\bW_2,\bW_2]\;+\;\Wcal_b\,J^{(2)}_b[\bW_2,\bV],\\
  \Ccal^{(1\perp)}_b &= \tfrac{2}{\la}\,\Wcal_b\,J^{(0)}_b[\bP,\bP],\\
  \Ccal^{(2)}_b &= \tfrac{1}{\la}\,\Wcal_b\,J^{(2)}_b[\bW_2,\bV]\;+\;\tfrac12\,\Wcal_b\,J^{(2)}_b[\bV,\bV],\\
  \Ccal^{(3)}_b &= \tfrac{1}{2\la}\,\Wcal_b\,J^{(2)}_b[\bV,\bV].
  \end{aligned}
\end{equation}
In the gradient channel the placement with a soft undifferentiated leg is suppressed by $\kappa^2$
at every $\la$, by \cref{eq:lin:Ptail}, and only the soft-$\sigma$ placement survives, with
$F_1\to\kappa/2$. In the two channels with one odd leg the clock is a sum of two structurally
different terms, one in which the soft leg is a $\sigma$ leg and a \emph{dressed-leg clock} in
which the soft line is a velocity leg and the $\sigma$'s sit hard. Exactly one leg is soft in every
term, and the soft-leg factor $\Wcal_b$ is common to all channels and all placements, contains no
quadrature, and is exact in $\la$: \emph{every squeezed clock in this paper is one closed-form soft
leg times one hard vertex integral}. Because \cref{eq:sq:C} carries a single pair of powers, only
one frequency ever appears. For a heavy field the two branches have equal modulus and combine into
the observable clock
\begin{equation}
  \label{eq:sq:clock}
  \boxed{\ \
  S^{(X)}\;\simeq_{\kappa\ll1}\;\Ncal_X\sqrt\kappa\;\Acal_X\,\sin\!\big[\muf\ln\kappa-\delta_X\big],
  \qquad
  \Acal_X e^{\ii\delta_X}=-\ii\big(\Ccal^{(X)}_++\Ccal^{(X)*}_-\big),\ \ }
\end{equation}
since $\Real[\Ccal_-\kappa^{\frac12+\ii\muf}]=\Real[\Ccal_-^*\kappa^{\frac12-\ii\muf}]$; a real
observable determines only $\Ccal_++\Ccal_-^*$. The exchange order shows up in the $\la$-scaling and in the $\muf$-dependence
of $\Acal_Xe^{\ii\delta_X}$, never in the frequency: it is $\muf$, not $2\muf$ or $3\muf$, at every
$n_\sigma$, which is the $\tfrac14$ theorem of \cref{sec:lin:tails} seen from the observable.

\subsection{Weak mixing: the clock coefficients in closed form}
\label{sec:sq:weak}

Weak mixing here means $\la^2\ll1$ at fixed mass, in the order of limits of
\cref{sec:lin:weak}: soft first, then small mixing. At this order the effective and the bare index
are interchangeable in every coefficient: $\n^2=\nbare^2-\la^2$ by \cref{eq:eft:nueff}, so that
trading one for the other changes a coefficient by a relative $\mathcal{O}(\la^2)$, which is the
order of the remainder of every row of \cref{eq:sq:weaktable}, the no-exchange one included, whose
coefficient is itself $\mathcal{O}(\la^2)$. We therefore write the bare index throughout this
subsection, $\nbare$ and, for a heavy field, $\mu\equiv-\ii\nbare=\sqrt{m^2/H^2-9/4}$, as in the
other weak-mixing parts of the paper (\cref{sec:lin:weak} and \cref{app:V2,app:double,app:dictionaries});
nowhere else does $\mu$ denote a mass, it is the $\sigma^3$ coupling of \cref{eq:eft:L3}. The
frequency of the clock \cref{eq:sq:clock} is not concerned: it is
$\muf$ exactly, at every $\la$, and the difference $\muf-\mu\simeq\la^2/2\mu$, negligible in a
coefficient, is visible in the phase $\muf\ln\kappa$ once $\kappa$ spans several decades. The
expansion of the tail coefficient is
\cref{eq:lin:Wcalweak}, whose two terms have opposite parities in $a$, so that the channel sum of
\cref{eq:lin:universalleg} keeps the even part of $\Wcal^{-a}_b$ together with the odd part of
$e^{a\pi\la/2}r_a$ and gives
\begin{equation}
  \label{eq:sq:Bweak}
  \boxed{\ \
  \Wcal_b=\la^2\,G_b\,\Scal(b\nbare)+\mathcal{O}(\la^4),
  \qquad
  \Scal(b\nbare)\equiv\Psi_b-\ii\varrho_1 ,\ \ }
\end{equation}
with $\Psi_b$ of \cref{eq:lin:V2closed} and $\varrho_1$ of \cref{eq:lin:rfrak}. The digammas
collapse to a single cotangent by the duplication and reflection formulas
\cref{eq:uf:digamma},
\begin{equation}
  \label{eq:sq:Sclosed}
  \Scal(x)=-\frac\pi2\cot\!\Big(\frac\pi4+\frac{\pi x}{2}\Big)-\frac{\ii\pi}{2}\,,
\end{equation}
so that for a heavy field the two branches differ greatly in size, $\Scal(\ii\mu)=\mathcal{O}(e^{-\pi\mu})$
while $\Scal(-\ii\mu)\to-\ii\pi$, and their observable combination is a single pair of digammas,
$\Scal(\ii\mu)+\Scal(-\ii\mu)^*=-\pi(1-\ii\sinh\pi\mu)/\cosh\pi\mu$. Since the free rotated
mode $\Vv_1$ has tail coefficients
exactly $G_b$, $\Wcal_b/\la^2=G_b\Scal(b\nbare)$ is that Bessel mode with its two non-analytic branches
re-weighted by $\Scal(\pm\nbare)$: the transfer function of a soft $\sigma$ leg with one mixing
insertion.

The hard legs are $\bW_2\to2$, $\bP\to2(1+\tfrac\xi2)$ by \cref{eq:lin:W2chan} and, by
\cref{eq:lin:sigmapower}, $\bV\to4\la\,v$ with
\begin{equation}
  \label{eq:sq:vdef}
  v(\xi)\;\equiv\;-\tfrac12\,\Vv_2(\xi)-\ii\varrho_1\,\Vv_1(\xi) ,
\end{equation}
the one-insertion dressed $\sigma$ leg per unit $4\la$, so the hard integrals of
\cref{eq:sq:family} reduce to
\begin{equation}
  \label{eq:sq:hardweak}
  \begin{gathered}
  J^{(2)}_b[\bW_2,\bW_2]\to4\Gamma(\tfrac52-b\nbare),\qquad
  J^{(2)}_b[\bW_2,\bV]\to8\la\,\Hcal_b,\qquad
  J^{(2)}_b[\bV,\bV]\to16\la^2\,\Tcal_b,\\[2pt]
  J^{(0)}_b[\bP,\bP]\to4\big[\Gamma(\tfrac12-b\nbare)+\Gamma(\tfrac32-b\nbare)
  +\tfrac14\Gamma(\tfrac52-b\nbare)\big],
  \end{gathered}
\end{equation}
where the two moments of the one-insertion dressed $\sigma$ leg,
\begin{equation}
  \label{eq:sq:Hdef}
  \Hcal_b\equiv J^{(2)}_b[v,1]=\int_0^\infty\!\dd\xi\;\xi^{\frac32-b\nbare}e^{-\xi}\,v(\xi),
  \qquad
  \Tcal_b\equiv J^{(2)}_b[v,v]=\int_0^\infty\!\dd\xi\;\xi^{\frac32-b\nbare}e^{-\xi}\,v(\xi)^2 ,
\end{equation}
are the only new objects the weak-mixing limit needs beyond $\Gamma$ functions: one dressed $\sigma$
leg is used as a tail when it is soft and as a moment when it is hard. The dressed-leg clocks are
$\la^2$-suppressed relative to the soft-$\sigma$ clocks, and the leading coefficients are those of
\cref{eq:sq:weaktable}: one soft leg and six hard integrals, with one power of $\la$ per mixing
insertion for every channel with a $\sigma$ leg. The no-exchange channel is the exception, its
clock being $\la^2$ against an analytic equilateral background of order $\la^0$, \cref{eq:bisp:lam0},
so that there the clock dominates only for $\kappa\ll\la^4$, whereas for $n_\sigma\ge1$ the clock
and the background carry the same power of $\la$ and the collider signal is visible at weak mixing
for $\kappa\ll1$ with no $\la$ penalty.
\begin{table}[t]
\centering\small
\renewcommand{\arraystretch}{1.35}
\begin{tabular}{@{}llll@{}}
\toprule
channel & $\Ccal^{(X)}_b$ at $\la\ll1$ & power & $\Acal_X/\la^{q_X}$ at $\mu\gg1$\\
\midrule
$(0)$ & $2\,\la^2\,G_b\Scal(b\nbare)\,\Gamma(\tfrac52-b\nbare)$ & $\la^2$ & $2\pi^{3/2}\,\mu^{3/2}e^{-\pi\mu}$\\
$(1\parallel)$ & $2\,\la\,G_b\Scal(b\nbare)\,\Gamma(\tfrac52-b\nbare)$ & $\la$ & $2\pi^{3/2}\,\mu^{3/2}e^{-\pi\mu}$\\
$(1\perp)$ & $8\,\la\,G_b\Scal(b\nbare)\,\big[\Gamma(\tfrac12-b\nbare)+\Gamma(\tfrac32-b\nbare)+\tfrac14\Gamma(\tfrac52-b\nbare)\big]$ & $\la$ & $2\pi^{3/2}\,\mu^{3/2}e^{-\pi\mu}$\\
$({\rm LI})$ & $4\,\la\,G_b\Scal(b\nbare)\,\Gamma(\tfrac72-b\nbare)/(\tfrac12-b\nbare)$ & $\la$ & $4\pi^{3/2}\,\mu^{3/2}e^{-\pi\mu}$\\
$(2)$ & $8\,\la^2\,G_b\Scal(b\nbare)\,\Hcal_b$, \cref{eq:sq:H} & $\la^2$ & $4\pi^{3/2}\,\mu^{-1/2}e^{-\pi\mu}$\\
$(3)$ & $8\,\la^3\,G_b\Scal(b\nbare)\,\Tcal_b$, \cref{eq:sq:Hdef,eq:sq:Texact} & $\la^3$ & $2\pi^{3/2}\,\mu^{-5/2}e^{-\pi\mu}$\\
\bottomrule
\end{tabular}
\caption{Weak-mixing squeezed clock coefficients of the five channels and of the Lorentz-invariant
combination, each with a relative $\mathcal{O}(\la^2)$ remainder. One soft leg, six hard integrals;
the third column is the power $\la^{q_X}$ of the physical shape, and the last the large-mass limit
of the observable amplitude $\Acal_X=|\Ccal^{(X)}_++\Ccal^{(X)*}_-|$ of \cref{eq:sq:clock} per unit
$\la^{q_X}$: every channel carries the perturbative Boltzmann factor $e^{-\pi\mu}$, with a power of
$\mu$ that falls by $\mu^{-2}$ per hard $\sigma$ leg. The soft factor is
$G_b=\Gamma(2b\nbare)/\Gamma(\tfrac12+b\nbare)$ of \cref{eq:lin:Wcalweak}, $\Scal$ the transcendental scalar
\cref{eq:sq:Sclosed}, $\Hcal_b$ the double-exchange moment in closed form \cref{eq:sq:H}, and $\Tcal_b$
the triple-exchange moment \cref{eq:sq:Hdef}, in closed form in \cref{eq:sq:Texact}.}
\label{eq:sq:weaktable}
\end{table}

\paragraph{No exchange.} The first row of \cref{eq:sq:weaktable} with \cref{eq:sq:clock} is the
weak-mixing collider clock of the $\dot\pic^{\,3}$ operator, fixed by $\tfrac12\times4=2$ and the
closed form \cref{eq:sq:Bweak} without any quadrature, and confirmed by the end-to-end squeezed fits
of the exact shapes (\cref{app:numerics}). It has no fixed-order counterpart in the literature, its
diagram being the contact vertex with two mixing insertions on the soft leg, and first appeared in
Ref.~\cite{Pinol:2026xnl}. In the observable combination the two branches of $\Scal$ collapse into the
pair of digammas noted after \cref{eq:sq:Sclosed},
\begin{equation}
  \label{eq:sq:C0obs}
  \boxed{\ \
  -\frac{\Ccal^{(0)}_++\Ccal^{(0)*}_-}{\la^2}=2\pi\,G_+\,\Gamma(\tfrac52-\ii\mu)\,
  \frac{1-\ii\sinh\pi\mu}{\cosh\pi\mu} ,
  \quad
  \frac{\Acal_0}{\la^2}\;\xrightarrow[\ \mu\gg1\ ]{}\;2\pi^{3/2}\,\mu^{3/2}\,e^{-\pi\mu} ,\ \ }
\end{equation}
the limit following from \cref{eq:uf:gammamod}: for a heavy field the amplitude carries the
perturbative Boltzmann factor $e^{-\pi\mu}$, from $|\Gamma(2\ii\mu)|$, $|\Gamma(\tfrac12+\ii\mu)|^{-1}$
and the vertex $|\Gamma(\tfrac52-\ii\mu)|$, with the power $\mu^{3/2}$ of \cref{eq:sq:weaktable}.

\paragraph{Single exchange.} The boost-invariant operator
$-\tfrac{1}{2\Lambda}\partial_\mu\pic\partial^\mu\pic\,\sigma$ sits at $\Lambda_1^{-1}=\Lambda^{-1}$,
$\Lambda_2^{-1}=-\Lambda^{-1}$, i.e.\ $\tilde M_3^3=0$, since in mostly-plus signature
$\partial_\mu\pic\partial^\mu\pic=-\dot\pic^{\,2}+(\partial_i\pic)^2/a^2$. Because
$\Ncal_{1\parallel}$ carries a relative minus sign, the substitution makes the two prefactors equal,
$\Ncal_{1\parallel},\Ncal_{1\perp}\to\Ncal_{\rm LI}=(64\pi R^{3/2}\Dz)^{-1}H/\Lambda$, so that
$S^{\rm LI}$ is the sum of the two brackets with a common coupling; had either sign in
\cref{sec:bisp:master} been wrong, the Lorentz-invariant operator would not produce a shape that is
the sum of its pieces. The LI row of \cref{eq:sq:weaktable} follows from
$\Gamma(A+2)+2\Gamma(A+1)+2\Gamma(A)=\Gamma(A+3)/A$ at $A=\tfrac12-b\nbare$; the pole of
$\Gamma(\tfrac12-b\nbare)$ in the gradient rows is the threshold $\nbare<\tfrac12$ that
\cref{eq:rules:convergence} imposes on every channel, visible here because $\bP$ tends to a non-zero
constant at $\xi\to0$. In the observable combination of \cref{eq:sq:clock} the two branches of
$\Scal$ recombine into the pair of digammas noted after \cref{eq:sq:Sclosed}, so that for the
Lorentz-invariant combination at the symmetry point $\Lambda^{-1}=\Lambda_1^{-1}=\rho/\fpi^2$,
\cref{eq:sq:weaktable} gives
\begin{equation}
  \label{eq:sq:Gclosed}
  \boxed{\ \
  \begin{gathered}
  -\frac{\Ccal^{\rm LI}_++\Ccal^{{\rm LI}*}_-}{\la}=
  2\sqrt\pi\;4^{\ii\mu}\,\Gamma(\ii\mu)\,\Gamma(\tfrac12-\ii\mu)\,
  (\tfrac52-\ii\mu)(\tfrac32-\ii\mu)\,\frac{1-\ii\sinh\pi\mu}{\cosh\pi\mu} ,\\[3pt]
  \frac{\Acal_{\rm LI}}{\la}\;\xrightarrow[\ \mu\gg1\ ]{}\;4\pi^{3/2}\,\mu^{3/2}\,e^{-\pi\mu} .
  \end{gathered}\ \ }
\end{equation}
This coincides identically, in modulus and phase and at every $\mu$, with the squeezed limits of
Refs.~\cite{Arkani-Hamed:2015bza,Qin:2023ejc} and with the single-flavour case of
Ref.~\cite{Pinol:2021aun}, which are thus particular cases of \cref{eq:sq:weaktable}; the maps
between their conventions and ours are in \cref{app:dictionaries}. The large-mass amplitude has
the same power of $\mu$ as no exchange, each of the two brackets contributing one half of it: the
extra terms of the gradient bracket are down by $1/\mu$, so that the velocity and gradient channels
have the same large-mass amplitude, $2\pi^{3/2}\mu^{3/2}e^{-\pi\mu}$ per unit $\la$, and differ at
that order only in phase.

\paragraph{Double exchange.} Both parities of $v$ are needed here, and all four pairings of the two
$\sigma$ legs survive at the same order: truncating $v$ to its odd$\times$odd part
$-\ii\varrho_1\Vv_1$ costs $11\%$ in the amplitude $\Acal_2$ and $0.91$~rad in the phase $\delta_2$
at $\mu=1.7$, the difference between agreeing and disagreeing with a collider template. The
odd$\times$odd moment is a moment of a Macdonald function, \cref{eq:uf:macdonald},
$J^{(2)}_b[\Vv_1,1]=\Gamma(2-2b\nbare)/\Gamma(\tfrac52-b\nbare)$; the even$\times$even moment reduces,
through the Pfaff form \cref{eq:lin:pfaff} of the weight, Euler's integral and a ${}_3F_2$ at unit
argument, to a degenerate series that Gauss's theorem cannot sum but which the two-term relation
\cref{eq:uf:threeftwo} does (\cref{app:double}):
\begin{equation}
  \label{eq:sq:H}
  \boxed{\ \
  \Hcal_b=\frac{(\tfrac52-b\nbare)\,\Gamma(\tfrac12-b\nbare)}{2\,(\tfrac32-b\nbare)}
  \;+\;\frac{2\,\Gamma(1-2b\nbare)}{(\tfrac32-b\nbare)\,\Gamma(\tfrac12-b\nbare)}
  \Big[\Scal(b\nbare)-\pi\tan(\pi b\nbare)\Big] ,\ \ }
\end{equation}
everything being $\Gamma$'s, digammas and a tangent, so that the double-exchange clock
$\Ccal^{(2)}_b/\la^2=8G_b\Scal(b\nbare)\Hcal_b$ of \cref{eq:sq:weaktable} is quadratic in the scalar
$\Scal$, with a linear term, where no and single exchange are linear in it. The route requires
$\Real\,\nbare<\tfrac12$ on the $b=+$ branch, which every heavy field satisfies. In the observable
combination of \cref{eq:sq:clock} the digammas again collapse,
\begin{equation}
  \label{eq:sq:C2obs}
  \boxed{\ \
  \begin{gathered}
  -\frac{\Ccal^{(2)}_++\Ccal^{(2)*}_-}{\la^2}=
  -2\ii\sqrt\pi\;4^{\ii\mu}\,\Gamma(\ii\mu)\,\Gamma(\tfrac12-\ii\mu)\,
  \frac{\tfrac52-\ii\mu}{\tfrac32-\ii\mu}\,\big[\tanh\pi\mu+\ii\,\sech\pi\mu\big] ,\\[3pt]
  \frac{\Acal_2}{\la^2}\;\xrightarrow[\ \mu\gg1\ ]{}\;4\pi^{3/2}\,\mu^{-1/2}\,e^{-\pi\mu} ,
  \end{gathered}\ \ }
\end{equation}
and this coincides identically, in modulus and phase and at every $\mu$, with the sum of the four
seed integrals of Ref.~\cite{Aoki:2024uyi}, the fixed-order double-exchange result with two mixing
insertions and one cubic vertex (\cref{app:dictionaries}). The large-mass amplitude sits two powers
of $\mu$ below the single-exchange law: the hard $\sigma$ leg has been integrated out, and the
reason is made explicit in the triple-exchange paragraph below.

\begin{figure}[t]
\centering
\includegraphics[width=\textwidth]{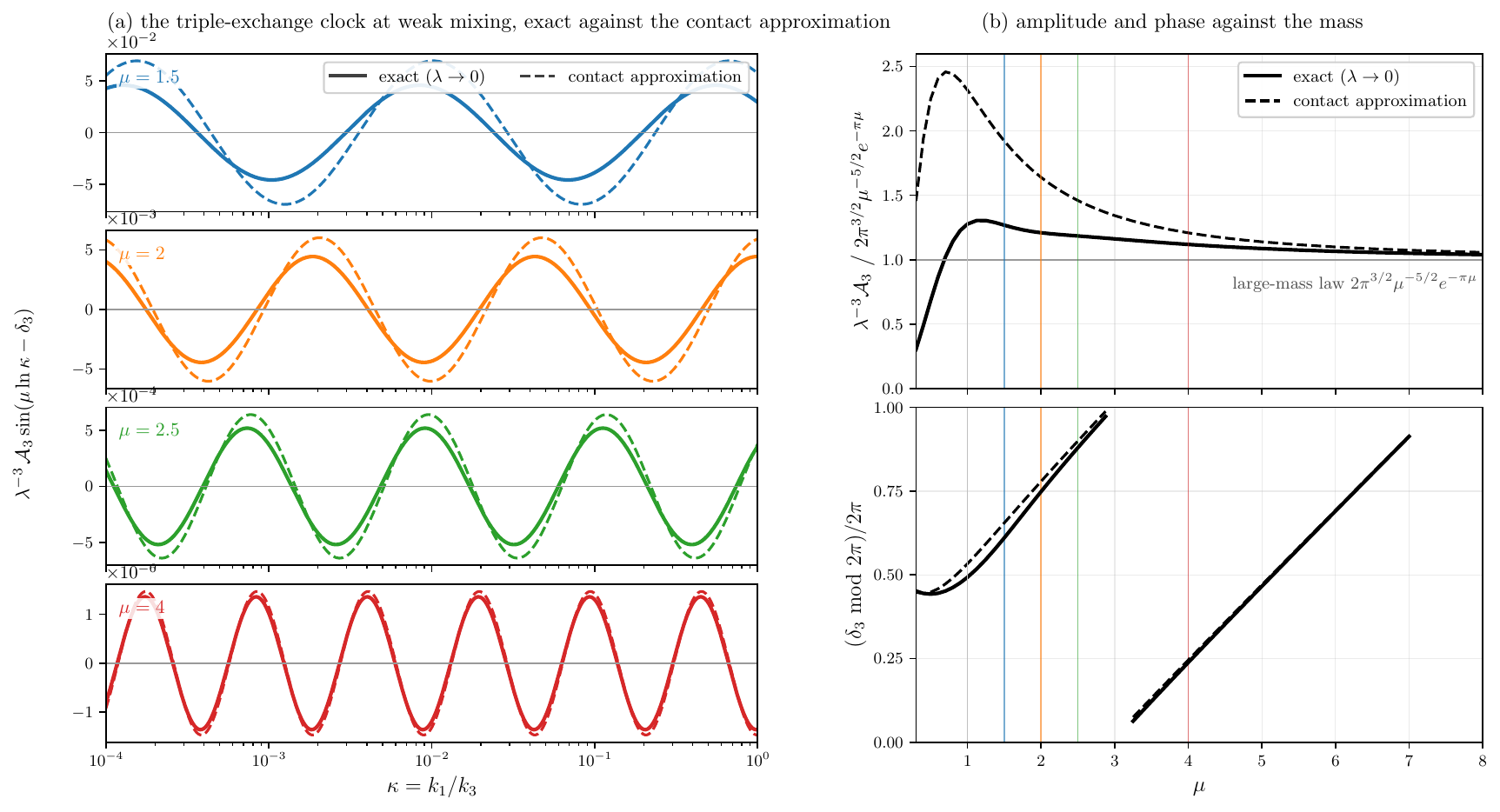}
\caption{The triple-exchange collider clock at weak mixing for a heavy field of mass parameter
$\mu=\sqrt{m^2/H^2-9/4}$. (a) The clock $\la^{-3}\Acal_3\sin(\mu\ln\kappa-\delta_3)$ at
$\mu=1.5$, $2$, $2.5$, $4$: exact, \cref{eq:sq:A3exact} (solid), against the contact approximation of
\cref{eq:sq:Tclosed} (dashed). (b) Amplitude, in units of the large-mass law of \cref{eq:sq:A3exact},
and phase, modulo $2\pi$ and in turns, against $\mu$ for the exact coefficient and the contact
approximation: the two agree above $\mu\simeq1$ and the approximation fails below.}
\label{fig:sq:triple}
\end{figure}

\paragraph{Triple exchange.} The last row of \cref{eq:sq:weaktable} is the one coefficient we
have not written in closed form exactly, but for an effectively heavy field its leading behaviour
does close, and the reason is in the $\sigma$ leg itself. Inserting \cref{eq:lin:V2closed} into
\cref{eq:sq:vdef}, the one-insertion leg is the free rotated mode with its two clock branches
re-weighted by $\Scal$, plus an entire function,
\begin{equation}
  \label{eq:sq:vsplit}
  v(\xi)=\sum_{b=\pm}\Scal(b\nbare)\,\Vv_1^{(b)}(\xi)\;-\;\frac{2}{1-4\nbare^2}\;{}_2F_2\big(1,1;\tfrac32-\nbare,\tfrac32+\nbare;\xi\big) ,
\end{equation}
with $\Vv_1^{(b)}$ the branch halves of \cref{eq:uf:V0halves}. For a heavy field the two branches
carry the coefficients $G_b\Scal(b\nbare)$ of \cref{eq:sq:Bweak}, of size $e^{-\pi\mu/2}$, and their
moments against $\xi^{3/2+\ii\mu}e^{-\xi}$ are either suppressed once more, the branch
$\xi^{-1/2+\ii\mu}$ oscillating at the frequency of the measure, or carry the doubly suppressed
$G_+\Scal(\ii\mu)=\mathcal{O}(e^{-3\pi\mu/2})$; the entire part, on the contrary, is the heavy
field integrated out. Its value at the origin, $-2/(1-4\nbare^2)=-2/(1+4\mu^2)\simeq-1/2\mu^2$, is the
contact term of the heavy propagator, and its Taylor coefficients fall as $\mu^{-2}$ per power of
$\xi$. Splitting $v$ into that constant and the rest, bilinearity gives exactly
\begin{equation}
  \label{eq:sq:Tclosed}
  \Tcal_b=-\frac{4\,\Hcal_b}{1-4\nbare^2}-\frac{4\,\Gamma(\tfrac52-b\nbare)}{(1-4\nbare^2)^2}
  +J^{(2)}_b\Big[v+\tfrac{2}{1-4\nbare^2}\,,\,v+\tfrac{2}{1-4\nbare^2}\Big] ,
\end{equation}
with $\Hcal_b$ the closed form \cref{eq:sq:H} and $J^{(2)}_b[1,1]=\Gamma(\tfrac52-b\nbare)$. The last
term is the moment of the squared deviation of the leg from its contact value, of relative order
$\mu^{-2}$, and dropping it gives the contact approximation of the moment, on which we comment
below. The moment itself closes exactly, at every mass: writing the one-insertion leg as the
Laplace transform of a pair of Legendre functions, the three moments that $\Tcal_b$ is made of
are Gauss-summable one by one (\cref{app:triple}). With $d=\tfrac12-b\nbare$, $h=1+\tfrac d2$ and
$c=\Scal(-b\nbare)$,
\begin{equation}
  \label{eq:sq:Texact}
  \boxed{\ \
  \begin{gathered}
  \Tcal_b=\mathsf{Q}(d)+2c\,\mathsf{R}(d)+c^2\,\mathsf{P}(d),\qquad
  \mathsf{P}(d)=2^{d-1}\sqrt\pi\;\frac{\Gamma(\tfrac{d+1}2)\,\Gamma(\tfrac{3d}2)\,\Gamma(1-\tfrac d2)}{\Gamma(1+\tfrac d2)},\\[3pt]
  \mathsf{Q}(d)=\frac{h^2\,\Gamma(d)}{d\,(d+1)^3}\;
  {}_6F_5\Big(\begin{matrix}1,1,1,1,h+1,h+1\\ d+2,d+2,d+2,h,h\end{matrix};1\Big),\\[3pt]
  \mathsf{R}(d)=\Gamma(d)\Big[J_1(d)-\frac{d}{2(d+1)}J_2(d)\Big],
  \end{gathered}\ \ }
\end{equation}
where $J_{1,2}$ are the combinations of ${}_4F_3$ and ${}_3F_2$ functions at unit argument of
\cref{eq:tri:J}: a finite combination of $\Gamma$ functions and ordinary hypergeometric functions,
every series absolutely convergent, with no large-mass expansion. The observable combination is
simpler still: on the physical line $v^*=v+\ii\pi\Vv_1$, and the mixed moment $\mathsf{R}$ cancels
identically between the two branches of \cref{eq:sq:clock}, leaving one hypergeometric function,
\begin{equation}
  \label{eq:sq:A3exact}
  \boxed{\ \
  \begin{gathered}
  \frac{\Acal_3e^{\ii\delta_3}}{\la^3}=-8\ii\,G_+\Big[-\pi\chi\,\mathsf{Q}(d_+)+\frac{\pi^3}{2\cos\pi\nbare}\,\mathsf{P}(d_+)\Big],
  \qquad \chi=\cot\Big(\frac\pi4+\frac{\pi\nbare}2\Big),\\[3pt]
  d_+=\tfrac12-\nbare,\qquad
  \frac{\Acal_3}{\la^3}\;\xrightarrow[\ \mu\gg1\ ]{}\;2\pi^{3/2}\,\mu^{-5/2}\,e^{-\pi\mu} ,
  \end{gathered}\ \ }
\end{equation}
the last column of \cref{eq:sq:weaktable}, since ${}_6F_5\to1+\mathcal{O}(\mu^{-3})$ and
$|\mathsf{P}|=\mathcal{O}(e^{-\pi\mu})$ at large mass. To our knowledge,
\cref{eq:sq:Texact,eq:sq:A3exact} are the first closed forms for the squeezed collider signal of the
$\sigma^3$ interaction, obtained here at weak mixing. In the large-mass limit $\Hcal_b$ is itself the
contact value times the vertex $\Gamma$, $\Hcal_b\to-2\Gamma(\tfrac52-b\nbare)/(1-4\nbare^2)$, and the
contact approximation of \cref{eq:sq:Tclosed}, $\Tcal_-\to4\Gamma(\tfrac52+\ii\mu)/(1+4\mu^2)^2$ on
the dominant branch, reproduces the same law; \cref{eq:sq:A3exact} does not require $\mu$ to be
large.
\Cref{fig:sq:triple} shows how the contact approximation performs against the exact coefficient:
its modulus is $1.52$, $1.36$, $1.23$ and $1.08$ times the exact one at $\mu=1.5$, $2$, $2.5$ and $4$,
the phase being off by at most $0.3$~rad, and it fails below $\mu\simeq1$, where the branches are no
longer subdominant; \cref{eq:sq:A3exact} agrees with the quadrature of \cref{eq:sq:Hdef} at every
mass, to $10^{-32}$ (\cref{app:numerics}). No fixed-order result exists for this channel; it rests on the prefactor
ladder \cref{eq:bisp:ladder}, three of whose four rungs are anchored above, and on the
strong-mixing checks of \cref{sec:sq:strong,app:numerics}.

\subsection{Strong mixing across the \texorpdfstring{$(\la,\muf)$}{(lambda, mu_eff)} plane}
\label{sec:sq:strong}

\begin{figure}[t]
\centering
\includegraphics[width=\textwidth]{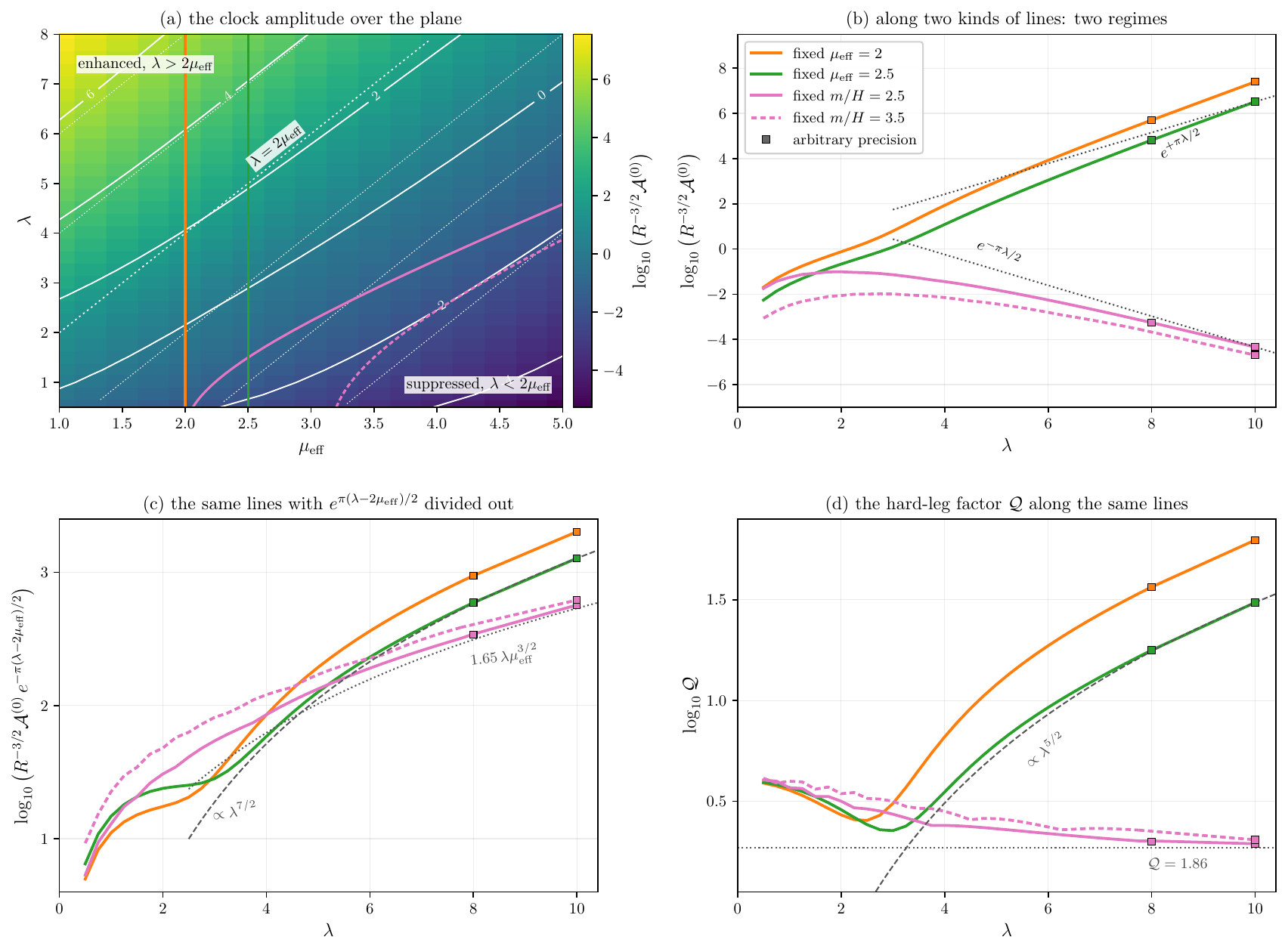}
\caption{The no-exchange collider clock across the $(\la,\muf)$ plane, \cref{sec:sq:strong}. (a) The
physical clock amplitude $R^{-3/2}\Acal^{(0)}$ on $\la\le8$, $\muf\le5$, with its contours (solid
white). The dotted straight lines are the lines $\la-2\muf=$ const of the exponent in
\cref{eq:sq:planeexp}, the thicker one $\la=2\muf$: the contours align with them at large
$\la/\muf$, which displays the exponent, and are only slightly tilted with respect to them already
at $\la=2\muf$, by the power-law prefactor that panel (c) isolates. The two verticals are the lines
of fixed effective mass $\muf=2$ and $2.5$ of the other panels, in the mass colours of
\cref{fig:sq:triple}, and the pink curves two Lagrangians of fixed bare mass, $m/H=2.5$ (solid) and
$3.5$ (dotted). (b) The same amplitude along the two verticals and along the two Lagrangians: the
former rise as $e^{\pi\la/2}$, the latter peak near $\la\simeq2$ and fall as $e^{-\pi\la/2}$, the
dotted references being anchored on the $\la=10$ points. (c) The same four lines once
$e^{\pi(\la-2\muf)/2}$ is divided out: what remains varies by powers only, over two decades, the
near-diagonal law \cref{eq:sq:diagonal} and $\la^{7/2}$ at fixed $\muf$, whereas the amplitudes of
panel (b) span up to twelve decades. (d) The hard-leg factor $\Qcal$ of \cref{eq:sq:planeB}, the one
factor not in closed form, along the same lines, with its two measured laws, $\Qcal\to1.86$ along
fixed bare mass and $\Qcal\propto\la^{5/2}$ at fixed $\muf$. In (b)--(d) the lines are the
double-precision engine up to $\la=8$ and the squares at $\la=8$ and $10$ the arbitrary-precision
engine, which carries the four lines to $\la=24$ with the same exponent (\cref{app:numerics}).}
\label{fig:sq:plane}
\end{figure}

At strong mixing the collider clock depends on two numbers, the mixing $\la$ and the effective mass
$\muf$, and nothing forces a hierarchy between them. We therefore treat them as the two coordinates
of a plane and ask one question: how large is the physical clock amplitude $R^{-3/2}\Acal$ of
\cref{eq:sq:clock} at each point? Two kinds of path through the plane matter
(\cref{fig:sq:plane}a). A given Lagrangian has a fixed bare mass $m$ and moves, as its mixing
grows, along the curve $\muf=\sqrt{\la^2+m^2/H^2-9/4}$, which approaches the diagonal $\la=\muf$
from the right: at strong mixing the effective mass of any fixed-mass field is set by the mixing
itself. Fixing the effective mass instead, as Ref.~\cite{Pinol:2026xnl} did, is a vertical line,
reached by driving the bare $m^2$ negative as $\la$ grows. With $\delta=\la-\muf$ of
\cref{eq:lin:delta}, the diagonal is $\delta=0$, fixed effective mass is $\delta\to\infty$, and the
region below the diagonal, $\muf>\la$, is where a field weighs more than it mixes.

\paragraph{The answer.} Over the whole plane,
\begin{equation}
  \label{eq:sq:planeexp}
  \boxed{\ \
  R^{-3/2}\Acal\;\simeq\;c\,\la^{P}\muf^{Q}\;e^{\pi(\la-2\muf)/2} ,\ \ }
\end{equation}
one exponential, with a prefactor that varies slowly (\cref{fig:sq:plane}): the single exponent
$\tfrac\pi2(\la-2\muf)$ interpolates between every regime studied before. It reads: the mixing
enhances the clock by $e^{\pi\la/2}$, the mass suppresses it by $e^{-\pi\muf}$, and the two balance
on the line $\la=2\muf$, not on the diagonal. At $\la\to0$ the exponent is the perturbative
Boltzmann factor $e^{-\pi\muf}$ of Ref.~\cite{Arkani-Hamed:2015bza}; at fixed effective mass and
$\la\to\infty$ it is the $e^{\pi\la/2}$ of Ref.~\cite{Pinol:2026xnl}; along the diagonal, hence for
every Lagrangian of fixed bare mass once its mixing is strong, the net is $e^{-\pi\muf/2}$, half the
Boltzmann suppression of the perturbative collider, the result of Ref.~\cite{Belrhali:2026uxn}; and
below the line $\la=2\muf$, in particular below the diagonal, the clock is suppressed as
$e^{-\pi(2\muf-\la)/2}$, ever more strongly as the field decouples. The rest of this subsection
explains where each factor comes from; the numbers behind every statement are collected in
\cref{app:numerics}.

\paragraph{Three factors.} We work with the no-exchange channel, whose clock coefficient
\cref{eq:sq:family} is $\Ccal_b=\tfrac12\Wcal_bJ^{(2)}_b[\bW_2,\bW_2]$: one soft leg, whose clock
tail is $\Wcal_b$, and two hard legs joined at the vertex. The hard integral contains the vertex
time integral. Writing each hard leg as its integral over $u$ and doing the $\xi$-integral first
with \cref{eq:uf:schwinger},
\begin{equation}
  \label{eq:sq:Jgamma}
  J^{(2)}_b\big[\mathbb{K},\mathbb{K}'\big]
  = \Gamma\big(\tfrac52-b\n\big)\;\hat J_b\big[\mathbb{K},\mathbb{K}'\big] ,
\end{equation}
where $\hat J_b$ is the pair of legs integrated over their two variables against
$(1+u+u')^{-5/2+b\n}$, and the $\Gamma$ is the time integral of the vertex against its three
plane waves, the same $\Gamma$ that appears in every row of \cref{eq:sq:weaktable}. Each branch of
the clock is therefore a product of three factors: the vertex, the soft leg and the hard legs.

\paragraph{The vertex factors out.} For a heavy field, $\n=\ii\muf$, the two branches carry
$\Gamma(\tfrac52-\ii\muf)$ and $\Gamma(\tfrac52+\ii\muf)$, complex conjugates of each other. In the
observable combination of \cref{eq:sq:clock}, which conjugates the $b=-$ branch, they become one
and the same factor,
\begin{equation}
  \label{eq:sq:planefact}
  \boxed{\ \
  \Acal e^{\ii\delta_A}=-\ii\big(\Ccal_++\Ccal_-^*\big)
  =-\frac{\ii}{2}\,\Gamma\big(\tfrac52-\ii\muf\big)\,\Wcal_-^{*}
  \Big[\hat J_-^{*}+\frac{\Wcal_+}{\Wcal_-^{*}}\,\hat J_+\Big] ,\ \ }
\end{equation}
exactly, everywhere in the plane. Its modulus is $\tfrac12|\Gamma(\tfrac52+\ii\muf)|\simeq
\sqrt{\pi/2}\,\muf^{2}e^{-\pi\muf/2}$: the vertex pays half a Boltzmann factor for producing the
pair of heavy quanta, at the time $k/a\sim\meff$, and it pays it at weak mixing and strong mixing
alike.

\paragraph{Normalising to the spectrum.} The remaining bracket contains one soft leg and two hard
legs, each dressed by the Boltzmann-weighted boundary values $e^{a\pi\la/2}r_a$. The power
spectrum is built from the same weights, $R=\tfrac12\sum_ae^{a\pi\la/2}|r_a|^2$ of \cref{sec:lin:R}:
it is the norm of a dressed leg. The physical amplitude divides by $R^{3/2}$, one $\sqrt R$ per
leg, so that
\begin{equation}
  \label{eq:sq:planeB}
  R^{-3/2}\Acal=\underbrace{\tfrac12\,\big|\Gamma(\tfrac52+\ii\muf)\big|}_{\text{vertex}}
  \;\times\;\underbrace{R^{-1/2}\big|\Wcal_-\big|}_{\text{soft leg}}
  \;\times\;\underbrace{\Qcal}_{\text{hard legs}} ,
  \qquad
  \Qcal\equiv R^{-1}\Big|\hat J_-^{*}+\frac{\Wcal_+}{\Wcal_-^{*}}\hat J_+\Big| ,
\end{equation}
and each factor can be sized on its own.

\paragraph{The soft leg.} Its normalised clock tail is the closed form \cref{eq:lin:softfree},
$R^{-1/2}|\Wcal_-|\to(\la/\sqrt{2\muf})\,e^{\pi(\la-\muf)/2}$, in which two exponentials are
visible. The $e^{-\pi\muf/2}$ is what the heavy branch of any heavy mode
costs, the same factor that the free mode's tail coefficient $G_b$ carries in \cref{eq:sq:weaktable};
the $e^{\pi\la/2}$ is the resummed mixing, which enhances the clock tail of a soft leg by the
Boltzmann weight of one channel. Their product cancels on the diagonal, which is why the soft leg
normalised as a bispectrum is a ridge there, \cref{eq:lin:ridge} and \cref{fig:lin:ingredients}a.

\paragraph{The hard legs.} At weak mixing they do nothing: $\bW_2\to2$ and $\Qcal\to4$. At strong
mixing each normalised hard leg is read in its own clock tail over the range the vertex samples
and carries $e^{\pi\delta/2}$, \cref{eq:lin:hardleg} and \cref{fig:lin:ingredients}b, so the two
terms of the bracket in \cref{eq:sq:planeB} are individually as large as $e^{\pi(\la-2\muf)}$, and
they are equally large: inserting the tails $\bW_2\simeq4\sum_b\Wcal_b\xi^{-1/2-b\n}$ into the hard
integrals, the product of two legs oscillates at the frequencies $\pm2\muf$ and $0$, the vertex
integral turns each into a $\Gamma$ whose modulus falls as $e^{-\pi\muf/2}$ per unit of frequency,
and it reads the $b=-$ branch at frequency $3\muf$ but the $b=+$ branch at $\muf$; the ratio of the
two hard integrals is then $e^{+\pi\muf}$, which undoes exactly the $e^{-\pi\muf}$ between the two
soft branches in \cref{eq:lin:Wplane}. The observable is the difference of two numbers of equal
size, and what survives that difference is the one quantity we cannot write in closed form. We
find that it carries no exponential at all: over the grid $\la\le8$, $\muf\le5$, where the
amplitude spans thirteen decades, $\Qcal$ moves between $1$ and $200$; along every curve of fixed
bare mass it tends to the same constant, $\Qcal\to1.86$, and at fixed effective mass it grows as a
power, $\Qcal\propto\la^{5/2}$ (\cref{fig:sq:plane}d and \cref{app:numerics}). \Cref{fig:sq:plane}c shows the consequence:
once $e^{\pi(\la-2\muf)/2}$ is divided out of the amplitude, the lines of fixed effective mass and
of fixed bare mass, which \cref{fig:sq:plane}b shows rising and falling with opposite exponentials,
become slowly varying and almost parallel. Nothing but this factor could have changed the
exponent, so the exponent of the clock is that of the vertex and of the soft leg alone,
$-\tfrac\pi2\muf+\tfrac\pi2(\la-\muf)=\tfrac\pi2(\la-2\muf)$, which is \cref{eq:sq:planeexp}. In words:
the resummed mixing enhances the clock tail of the soft leg by $e^{\pi\la/2}$ and does nothing
exponential to the hard legs, and the mass is paid twice at half price, once by the vertex and once
by the soft leg, exactly as in the perturbative collider, where the same two factors make up its
$e^{-\pi\muf}$.

\paragraph{Reading the plane.} Near the diagonal, where $\Qcal\simeq1.86$, the three factors
give an explicit law for a Lagrangian of fixed bare mass at strong mixing,
\begin{equation}
  \label{eq:sq:diagonal}
  R^{-3/2}\Acal\;\simeq\;1.65\;\la\,\muf^{3/2}\,e^{\pi(\la-2\muf)/2}
  \qquad(\la\simeq\muf\gg1),
\end{equation}
where the one measured number is $\Qcal$. Such a signal rises with the mixing while the mixing is
weak, peaks at the bottom of the strong-mixing regime, $\la\simeq1$--$2$, and then falls as
$e^{-\pi\la/2}$ rather than as the $e^{-\pi\la}$ a perturbative count would give
(\cref{fig:sq:plane}b): there is an optimal mixing for the collider signal of a given model, at the
bottom of the strong-mixing regime. This half Boltzmann factor along a line of fixed bare mass is
the result of Ref.~\cite{Belrhali:2026uxn}, obtained there at strong mixing; it is the diagonal
slice of \cref{eq:sq:planeexp}. At fixed effective mass the clock grows as $\la^{7/2}e^{\pi\la/2}$:
the exponential found in Ref.~\cite{Pinol:2026xnl}, the vertical slice of \cref{eq:sq:planeexp}, with
a larger power, the $\la^{5/2}$ of the hard legs times the $\la$ of the soft leg; we regard the
powers as measured and do not build on them. The exponent of \cref{eq:sq:planeexp} was in fact
already visible in Ref.~\cite{Pinol:2026xnl}: its estimate of the squeezed $f_{\rm NL}$, taken at
$\muf\gg1$, carries $e^{\pi(\la-2\muf)/2}$, but that estimate was derived for $\la\gg\muf\gg1$ only,
whereas \cref{eq:sq:planeexp} holds over the whole plane, the two published statements being two
slices of it. Below the diagonal, finally, the exponent turns into the perturbative $e^{-\pi\muf}$
as $\la/\muf\to0$, with no discontinuity anywhere: \cref{eq:sq:planeexp} joins the weak-mixing rows
of \cref{eq:sq:weaktable}, whose first row gives $\Acal\propto\la^2\muf^{3/2}e^{-\pi\muf}$, to the
strong-mixing lines, and the perturbative Boltzmann factor is simply the $\la\to0$ edge of the
plane.

\begin{figure}[t]
\centering
\includegraphics[width=\textwidth]{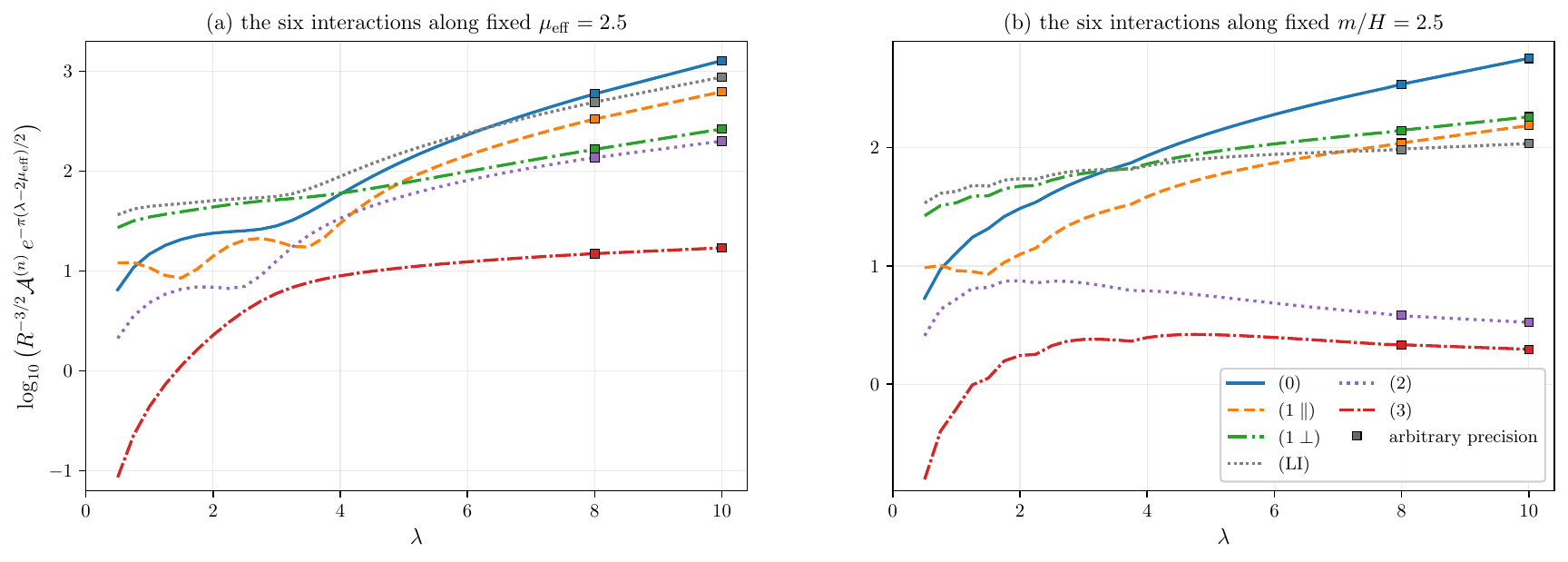}
\caption{The six interactions with the common exponent divided out: $R^{-3/2}\Acal^{(n)}e^{-\pi(\la-2\muf)/2}$
in decades, (a) along the vertical of fixed effective mass $\muf=2.5$ and (b) along the Lagrangian
of fixed bare mass $m/H=2.5$, from the double-precision engine up to $\la=8$ and the
arbitrary-precision points (squares) at $\la=8$ and $10$. The exponent of \cref{eq:sq:planeexp} is
common to all six channels, which differ by powers only.}
\label{fig:sq:planech}
\end{figure}

\paragraph{The other channels.} Two of the three factors are common to all six interactions: the
soft leg, since a velocity leg and a $\sigma$ leg have the same clock tail $\Wcal_b$ by the
$\tfrac14$ theorem of \cref{sec:lin:tails}, and the vertex, whose time integral is
$\Gamma(\tfrac52-b\n)$ for the four vertices with two time derivatives and $\Gamma(\tfrac12-b\n)$
for the gradient one, with the same $e^{-\pi\muf/2}$. Only the hard legs differ, and
\cref{fig:sq:planech} shows that they carry no exponential in any channel: once $e^{\pi(\la-2\muf)/2}$
is divided out, the six amplitudes along fixed $\muf$ and along fixed bare mass all vary by powers
only, and the channels differ by powers, at fixed $\muf$ the $\la^{-n_\sigma}$ of
\cref{eq:bisp:prop}. The exponent $\tfrac\pi2(\la-2\muf)$ is a property of the resummed mixing, not
of a vertex.

What is exact in this subsection is \cref{eq:lin:Rplane,eq:lin:Wplane,eq:lin:softfree,eq:sq:planefact,eq:sq:planeB};
what is derived, in the approximation that the hard legs are read in their clock tails, is that the
two terms of $\Qcal$ have equal moduli; what is measured, on a grid $\la\le8$, $\muf\le5$ and along
four lines carried to $\la=24$ at arbitrary precision, is that $\Qcal$ carries no exponential, with
the weak-mixing edge of the plane as its analytic anchor.

\subsection{Cosmological collider signals of all interactions}
\label{sec:sq:pheno}

\begin{figure}[p]
\centering
\includegraphics[width=0.97\textwidth]{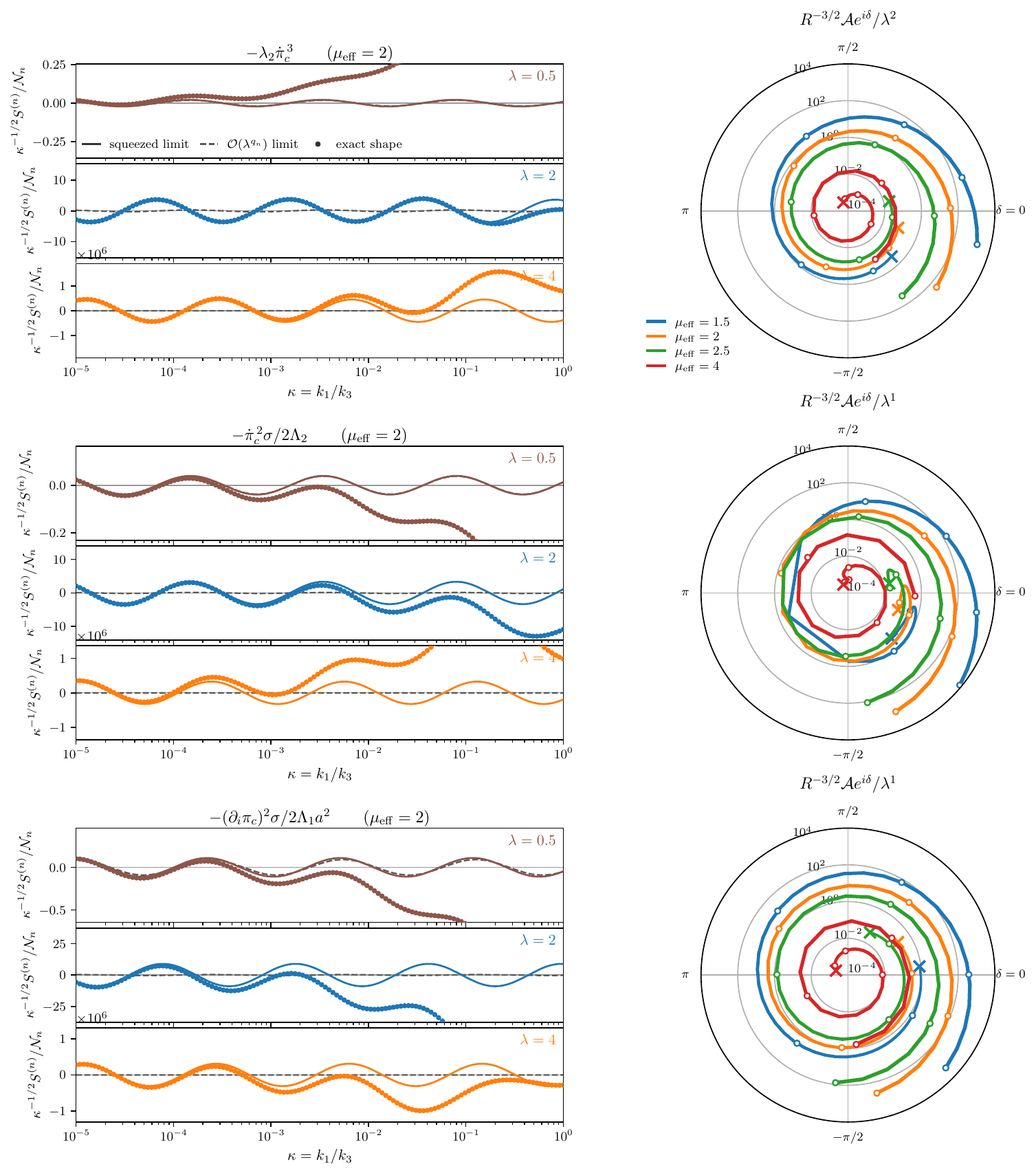}
\caption{The cosmological collider signal of the no-exchange and single-exchange interactions.
\emph{Left of each row}: $\kappa^{-1/2}S^{(n)}/\Ncal_n$ on isosceles configurations
($k_2=k_3$, $\kappa=k_1/k_3$) at $\muf=2$, for $\la=0.5$, $2$, $4$; points are the exact shape of
\cref{eq:bisp:master,eq:bisp:perp}, solid lines the closed-form squeezed limit of
\cref{sec:squeezed}, dashed lines its $\mathcal{O}(\la^{q_n})$ weak-mixing limit of
\cref{sec:sq:weak}. The vertical scale is set by the clock, so the analytic background leaves the
frame at large $\kappa$. \emph{Right}: the physical complex clock amplitude
$R^{-3/2}\Acal^{(n)} e^{\ii\delta_n}/\la^{q_n}$ as a parametric curve of $\la\in[0.05,6]$ (thickness
increases with $\la$) for $\muf=1.5$, $2$, $2.5$, $4$; open circles mark $\la=1,\dots,6$ and crosses
the $\mathcal{O}(\la^{q_n})$ prediction, with $q_n$ the weak-mixing power of
\cref{eq:sq:weaktable}. The radial scale is logarithmic.}
\label{fig:sq:clock1}
\end{figure}

\begin{figure}[p]
\centering
\includegraphics[width=0.97\textwidth]{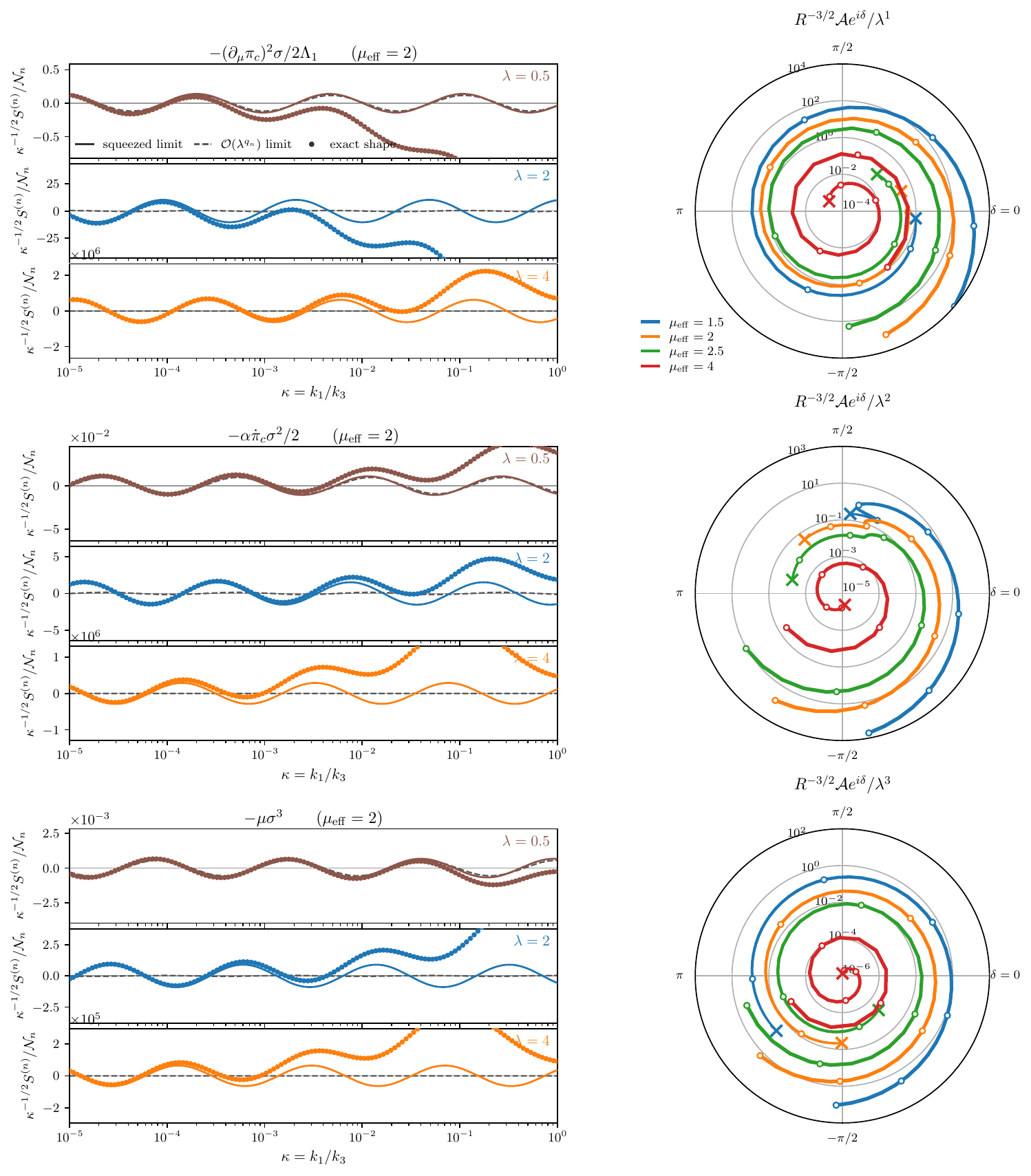}
\caption{As \cref{fig:sq:clock1}, for the boost-invariant single-exchange combination and for
double and triple exchange.}
\label{fig:sq:clock2}
\end{figure}

\Cref{fig:sq:clock1,fig:sq:clock2} put the closed forms of this section next to the exact shapes,
in the format of Fig.~2 of Ref.~\cite{Pinol:2026xnl}, for the six interactions. The left column of
each row shows
$\kappa^{-1/2}S^{(n)}/\Ncal_n$, the bracket of \cref{eq:bisp:master} with its collider scaling
divided out, on isosceles configurations $k_2=k_3$ over five decades in $\kappa=k_1/k_3$, at
$\muf=2$ and for $\la=0.5,2,4$: points are the exact shape, the solid line the closed-form squeezed
limit, the dashed line its weak-mixing limit. The right column shows the complex clock amplitude
$R^{-3/2}\Acal^{(n)}e^{\ii\delta_n}/\la^{q_n}$ as a parametric curve in $\la\in[0.05,6]$ for four
values of $\muf$, divided by the weak-mixing power $q_n=1,2,3$ of \cref{eq:sq:weaktable} that makes
the $\la\to0$ limit finite.

Three things are read off these figures. The closed-form squeezed limit captures the exact shape
over four decades in $\kappa$, the residual being the analytic background of \cref{sec:sq:softleg},
which enters with relative powers of $\sqrt\kappa$ and of $\kappa\la$, so that the agreement
improves as $\sqrt\kappa$ towards the squeezed corner, while the collider oscillations themselves
invade mildly squeezed configurations, $\kappa\simeq0.1$--$0.3$, at moderate mixing. The frequency
is $\muf$ at every exchange order, set by the effective mass and not by the
bare one, but the phase is not universal: the three mixing strengths sit at visibly different
phases at fixed $\muf$, and so do the six channels at fixed $\la$, which is the content of the
coefficients $\Ccal_b$ and what the polar panels display, each curve spiralling outwards and
rotating as $\la$ grows. And the weak-mixing closed forms of \cref{sec:sq:weak}, the crosses of the
polar panels and the dashed lines of the left ones, are excellent at $\la=0.5$ and useless beyond
$\la\simeq2$: the exact clock coefficient has left them by an order of magnitude at $\la=2$ and by
five at $\la=4$, which in the $R^{-3/2}$-normalised amplitude is a factor of a few and two orders of
magnitude respectively, and the phase is wrong by radians, except for the gradient channel, where
at $\la=4$ it is wrong by only $0.3$~rad. The perturbative treatment of the mixing therefore does
not merely mis-normalise the collider signal at strong mixing, it mis-predicts the oscillation it
describes.

The plane of \cref{sec:sq:strong} is where these figures meet the asymptotics: at fixed bare mass
the clock rises, peaks between $\la\simeq1$ and $\la\simeq2$ depending on the channel, and then
falls as $e^{-\pi\muf/2}\simeq e^{-\pi\la/2}$, while at fixed effective mass it grows as
$e^{\pi\la/2}$; there is thus an optimal mixing strength for a collider search of a given
Lagrangian, at the bottom of the strong-mixing regime. The channel dependence is in the powers,
not in the exponential (\cref{fig:sq:planech}), and the proportionality \cref{eq:bisp:prop} is a
statement about $\la\gtrsim8$: at the moderate mixings of these figures the six curves are neither
proportional nor ordered as that counting would have them, the same warning as in
\cref{sec:bisp:strong} and the reason the cosines between channels of \cref{sec:bisp:pheno} still
separate them.

\section{A first confrontation with Planck data}
\label{sec:data}

The shapes of this paper are exactly scale invariant, which makes a first confrontation with the
cosmic microwave background unusually simple. Philcox~\cite{Philcox:2026njr} has measured, from the
Planck PR4 temperature and polarisation maps, the primordial shape function itself: the bispectrum
amplitude on a grid of $171$ triangle configurations, labelled by $x=k_1/k_3$ and $y=k_2/k_3$ with
$10^{-3}\le x\le y\le1$, together with the $171\times171$ covariance of these numbers calibrated on
$400$ simulations. This binned measurement $\hat S(x,y)$ can be compared with any of our templates
directly, in the primordial variables, without transfer functions, masks or a map-level likelihood;
the price is that the information is compressed into $171$ bins, which the release documents to cost
between a few and ten per cent of the sensitivity to standard templates.

\paragraph{Method.} For a template $S(x,y)$ normalised to unity at the equilateral point, the
best-fitting amplitude and its error are given by the weighted linear regression of the data on the
template,
\begin{equation}
  \label{eq:data:estimator}
  \hat f_{\rm NL}=\frac{S^{\rm T}C^{-1}\hat S}{S^{\rm T}C^{-1}S},
  \qquad
  \sigma(f_{\rm NL})=\big(S^{\rm T}C^{-1}S\big)^{-1/2},
\end{equation}
$f_{\rm NL}$ being the amplitude in the equilateral configuration, the convention of the release.
As a check of the pipeline, the local, equilateral and orthogonal templates return
$\sigma(f_{\rm NL})=5.6$, $47$ and $25$, against $5.0$, $46$ and $21$ from the full Planck PR4
analysis~\cite{Jung:2025nss}: a loss of $12\%$, $2\%$ and $19\%$, the first and last somewhat above
the range the release quotes. Two properties of $C$ bear on the numbers below and neither is
resolved here: it is a hybrid estimate, with the correlation structure of the inverse Fisher matrix
and an amplitude calibrated on $400$ simulations, so that with $171$ bins the inversion of a purely
empirical covariance would carry a bias of order $N_{\rm bin}/N_{\rm sim}$, and we use the released
matrix as published; and the $12\%$ and $19\%$ losses are unexplained, while our templates are
weighted towards the squeezed corner more than the standard ones, so the sensitivity we quote is
indicative. Both points are among the items left to Ref.~\cite{PPRW}. Since
\cref{eq:data:estimator} is a pair of matrix products, the whole cost is the evaluation of the
templates, one kernel tabulation per point of the parameter plane shared by all six channels
(\cref{sec:implementation}), and we scan the plane adaptively in three passes: a coarse pass over
$\la\in[0.1,6]$, $\muf\in[1,4]$ at $\Delta\la=\Delta\muf=0.1$ ($1860$ points, eight minutes on two
cores), then a $\Delta=0.03$ and a $\Delta=0.01$ lattice restricted to the cells the previous pass
left above $1\sigma$ and $2\sigma$ respectively, with one cell of margin; the three passes contain
$11\,160$, $64\,590$ and $20\,304$ templates, close to $10^5$ in all for about three CPU-hours.
For comparison, Ref.~\cite{Philcox:2026tjj} reports of the order of $10^5$ CPU-hours for $4\times10^4$
CosmoFlow templates: the gain is five orders of magnitude per template, which is what makes the
strongly mixed plane scannable at the resolution of a ridge. The cosmology is held fixed, one
template is fitted at a time, and the likelihood is Gaussian. That the scan resolves what it finds must be checked, since
a ridge narrower than the grid is sampled on its flanks and its height underestimated:
\begin{center}
\begin{tabular}{@{}lccc@{}}
\toprule
$\max|\hat f_{\rm NL}/\sigma|$ & $\Delta=0.1$ & $\Delta=0.03$ & $\Delta=0.01$ \\
\midrule
$(0)$: $\dot\pic^{\,3}$         & $2.579$ & $2.759$ & $2.760$ \\
$(1\parallel)$: $\dot\pic^2\sigma$ & $1.180$ & $2.652$ & $2.696$ \\
$(3)$: $\sigma^3$               & $2.506$ & $2.690$ & $2.693$ \\
\midrule
$(1\perp)$, $(\rm LI)$, $(2)$   & $1.483$, $1.636$, $1.048$ & $1.483$, $1.636$, $1.055$ & --- \\
\bottomrule
\end{tabular}
\end{center}
Two ridges are converged at $\Delta=0.03$, but the single-exchange ridge is narrower than the coarse
grid, which straddled it and returned $1.2\sigma$ where the true maximum is $2.7\sigma$; it appears
only because the refinement was triggered at $1\sigma$. The three remaining channels have no cell
above $2\sigma$ and smooth maps, so for them the first two passes are final. A ridge that the coarse
pass samples below $1\sigma$ everywhere would still be missed.

\begin{figure}[t]
\centering
\includegraphics[width=\textwidth]{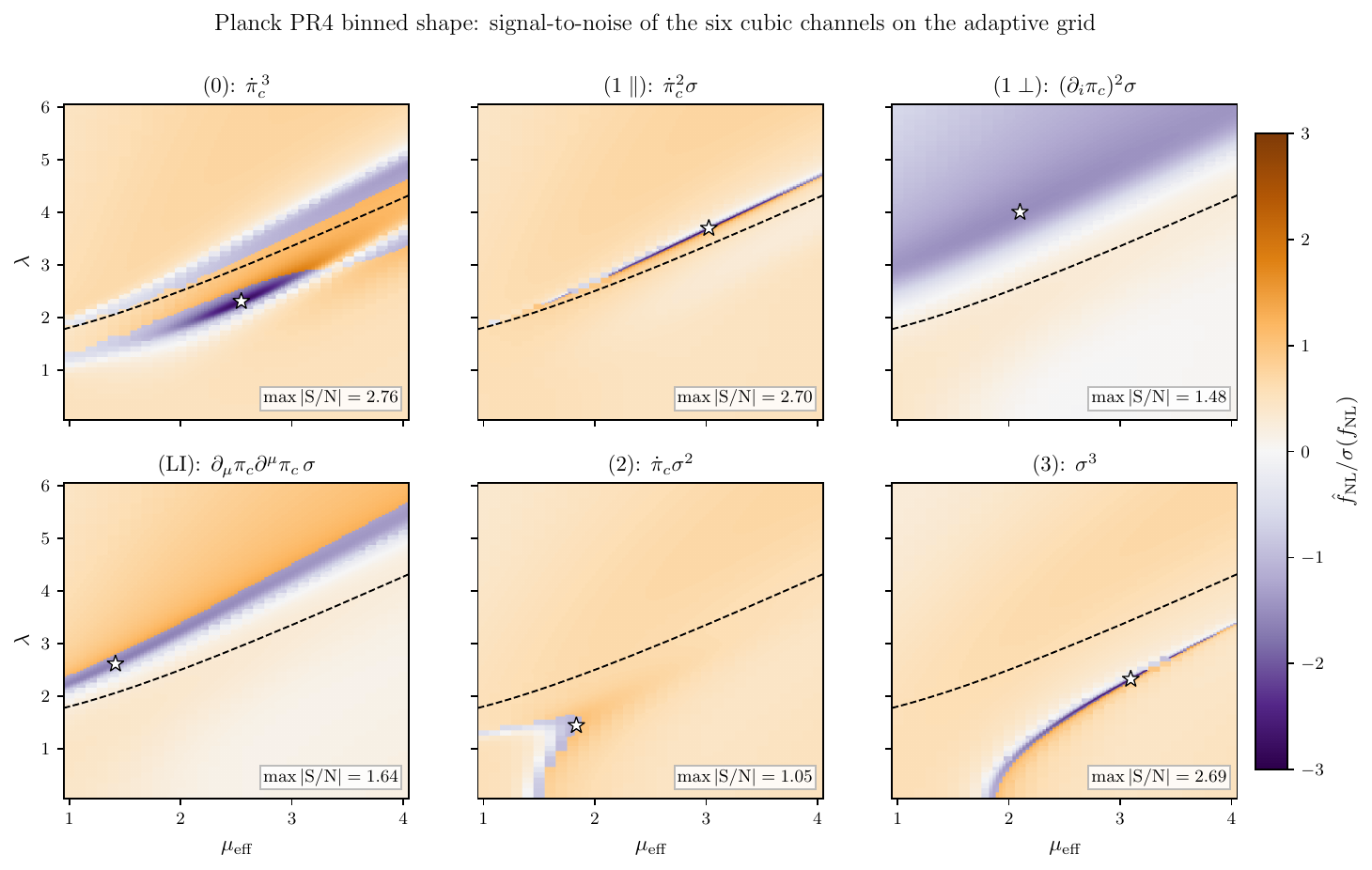}
\caption{Signal-to-noise of the six cubic channels against the Planck PR4 binned shape measurement
of Ref.~\cite{Philcox:2026njr}. Each region is drawn with the smallest pixel actually searched in
the data: $\Delta\la=\Delta\muf=0.1$ in the coarse pass, $0.03$ where the medium pass refined it
(cells above $1\sigma$) and $0.01$ where the fine pass did (cells above $2\sigma$), the three
lattices being nested. Stars mark the maxima, boxes their values. Dashed: the bare-massless line
$\la=\sqrt{9/4+\muf^2}$. The ridges follow each channel's own decorrelation band; those of
$(1\parallel)$ and $(3)$ are narrower than the coarse grid and are resolved only by the
refinement.}
\label{fig:data:six}
\end{figure}

\paragraph{Result.} \Cref{fig:data:six} shows the signal-to-noise $\hat f_{\rm NL}/\sigma(f_{\rm NL})$
of the six channels, each region drawn at the resolution at which it was searched. Away from the
decorrelation bands the templates are close to the equilateral shape, the data are consistent with
zero, and the maps are featureless: this is the regime in which the analysis returns bounds, a
representative set at $(\la,\muf)=(0.5,2)$ being $\lambda_2H^2<0.16$, $H/\Lambda_2<1.7$,
$H/\Lambda_1<0.53$, $|\alpha|<14$ and $\gthree/H<153$ at $95\%$ confidence, the first at the level
where the perturbative single-field bound sits. Along the bands, where a shape decorrelates from the
equilateral template and changes sign (\cref{sec:bisp:pheno}), narrow ridges of preferred amplitude
appear, at a position that differs from channel to channel because each band follows its own
decorrelation locus. Three channels reach comparable excursions there: $2.76\sigma$ for no exchange
at $(\la,\muf)=(2.31,2.54)$ with $\hat f_{\rm NL}=-121\pm44$, $2.70\sigma$ for single exchange at
$(3.69,3.02)$ with $-162\pm60$, and $2.69\sigma$ for triple exchange at $(2.33,3.10)$ with
$-154\pm57$. These are the largest excursions of a scan over tens of thousands of correlated
templates, not significances (see below). The other three channels stay below $1.7\sigma$
everywhere.

Two properties of these ridges matter for how the numbers should be read. First, the crest is
degenerate: for the no-exchange channel the signal-to-noise stays within one per cent of its maximum
over a stretch $\la\in[2.25,2.37]$, $\muf\in[2.48,2.62]$ of the crest, while the amplitude and its
error slide together from $\hat f_{\rm NL}=-138\pm50$ to $-105\pm38$; only the ratio is a property
of the ridge, and quoting $\hat f_{\rm NL}$ requires naming the point.\footnote{The first constraint
of this kind in the $(\la,\muf)$ plane, for the $\dot\pic^{\,3}$ interaction on the same data, was
obtained by O.~Philcox, whom we thank for sharing it with us in a private communication:
$f_{\rm NL}=-110\pm40$ at $(\la,\muf)=(2.36,2.60)$. Our results are consistent with his: our own
value at his point is $-106\pm39$, i.e.\ $2.74\sigma$.} Second, the crest of
the $\sigma^3$ channel crosses a zero of the equilateral value $S(k,k,k)$, where $\hat f_{\rm NL}$
and $\sigma$ both pass through zero with their ratio finite; the meaningful outputs there are the
signal-to-noise and the bound on the coupling. Read with that caveat, the three crests correspond to
\begin{center}\small
\begin{tabular}{@{}lccc@{}}
\toprule
 & $(0)$: $\dot\pic^{\,3}$ & $(1\parallel)$: $\dot\pic^{\,2}\sigma$ & $(3)$: $\sigma^3$ \\
\midrule
$\la=\rho/H$                      & $2.31$        & $3.69$        & $2.33$ \\
$\muf$                            & $2.54$        & $3.02$        & $3.10$ \\
$\meff/H=\sqrt{\muf^2+9/4}$       & $2.95$        & $3.37$        & $3.44$ \\
$m^2/H^2=\muf^2+9/4-\la^2$        & $+3.37$       & $-2.25$       & $+6.43$ \\
$\hat f_{\rm NL}$                 & $-121\pm44$   & $-162\pm60$   & $-154\pm57$ \\
$|\hat f_{\rm NL}/\sigma|$        & $2.76$        & $2.70$        & $2.69$ \\
cubic coupling                    & $\lambda_2H^2=-1.7\pm0.6$ & $H/\Lambda_2=+3.0\pm1.1$ & $\gthree/H=+55\pm21$ \\
extrapolated bound, \cref{eq:eft:strongbounds} & $\lesssim1.9$ & $\lesssim2.7$ & $\lesssim1.9$ \\
\bottomrule
\end{tabular}
\end{center}
where the coupling is the fitted amplitude divided by the predicted amplitude per unit coupling
$f^{(n)}_{\rm NL}/c_n$ of \cref{eq:bisp:reduced} at the same point. All three points have
$\meff\simeq3H$ and $\la\simeq2$--$4$: the fits live in the strongly mixed, effectively heavy
corner of the plane, which the perturbative expansion in $\la$ does not describe, and the
single-exchange point has $m^2<0$, above the bare-massless line, where the field is stabilised by
the mixing alone and only a resummed treatment applies. The last line sets the fitted couplings
against the perturbative-mixing bound extrapolated to their mixing, $\la^{3/4}$ of
\cref{eq:eft:strongbounds}: the no-exchange fit sits just below it, single exchange just above, and
triple exchange far above, the last because its crest crosses a zero of $S^{(3)}(k,k,k)$. We
record this as a curiosity and draw nothing from it, since we do not believe the bound to hold in
this regime (\cref{sec:eft:validity}).

\begin{figure}[t]
\centering
\includegraphics[width=\textwidth]{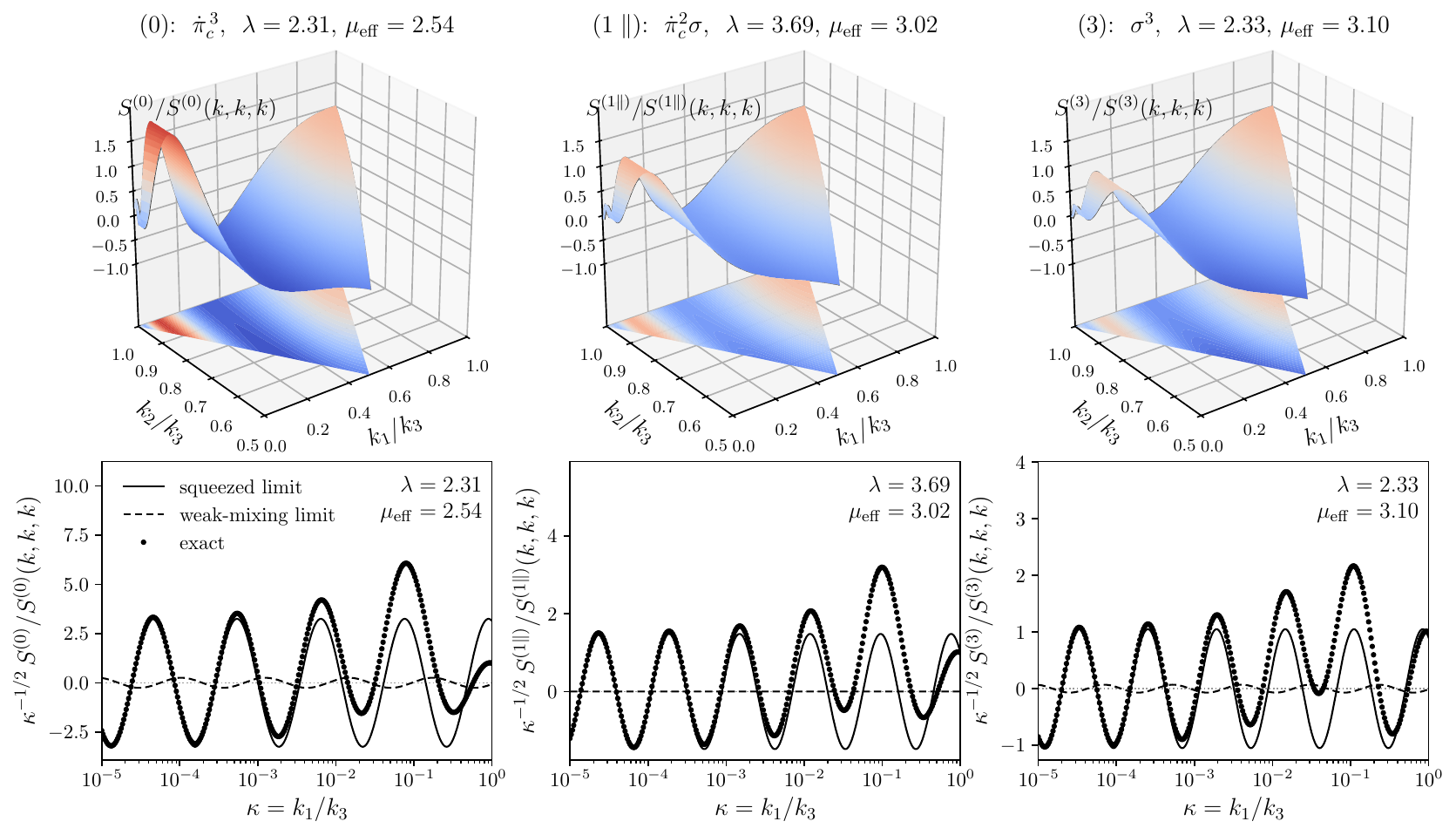}
\caption{The three best-fitting shapes: no exchange at $(\la,\muf)=(2.31,2.54)$, single exchange at
$(3.69,3.02)$ and triple exchange at $(2.33,3.10)$, each normalised to its own equilateral value
$S^{(n)}(k,k,k)$. Top: the shape over triangle configurations, on a linear scale; the same field is
projected below the surface, where the altitude carries no meaning and only the colour does. Bottom:
isosceles configurations $k_2=k_3$ against the squeezing parameter $\kappa=k_1/k_3$ on a logarithmic
scale, with the $\sqrt\kappa$ envelope divided out; dots are the exact shape, the solid line the
closed-form squeezed limit of \cref{sec:squeezed}, and the dashed line its weak-mixing limit, which
at these mixings is smaller by one to three orders of magnitude.}
\label{fig:data:shapes}
\end{figure}

\paragraph{What the preferred shapes look like.} \Cref{fig:data:shapes} draws the three of them,
and they look alike: strongly non-equilateral over the triangle, rising to nearly twice the
equilateral value towards folded configurations, changing sign in the interior and vanishing in the
squeezed corner. The Fisher correlations between the three templates, $S_a^{\rm T}C^{-1}S_b$
suitably normalised, are $0.91$, $0.93$ and $0.95$: the three excursions are one feature of the
data seen through three nearly parallel templates, as \cref{sec:bisp:strong} anticipates. On
isosceles configurations the collider oscillation in $\ln\kappa$ is visible over four decades, and
the closed form of \cref{sec:squeezed} reproduces it once $\kappa\lesssim10^{-3}$; its weak-mixing
limit, also drawn, misses the amplitude by a factor $12.6$, $1.7\times10^3$ and $14.9$ respectively
(the middle one at $\la=3.69$, the most strongly mixed of the three) and the phase by about two
radians. Nothing about these shapes is perturbative in the mixing,
which is why the resummation of \cref{sec:linear} is what makes this comparison possible at all.

\paragraph{Limitations of the current approach.} None of these ridges is a detection, and the
exercise has been kept deliberately simple; its limitations are the following. The quoted
signal-to-noise is the largest excursion in a scan over tens of thousands of correlated templates,
not a physical significance: converting it into a probability requires a look-elsewhere correction,
which lowers it, and Ref.~\cite{Suman:2025vuf} finds the effect substantial for collider signals in
Planck; we have not attempted that correction. The channels are fitted one at a time, each with a
free amplitude, whereas the theory predicts their sum: a proper analysis is joint over the couplings,
and the non-linearly realised symmetry, which we have not enforced, ties one of them to the mixing,
$H/\Lambda_1=2\pi\Dzo\la$ (\cref{sec:eft:lagrangians}), so that the gradient channel enters every
point of the plane with a fixed amplitude rather than a fitted one, and the other five ride on top
of it. The cosmology and the standard templates are held fixed rather than marginalised over, no
model comparison is made, and the scan does not reach beyond $\la\simeq6$, where the exact shapes
are also most different from the CosmoFlow templates~\cite{Werth:2024aui} used in
Ref.~\cite{Philcox:2026tjj}. All of this, together with a cross-check of the shapes against
CosmoFlow, is the subject of Ref.~\cite{PPRW}. Nor is this the first search for collider shapes in
cosmological data: Refs.~\cite{Sohn:2024xzd} and \cite{Cabass:2024wob} have constrained oscillating
primordial templates in the Planck maps and in the BOSS galaxy clustering respectively, with
perturbative templates and over a different range of masses. What this exercise does establish is
that Planck constrains the couplings of the theory at the level quoted above, and that no point of
the scanned region of the $(\la,\muf)$ plane is preferred at more than $2.8\sigma$ before any
look-elsewhere correction, which such a correction can only reduce.

\section{Conclusions}
\label{sec:conclusion}

In this paper, we have computed the primordial bispectra generated by every cubic interaction of
dimension four or less of the effective field theory of inflationary fluctuations extended by a
massive isocurvature scalar, at unit sound speed, with the curvature--isocurvature quadratic mixing
$\rho\,\dot\pic\sigma$ resummed to all orders in $\la=\rho/H$. Ref.~\cite{Pinol:2026xnl} treated the
no-exchange interaction $\dot\pic^{\,3}$; here the construction covers the four exchange
orders $n_\sigma=0,1,2,3$, with the derivations in full and the diagrammatic rules of the resummed
theory made explicit.

The structural content is the following. Placing the mixing in the free Hamiltonian and recasting
the exact linear solutions of Ref.~\cite{Huenupi:2026abj} as a single integral over one weight $\omega_a(u)$,
each channel mode collapses onto a single dressed plane wave, so that every scale-invariant
tree-level bispectrum is one Schwinger-parameter integral over products of three dressed legs, all
of which descend from one tabulated seed function $\Wn^a_0(\beta)$
(\cref{eq:bisp:master,eq:bisp:perp}). What perturbation theory calls single, double and triple
exchange are contact diagrams of the resummed theory, labelled by the number of $\sigma$ legs at the
vertex, and the four exchange orders are one calculation: a universal measure $\xi^2e^{-\xi}$, one
dressed leg per field, and a prefactor ladder. The exact relation $\Vv^a=\tfrac14(\Wn^a_2-\Wn^a_0)$
between the isocurvature and velocity kernels makes the soft tails of the two kinds of legs
proportional at every $\la$, so that the collider frequency is the single dressed $\muf$ at every
exchange order, and every squeezed clock is one closed-form soft leg times one hard integral.

At weak mixing the clock coefficients are organised by one transcendental scalar $\Scal$, and single
and double exchange reduce analytically, in amplitude and phase, to the perturbative results of
Refs.~\cite{Arkani-Hamed:2015bza,Qin:2023ejc,Pinol:2021aun,Aoki:2024uyi}, the identity
$\Scal(\ii\muf)+\Scal(-\ii\muf)^*=-\pi(1-\ii\sinh\pi\muf)/\cosh\pi\muf$ being where the
$(1-\ii\sinh\pi\muf)$ of the literature comes from; for triple exchange, for which no perturbative
result exists, we obtain the clock coefficient in closed form at every mass, as a finite
combination of $\Gamma$ functions and ordinary hypergeometric functions at unit argument, the
observable combination reducing to a single ${}_6F_5$ (\cref{app:triple}); each hard $\sigma$ leg
integrated out contributes the contact term $-1/2\mu^2$ of its propagator, so that the amplitude
falls as $\mu^{-5/2}e^{-\pi\mu}$ at large mass. At strong mixing, treating $\la$ and $\muf$ as
independent coordinates, the observable clock amplitude factorises exactly into the vertex time
integral, the soft leg and the spectrum-normalised hard legs; the first two are closed forms with
the exponentials $e^{-\pi\muf/2}$ and $e^{\pi(\la-\muf)/2}$, the third carries no exponential, and
the amplitude scales as $e^{\pi(\la-2\muf)/2}$ over the plane: the perturbative $e^{-\pi\muf}$ at
weak mixing, half of it along every line of fixed bare mass, which approaches the diagonal
$\la=\muf$, and $e^{\pi\la/2}$ at fixed effective mass, the exponential found in
Ref.~\cite{Pinol:2026xnl}, whose power of $\la$ we do not confirm and do not replace. For
$\la\gg\muf$ all channels become proportional to one another with one factor $1/\la$ per
$\sigma$ leg, and along fixed bare mass the equilateral amplitude saturates. A first confrontation of
the six exact shapes with the Planck PR4 binned bispectrum shows a preference for strong mixing,
$\la\simeq2$--$4$, where the cosmological collider oscillations invade mildly squeezed configurations,
at $2.7$--$2.8\sigma$ before any look-elsewhere correction, and yields upper bounds on the
dimensionless cubic couplings, from $0.16$ to $153$ at a representative weakly mixed point.

Several questions are left open, each stated where it arises. The hard-leg factor $\Qcal$ of the
strong-mixing clock, a constant $1.86$ along fixed bare mass and $\propto\la^{5/2}$ at fixed
effective mass, is measured, not derived, and with it the statement that it carries no exponential;
a closed form for it would complete \cref{eq:sq:planeexp}. Light fields are covered only for $\n<\tfrac12$, the
continuation of the Schwinger integral beyond that threshold being stated but not executed
(\cref{sec:rules:contour}). The contour rotation behind the Schwinger reduction
is verified numerically but lacks a uniform proof, and the loop expansion parameter of the resummed
theory is not established, the tree-level cancellation of \cref{sec:sq:strong} showing that a naive
count of Boltzmann weights cannot be trusted.

\paragraph{Towards a fully analytical construction.} Every kernel of this paper descends from the
one seed $\Wn^a_0(\beta)$ by exact relations, derivatives and one algebraic identity
(\cref{sec:lin:kernels}), and the closed forms obtained in its limits, $r_a$, the soft tails, the
weak-mixing expansion, the strong-mixing ridge and plane wave, indicate that $\Wn^a_0$ is not far
from being a special function in its own right. The route is clear: $\Wn^a_0$ is the Laplace
transform of a weight which is itself hypergeometric, so that it obeys a linear differential
equation in $\beta$ whose power series at the origin, with connection data fixed by the boundary
values and tails already in hand, would make every kernel a convergent series at any mixing and
mass. Along that route one expects to meet the Laplace-space representation of
Refs.~\cite{Belrhali:2026ygh,Belrhali:2026jqe,Belrhali:2026uxn}, since $\Wn^a_0$ is, at the end,
one Laplace transform. The present construction
already delivers what such series would: scaling laws in every regime, closed forms at weak
mixing and in the squeezed limit, and a quadrature so cheap, milliseconds per configuration
(\cref{sec:implementation}), that it competes with, or wins over, nested series at fixed precision.

The rules of \cref{sec:rules} are not restricted to the bispectrum: trispectra with internal mixed
lines and one-loop corrections reduce in the same way to Schwinger integrals over the same kernels
(\cref{sec:rules:examples}). On the observational side, the complete analysis of the Planck data
with the exact shapes, including a look-elsewhere correction, a joint treatment of the couplings and
a cross-check against CosmoFlow, is the subject of Ref.~\cite{PPRW}; since every template costs one
tabulated seed function, the strongly mixed regime, where the signal is largest, can now be searched
for in the experimental data as systematically as the weakly mixed one.

\paragraph{Theoretical data and code.} What was, until now, the expensive part of a collider search at strong
mixing, the computation of the templates themselves, is now essentially free: the six exact shapes
over the whole $(\la,\muf)$ plane, the two Python modules that compute them and the scripts behind
every figure of this paper are public~\cite{Pinol:2026code}, together with an explorer that
evaluates any shape in the browser and compares it with the Planck data of \cref{sec:data}
(\url{https://lucaspinolcnrs.github.io/exact-collider/}). A search over the plane that required
tens of thousands of numerically integrated templates in Ref.~\cite{Philcox:2026tjj} can be repeated,
extended to any interaction and pushed into the squeezed limit at the cost of a few minutes on one
core.

\paragraph*{Acknowledgements.} I am grateful to Oliver Philcox, Diederik Roest and Denis Werth for
useful discussions.

\appendix
\section{Useful formulae}
\label{app:formulae}

This appendix collects, with a word on what each one does, the identities the main text uses. None
is new; we gather them so that the body can point at an equation rather than at a name, and so
that a reader who does not carry the hypergeometric literature in their head can follow every step.
They are all standard, and can be looked up in Chapters~5, 13, 15 and 16 of the NIST Digital
Library of Mathematical Functions~\cite{DLMF}, on the $\Gamma$ and digamma functions, the
confluent, the Gauss and the generalised hypergeometric functions respectively.
Throughout, $\hyp$ is the Gauss function ${}_2F_1$, and $(a)_n=\Gamma(a+n)/\Gamma(a)$.

\paragraph{Gamma and digamma.} Besides $\Gamma(z+1)=z\Gamma(z)$ we use the reflection and
duplication formulas,
\begin{equation}
  \Gamma(z)\Gamma(1-z)=\frac{\pi}{\sin\pi z},
  \qquad
  \Gamma(2z)=\frac{2^{2z-1}}{\sqrt\pi}\,\Gamma(z)\Gamma\big(z+\tfrac12\big),
\label{eq:uf:gamma}
\end{equation}
and their consequences for imaginary argument, which are what turn ratios of $\Gamma$'s into
Boltzmann factors:
\begin{equation}
  \big|\Gamma(\tfrac12+\ii y)\big|^2=\frac{\pi}{\cosh\pi y},
  \qquad
  \big|\Gamma(\ii y)\big|^2=\frac{\pi}{y\sinh\pi y},
  \qquad
  \big|\Gamma(1+\ii y)\big|^2=\frac{\pi y}{\sinh\pi y} ,
\label{eq:uf:gammamod}
\end{equation}
each obtained from the first of \cref{eq:uf:gamma} with $z=\tfrac12+\ii y$ or $z=\ii y$ together
with $\Gamma(\bar z)=\overline{\Gamma(z)}$. The digamma $\psi=\Gamma'/\Gamma$ inherits both,
\begin{equation}
  \begin{gathered}
  \psi(z)-\psi(1-z)=-\pi\cot\pi z,
  \qquad
  \psi(\tfrac12+z)-\psi(\tfrac12-z)=\pi\tan\pi z,\\[2pt]
  \psi(2z)=\tfrac12\big[\psi(z)+\psi(z+\tfrac12)\big]+\ln2 .
  \end{gathered}
\label{eq:uf:digamma}
\end{equation}

\paragraph{The Schwinger moment.} Every vertex time integral in this paper is reduced by
\begin{equation}
  \int_0^\infty\!\dd\xi\;\xi^{s-1}e^{-\xi(1+u)} = \frac{\Gamma(s)}{(1+u)^{s}},
  \qquad \Real\,s>0 ,
\label{eq:uf:schwinger}
\end{equation}
read from left to right to trade a power of the total energy for an integral, and from right to
left to do the $\xi$-integral of a moment once the $u$-integrals have been exposed.

\paragraph{An integral representation of the Tricomi function.} The confluent function $U$ has the
representation
\begin{equation}
  U(c,b,w)=\frac{1}{\Gamma(c)}\int_0^\infty\!\dd u\;u^{c-1}(1+u)^{b-c-1}e^{-wu} ,
  \qquad \Real\,c>0 ,
\label{eq:uf:tricomi}
\end{equation}
which is what turns the mode function of \cref{sec:lin:omega} into a superposition of plane waves:
the exponential $e^{-wu}$ is the plane wave, and everything else is the weight. The condition
$\Real\,c>0$ fails for the purely imaginary $c$ we need, and \cref{app:continuation} says in what
sense the right-hand side still defines the left.

\paragraph{Pfaff's transformation.} Substituting $u=t/(1-t)$, so that $u\in(0,\infty)$ becomes
$t\in(0,1)$, and using
\begin{equation}
  \hyp(A,B;C;-u) = (1-t)^{A}\,\hyp\big(A,\;C-B;\;C;\;t\big),
  \qquad t=\frac{u}{1+u} ,
\label{eq:uf:pfaff}
\end{equation}
maps the semi-infinite $u$-integrals of the linear theory onto the unit interval, where both
endpoint behaviours are explicit powers. This is the form the numerics uses.

\paragraph{The connection formula at large argument.} A $\hyp(A,B;C;-u)$ is defined by a series in
$u$ around the origin and says nothing directly about $u\to\infty$. The connection formula
re-expands it around the point at infinity, where the equation has the two exponents $-A$ and
$-B$, as a sum of two power series, one for each:
\begin{equation}
  \hyp(A,B;C;-u) \;=\; \frac{\Gamma(C)\Gamma(B-A)}{\Gamma(B)\Gamma(C-A)}\,u^{-A}\big[1+\mathcal{O}(u^{-1})\big]
  \;+\;\big(A\leftrightarrow B\big) ,
  \qquad u\to\infty ,
\label{eq:uf:connection}
\end{equation}
valid when $B-A$ is not an integer. In this paper $A,B=\tfrac12\mp\n$, so the two exponents are
$-\tfrac12\pm\n$, and \cref{eq:uf:connection} is the single source of the two non-analytic powers
that carry the collider signal.

\paragraph{Sums at unit argument.} When the argument is $1$ the Gauss function is a ratio of
$\Gamma$'s,
\begin{equation}
  \hyp(A,B;C;1)=\frac{\Gamma(C)\,\Gamma(C-A-B)}{\Gamma(C-A)\,\Gamma(C-B)},
  \qquad \Real(C-A-B)>0 ,
\label{eq:uf:gauss}
\end{equation}
and one step up the ladder Euler's integral converts a $\hyp$ against a power measure into a
${}_3F_2$ at unit argument,
\begin{equation}
  \int_0^1\!\dd t\;t^{c-1}(1-t)^{d-1}\,\hyp(A,B;C;t)
  = \frac{\Gamma(c)\Gamma(d)}{\Gamma(c+d)}\;{}_3F_2\big(A,B,c;\;C,c+d;\;1\big) .
\label{eq:uf:euler}
\end{equation}
A generic ${}_3F_2$ at unit argument is not a ratio of $\Gamma$'s; \cref{eq:uf:gauss} is recovered
when one numerator parameter equals one denominator parameter. The two-term relation
\begin{equation}
  \begin{gathered}
  {}_3F_2\big(A,B,c;\;C,c+d;\;1\big)
  = \frac{\Gamma(C)\,\Gamma(c+d)\,\Gamma(P)}{\Gamma(A)\,\Gamma(C-A+P)\,\Gamma(c+d-A)}\\[2pt]
  \times\;{}_3F_2\big(C-B,\;c+d-B,\;P;\;C-A+P,\;P+B;\;1\big),
  \end{gathered}
\label{eq:uf:threeftwo}
\end{equation}
with $P\equiv C+d-A-B$, the parametric excess (the sum of the lower parameters minus the sum of the
upper ones), is the one place where we need more than
\cref{eq:uf:gauss}: it trades a series whose parametric excess degenerates for one whose excess
does not, and \cref{app:double} uses it in exactly that way.

\paragraph{Small-$\beta$ behaviour of a one-sided integral.} If $f(u)\simeq c\,u^{-p}$ as
$u\to\infty$ with $p<1$, then the exponential in $\int_0^\infty\dd u\,f(u)e^{-\beta u}$ cuts the
integral off at $u\sim1/\beta$, and the small-$\beta$ behaviour is controlled by that tail alone:
\begin{equation}
  \int_0^\infty\!\dd u\;f(u)\,e^{-\beta u}
  \;\simeq_{\beta\to0}\;
  c\int_0^\infty\!\dd u\;u^{-p}e^{-\beta u} \;=\; c\,\Gamma(1-p)\,\beta^{\,p-1} ,
\label{eq:uf:watson}
\end{equation}
the corrections being one power of $\beta$ down for each further power of $1/u$ in the tail, plus
the analytic terms $\beta^{n}$ generated by the part of $f$ that is integrable on its own. The
statement holds term by term for a complex exponent $p$ with $\Real\,p<1$, which is the case used
throughout \cref{sec:lin:tails}. The same argument at the other end gives the large-$\beta$
behaviour from the behaviour at $u\to0$: if $f(u)\simeq c\,u^{q-1}$ there, with $\Real\,q>0$,
\begin{equation}
  \int_0^\infty\!\dd u\;f(u)\,e^{-\beta u}\;\simeq_{\beta\to\infty}\;c\,\Gamma(q)\,\beta^{-q} ,
  \qquad
  \int_0^\infty\!\dd u\;u^{q-1}\ln u\;e^{-\beta u}=\Gamma(q)\,\beta^{-q}\big[\psi(q)-\ln\beta\big] ,
\label{eq:uf:watsonbig}
\end{equation}
the second display being the logarithmic case, which at $q=1$ reads $-(\gamma+\ln\beta)/\beta$ and
is what \cref{app:V2} needs.

\paragraph{The two halves of a Macdonald function.} Because
$K_{\n}=\tfrac{\pi}{2}\big(I_{-\n}-I_{\n}\big)/\sin\pi\n$, the free rotated mode $\Vv_1$ of
\cref{eq:lin:V0} splits into two pieces, each carrying one of the two soft powers:
\begin{equation}
  \Vv_1=\Vv_1^{(+)}+\Vv_1^{(-)},
  \qquad
  \Vv_1^{(b)}(\beta)\equiv b\,\frac{\pi}{2\sin\pi\n}\,\frac{e^{\beta/2}}{\sqrt{\pi\beta}}\,
  I_{-b\n}\big(\tfrac\beta2\big)
  \;\simeq_{\beta\to0}\; G_b\,\beta^{-\frac12-b\n} ,
\label{eq:uf:V0halves}
\end{equation}
with $G_b$ of \cref{eq:lin:Wcalweak}. We use this splitting in \cref{app:V2} and in the triple-exchange paragraph of \cref{sec:sq:weak}.

\paragraph{A Macdonald moment.} Finally, because the exponential rate and the Bessel argument
coincide in $\Vv_1$ of \cref{eq:lin:V0}, its moments collapse to $\Gamma$'s:
\begin{equation}
  \int_0^\infty\!\dd x\;e^{-x/2}K_{\n}(x/2)\,x^{w-1}
  = \sqrt\pi\;\frac{\Gamma(w+\n)\,\Gamma(w-\n)}{\Gamma(w+\tfrac12)} ,
  \qquad \Real\,w>|\Real\,\n| .
\label{eq:uf:macdonald}
\end{equation}

\section{The analytic continuation at \texorpdfstring{$u\to0$}{u->0}}
\label{app:continuation}

The integral representation of the Tricomi function used in \cref{sec:lin:omega},
\begin{equation}
  U(c,b,w)=\frac{1}{\Gamma(c)}\int_0^\infty\!\dd u\;u^{c-1}(1+u)^{b-c-1}e^{-wu} ,
\label{eq:app:tricomi}
\end{equation}
holds for $\Real\,c>0$, whereas the resummation needs $c=z_a=\ii a\la/2$, purely imaginary. Let
$\varphi(u)$ be any function analytic at the origin and decaying at infinity, with Taylor
coefficients $\varphi_m$. Splitting at $u=1$ and subtracting $M$ Taylor terms,
\begin{equation}
  \label{eq:app:continuation}
  \frac{1}{\Gamma(c)}\int_0^\infty\!\!\dd u\,u^{c-1}\varphi(u)
  = \frac{1}{\Gamma(c)}\Bigg[\sum_{m=0}^{M-1}\frac{\varphi_m}{c+m}
  + \int_0^1\!\!\dd u\,u^{c-1}\Big(\varphi-\sum_{m<M}\varphi_mu^m\Big)
  + \int_1^\infty\!\!\dd u\,u^{c-1}\varphi\Bigg].
\end{equation}
The first bracket is explicit, the second is analytic for $\Real\,c>-M$, the third is entire in $c$;
for $M\ge1$ the right-hand side is analytic in a neighbourhood of $c=0$ and defines the left-hand
side there, the only singularity at $c=0$, the pole $\varphi_0/c$, being cancelled by the
$1/\Gamma(c)$ in front. \Cref{eq:app:continuation} is at once the proof of the
continuation, the algorithm a quadrature must implement, and the source of the weak-mixing
expansion: with $M=1$ and $1/\Gamma(c)=c+\gamma c^2+\mathcal{O}(c^3)$, the pole times the leading
$c$ leaves exactly $\varphi(0)$ and every other term carries a positive power of $c$, which is
\cref{eq:lin:distrib}. In the Pfaff variable $t=u/(1+u)$ of \cref{eq:lin:pfaff} the two endpoint
powers $t^{z_a-1}$ and $(1-t)^{-1/2-\n}$ are explicit; the first is treated by
\cref{eq:app:continuation}, the second is the clock tail, integrable for $\Real\beta>0$ because
$e^{-\beta u}=e^{-\beta t/(1-t)}$, and the source of the $\beta^{-1/2\mp\n}$ tails at $\beta\to0$.

\section{The weak-mixing \texorpdfstring{$\sigma$}{sigma} leg in closed form}
\label{app:V2}

This appendix derives the closed form \cref{eq:lin:V2closed} of $\Vv_2$, the coefficient of $z_a^2$
in the expansion \cref{eq:lin:Kexpand} of the $\sigma$ kernel. Writing the kernel as
$\Vv^a=I(z_a)/\Gamma(z_a)$ with $I(z)\equiv\int_0^\infty\dd u\,u^{z}(1+u)^{-z}
\hyp(\tfrac12-\nbare,\tfrac12+\nbare;1+2z;-u)\,e^{-\beta u}$, analytic at $z=0$ with $I(0)=\Vv_1$, and
$1/\Gamma(z)=z+\gamma z^2+\mathcal{O}(z^3)$,
\begin{equation}
  \label{eq:app:V2def}
  \begin{gathered}
  \Vv_2 = \gamma\Vv_1+I_1,
  \qquad
  I_1\equiv I'(0) = \int_0^\infty\!\!\dd u\;e^{-\beta u}\,g(u),\\[2pt]
  g(u) \equiv \ln\!\Big(\tfrac{u}{1+u}\Big)F(u)
     + 2\,\partial_c\hyp\big(\tfrac12-\nbare,\tfrac12+\nbare;c;-u\big)\big|_{c=1} ,
  \end{gathered}
\end{equation}
$F(u)\equiv\hyp(\tfrac12-\nbare,\tfrac12+\nbare;1;-u)$; the three sources of $z_a$-dependence in $\omega_a$
are $(1+u)^{-z_a}$, the third parameter $1+2z_a$ of the Gauss function, and $1/\Gamma(z_a)$. The
derivation has three steps: an equation, a series, and two constants read off the exact soft tail.

\paragraph{The equation.} $\Vv_2$ is a particular solution of the same second-order equation that
$\Vv_1$ solves, with a constant source,
\begin{equation}
  \label{eq:lin:V2ode}
  \mathcal{L}\,\Vv_1=0 ,\qquad \mathcal{L}\,\Vv_2=1 ,\qquad
  \mathcal{L}\;\equiv\;\beta^2\partial_\beta^2+(2\beta-\beta^2)\partial_\beta
  +\big(\tfrac14-\nbare^2-\beta\big) ,
\end{equation}
whose indicial roots at $\beta=0$ are exactly the two soft powers $-\tfrac12\mp\nbare$. Indeed, the Gauss
function satisfies $\mathcal{P}F\equiv u(1+u)F''+(1+2u)F'+(\tfrac14-\nbare^2)F=0$. Multiplying by
$e^{-\beta u}$, integrating over $u$ with $u\mapsto-\partial_\beta$ and one integration by parts per
$\partial_u$ (every boundary term carries a factor $u$ or $u^2$ and vanishes at both ends) converts
$\mathcal{P}F=0$ into $\partial_\beta[\mathcal{L}\Vv_1]=0$, and the constant vanishes because
$\Vv_1\to0$ at large $\beta$. The integrand $g$ of $I_1$ obeys the same homogeneous equation:
differentiating the Gauss equation in $c$ at $c=1$ gives $\mathcal{P}[\partial_cF]=-u\partial_uF$, a
Leibniz expansion using $u\partial_u\ln\tfrac{u}{1+u}=(1+u)^{-1}$ gives
$\mathcal{P}[\ln\tfrac{u}{1+u}F]=+2u\partial_uF$, and the two cancel in $g$. Hence
$\partial_\beta[\mathcal{L}I_1]=0$ as well, but now the constant is not zero: since
$g(u)=\ln u+\mathcal{O}(u\ln u)$ at $u\to0$, \cref{eq:uf:watsonbig} gives
$I_1=-(\gamma+\ln\beta)/\beta+\mathcal{O}(\ln\beta/\beta^2)$ at large $\beta$, in which the only piece
of $\mathcal{L}I_1$ that survives is $-\beta(1+\beta\partial_\beta)I_1\to+1$. With
$\Vv_2=\gamma\Vv_1+I_1$ this is $\mathcal{L}\Vv_2=1$, with no condition on $\nbare$. The homogeneous
solutions of $\mathcal{L}$ at $\beta=0$ are the two branch halves $\Vv_1^{(\pm)}$ of $\Vv_1$,
\cref{eq:uf:V0halves}, so that
\begin{equation}
  \label{eq:app:V2split}
  \Vv_2 \;=\; \sum_{b=\pm}\big(\text{constant}\big)_b\,\Vv_1^{(b)} \;+\; P ,
\end{equation}
with $P$ any particular solution and two constants to be fixed.

\paragraph{The particular solution, by a two-term recursion.} Because the source is a constant and
the indicial roots $-\tfrac12\mp\nbare$ are not integers, $\mathcal{L}$ admits an \emph{analytic}
particular solution $P=\sum_{m\ge0}c_m\beta^m$. Acting on a single power,
\begin{equation}
  \label{eq:app:Lpower}
  \mathcal{L}\,\beta^m=\big[(m+\tfrac12)^2-\nbare^2\big]\beta^m-(m+1)\,\beta^{m+1} ,
\end{equation}
so matching $\mathcal{L}P=1$ order by order gives $c_0=4/(1-4\nbare^2)$ from $\beta^0$ and
\begin{equation}
  \label{eq:app:crec}
  c_m=\frac{m}{(m+\tfrac12)^2-\nbare^2}\;c_{m-1}
  =\frac{m}{(m+\tfrac12-\nbare)(m+\tfrac12+\nbare)}\;c_{m-1}
  \qquad(m\ge1)
\end{equation}
from $\beta^m$. This is the recursion that builds a hypergeometric series,
$c_m=c_0\,m!/[(\tfrac32-\nbare)_m(\tfrac32+\nbare)_m]$, whence
\begin{equation}
  \label{eq:app:V2part}
  P(\beta) \;=\; \frac{4}{1-4\nbare^2}\;{}_2F_2\!\Big(1,1;\tfrac32-\nbare,\tfrac32+\nbare;\beta\Big) ,
\end{equation}
an entire function of $\beta$.

\paragraph{The two constants, from the exact soft tail.} The two branch powers
$\beta^{-1/2\mp\nbare}$ of $\Vv_2$ can come only from the homogeneous part of \cref{eq:app:V2split},
since $P$ is analytic at the origin, so the two constants are fixed by the leading soft behaviour
of $\Vv_2$, which is already known exactly. The soft tail of the $\sigma$ kernel is
\cref{eq:lin:tails}, $\Vv^a(\beta)\simeq\sum_b\Wcal^a_b\beta^{-1/2-b\nbare}$, valid at every $\la$, and
its coefficient expands as \cref{eq:lin:Wcalweak}, $\Wcal^a_b=G_b[z_a-2z_a^2\Psi_b]+\mathcal{O}(z_a^3)$.
Comparing order by order in $z_a$ with \cref{eq:lin:Kexpand}, the term linear in $z_a$ gives the
tail $G_b$ of $\Vv_1$, and the term in $z_a^2$ gives
\begin{equation}
  \label{eq:lin:V2tail}
  \Vv_2(\beta)\;\simeq_{\beta\ll1}\;-2\sum_{b=\pm}\Psi_b\,G_b\,\beta^{-\frac12-b\nbare} .
\end{equation}
Since $\Vv_1^{(b)}$ carries the tail $G_b\beta^{-1/2-b\nbare}$, \cref{eq:app:V2split} forces the
constants to be $-2\Psi_b$, and with \cref{eq:app:V2part},
\begin{equation}
  \label{eq:app:V2closed}
  \Vv_2(\beta)\;=\;-2\sum_{b=\pm}\Psi_b\,\Vv_1^{(b)}(\beta)
  \;+\;\frac{4}{1-4\nbare^2}\;{}_2F_2\!\Big(1,1;\tfrac32-\nbare,\tfrac32+\nbare;\beta\Big) ,
\end{equation}
which is \cref{eq:lin:V2closed} of the main text, the first term being $\Vv_1$ with its two halves
weighted by $-2\Psi_\pm$ instead of equally. Inserting \cref{eq:lin:V2tail} into \cref{eq:lin:sigmapower}
gives $\bV\simeq4\la\sum_bG_b[\Psi_b-\ii\varrho_1]\beta^{-1/2-b\nbare}$, the same coefficient as the
channel sum of \cref{eq:lin:Wcalweak}: the two orders of limits of \cref{sec:lin:weak} agree on
their overlap. The pairing is easy to invert by mistake: $\Psi_b$ carries $\psi(\tfrac12+b\nbare)$ and
must sit on the branch whose power is $\beta^{-1/2-b\nbare}$, i.e.\ on $\Vv_1^{(b)}$, since both come
from the same factor $1/\Gamma(\tfrac12+b\nbare+2z_a)$ of $\Wcal^a_b$.

\paragraph{Removable singularities and regroupings.} Written as \cref{eq:app:V2closed} the
right-hand side is singular, and $\Vv_2$ regular, at $\nbare=\tfrac12$ and at every
$\nbare\in\tfrac12+\mathbb{Z}_{\ge0}$: $\Psi_-$ and the prefactor $1/(1-4\nbare^2)$ both have poles at
$\nbare=\tfrac12$, and their residues cancel because $\Vv_1^{(-)}|_{\nbare=1/2}=-(e^\beta-1)/2\beta$ against
${}_2F_2(1,1;1,2;\beta)=(e^\beta-1)/\beta$. The limit is
\begin{equation}
  \label{eq:app:V2half}
  \Vv_2(\beta)\Big|_{\nbare=1/2}\;=\;-\,\frac{\ln\beta+e^{\beta}E_1(\beta)}{\beta} ,
\end{equation}
obtained by solving $\mathcal{L}\Vv_2=1$ at $\nbare=\tfrac12$, where $\mathcal{L}$ has the solutions
$1/\beta$ and $e^\beta/\beta$, by variation of parameters with the growing solution excluded; the
two branch powers collide there and the $b=-$ coefficient is replaced by a logarithm, the ordinary
Frobenius degeneracy. At integer $\nbare$ the split $\Vv_1=\Vv_1^{(+)}+\Vv_1^{(-)}$ is itself singular,
each half carrying $1/\sin\pi\nbare$, the pole being killed in \cref{eq:app:V2closed} by
$\Psi_+-\Psi_-=\pi\tan\pi\nbare\to0$; regrouping into sum and difference,
\begin{equation}
  \label{eq:app:V2sumdiff}
  \Vv_2(\beta)=-(\Psi_++\Psi_-)\,\Vv_1(\beta)
  -\pi\tan(\pi\nbare)\Big[\Vv_1^{(+)}(\beta)-\Vv_1^{(-)}(\beta)\Big]
  +\frac{4\,{}_2F_2\big(1,1;\tfrac32-\nbare,\tfrac32+\nbare;\beta\big)}{1-4\nbare^2},
\end{equation}
is manifestly regular there. In both representations the Bessel and ${}_2F_2$ pieces grow like
$e^{\beta}$ while $\Vv_2$ decays like $\ln\beta/\beta$, so that at fixed working precision the
evaluation loses about $\beta/\ln10$ significant digits; the implementation carries that many guard
digits.

The route above fixes the two constants from the same soft tail that \cref{sec:lin:tails}
obtained from $\Wcal^a_b$, so the agreement just noted is automatic. An independent check is the
moment identity: the moment $-\tfrac12\int_0^\infty\dd\xi\,\xi^{3/2-b\nbare}e^{-\xi}\Vv_2(\xi)$
evaluated from \cref{eq:app:V2closed} reproduces the even$\times$even half of the double-exchange
moment \cref{eq:sq:H}, which is obtained in \cref{app:double} without ever touching $\Vv_2$ as a
function of $\beta$; the two agree at heavy and light $\nbare$ and on both branches
(\cref{app:numerics}). Since that moment is what is matched to the perturbative double-exchange
result, this ties the kernel that the triple channel runs on to the one external anchor of the
family.

\section{The double-exchange moment}
\label{app:double}

We derive the closed form \cref{eq:sq:H} of $\Hcal_b=\int_0^\infty\dd\xi\,\xi^{3/2-b\nbare}e^{-\xi}v(\xi)$,
$v=-\tfrac12\Vv_2-\ii\varrho_1\Vv_1$, by a route that never uses the closed form of $\Vv_2$. One
shorthand
is used throughout, the exponent of the vertex time integral,
\begin{equation}
  \label{eq:app:sdef}
  s \;\equiv\; \tfrac52-b\nbare ,
\end{equation}
and the displays are written on the branch $b=+$, the other following by $\nbare\to-\nbare$.

\paragraph{The odd$\times$odd piece.} With $\Vv_1(\beta)=e^{\beta/2}K_{\nbare}(\beta/2)/\sqrt{\pi\beta}$
and the moment \cref{eq:uf:macdonald}, which collapses to $\Gamma$'s because the exponential rate
and the Bessel argument coincide,
\begin{equation}
  \label{eq:app:V0moment}
  \int_0^\infty\!\!\dd\xi\,\xi^{\frac32-b\nbare}e^{-\xi}\,\Vv_1(\xi)
  =\frac{\Gamma(2-b\nbare+\nbare)\,\Gamma(2-b\nbare-\nbare)}{\Gamma(\tfrac52-b\nbare)}
  =\frac{\Gamma(2s-3)}{\Gamma(s)} ,
\end{equation}
one of the two numerator $\Gamma$'s being $\Gamma(2)=1$.

\paragraph{The even$\times$even piece as a ${}_3F_2$.} For $\Real\,s>0$, the Schwinger moment
\cref{eq:uf:schwinger} turns the moment of the full $\sigma$ kernel into
$\int_0^\infty\dd\xi\,\xi^{s-1}e^{-\xi}\Vv^a(\xi)=\Gamma(s)\int_0^\infty\dd u\,\omega_a(u)\,u\,(1+u)^{-s}$.
In the Pfaff form \cref{eq:lin:pfaff}, $u=t/(1-t)$ and $(1+u)^{-s}=(1-t)^s$, and Euler's integral
\cref{eq:uf:euler} gives
\begin{equation}
  \label{eq:app:3F2}
  \int_0^\infty\!\!\dd u\,\omega_a(u)\,u\,(1+u)^{-s}
  = z_a\,\frac{\Gamma(s-\tfrac12-\nbare)}{\Gamma(2s-2+z_a)}\;
  {}_3F_2\big(s-2,s-2+2z_a,1+z_a;\,1+2z_a,2s-2+z_a;\,1\big).
\end{equation}
At $z_a=0$ the ${}_3F_2$ collapses to the Gauss-summable $\hyp(s-2,s-2;2s-2;1)$ of
\cref{eq:uf:gauss}, reproducing \cref{eq:app:V0moment}; the $\mathcal{O}(z_a)$ term of the
$\Gamma$-ratio times ${}_3F_2$ is the moment of $\Vv_2$. Expanding the four Pochhammer symbols
$(s-2+2z)_n$, $(1+z)_n$, $(1+2z)_n$, $(2s-2+z)_n$ to first order gives four series. Two of them are
parameter derivatives of Gauss's theorem \cref{eq:uf:gauss} and collapse at once. The remaining two,
which come from the parameters that degenerate to $1$, produce the same series with relative
coefficients $+1$ and $-2$ and leave one Euler-type sum,
\begin{equation}
  \label{eq:app:Y}
  \Ycal\;\equiv\;\sum_{n\ge1}\frac{(s-2)_n^2}{(2s-2)_n\,n!}\,H_n
  =\int_0^1\!\dd t\;\frac{\hyp(s-2,s-2;2s-2;1)-\hyp(s-2,s-2;2s-2;t)}{1-t},
\end{equation}
with $H_n=\psi(n+1)-\psi(1)$. Gauss's theorem cannot reach $\Ycal$, because the parameters one
would have to differentiate have degenerated to integers.

\paragraph{Removing the degeneracy.} The two-term relation \cref{eq:uf:threeftwo} does reach it.
Write $\Ycal$ as a derivative in a parameter that has been pushed off the integer,
$\Ycal=-\partial_\epsilon\,{}_3F_2(s-2,s-2,1;2s-2,1+\epsilon;1)|_{\epsilon=0}$. What
\cref{eq:uf:threeftwo} does is trade a ${}_3F_2$ at unit argument for another one with the same
parametric excess, and the excess here is $2+\epsilon$, independent of the mass; the image is
\begin{equation}
  \label{eq:app:image}
  \begin{gathered}
  {}_3F_2\big(s-2,s-2,1;\,2s-2,\,1+\epsilon;\,1\big)\\[2pt]
  =\frac{\Gamma(2s-2)\,\Gamma(1+\epsilon)\,\Gamma(2+\epsilon)}
  {\Gamma(s-2)\,\Gamma(s+\epsilon)\,\Gamma(3+\epsilon)}\;
  {}_3F_2\big(s,\ 3-s+\epsilon,\ 2+\epsilon;\ s+\epsilon,\ 3+\epsilon;\ 1\big).
  \end{gathered}
\end{equation}
The point of the image is that its degeneracy has moved onto a pair of \emph{integers}: at
$\epsilon=0$ the parameter $s$ cancels between numerator and denominator, leaving
$\hyp(3-s,2;3;1)$, whose associated function of $t$ is elementary,
$\hyp(3-s,2;3;t)=2\int_0^1\dd v\,v\,(1-tv)^{s-3}$, because the two remaining parameters are $2$ and
$3$ and not functions of the mass. Differentiating \cref{eq:app:image} in $\epsilon$ gives four
series again; three are parameter derivatives of Gauss's theorem at
$\hyp(3-s,2;3;1)=2/[(s-2)(s-1)]$, and the fourth, the one carrying $\psi(s+n)$, is
$\int_0^1\dd t\,t^{\,s-1}[F(1)-F(t)]/(1-t)$ with $F$ that elementary function, i.e.\ three Beta
integrals. Assembling,
\begin{equation}
  \label{eq:app:Yclosed}
  \Ycal=\frac{2s-3}{(s-2)(s-1)^2}
  \left\{s-\frac{2\,\Gamma(2s-4)}{\Gamma(s-2)^2}
  \Big[1-2\gamma-2\psi(s-2)\Big]\right\},
\end{equation}
and feeding this back into the four series of \cref{eq:app:3F2}, where the digammas collect into the
single scalar $\Scal(b\nbare)$ of \cref{eq:sq:Bweak}, gives \cref{eq:sq:H}.

The route has a domain: \cref{eq:app:image} carries $1/\Gamma(s-2)$ and its right-hand side diverges
as $s\to2^+$, so it is valid for $\Real\,s>2$, i.e.\ $\Real\,\nbare<\tfrac12$ on the $b=+$ branch, which
every heavy field satisfies. It is also the only route we found: none of the classical summation
theorems for a ${}_3F_2$ at unit argument applies to \cref{eq:app:Y}, since they all require either
a parametric excess that this series does not have or a relation between its parameters that the
mass would have to satisfy, and contiguous relations only reproduce the same class of sum. It is
specifically the image \cref{eq:app:image}, whose excess does not depend on the mass, that works,
because the whole mass dependence is carried by the parameter that cancels.

The result was verified against a direct two-stage quadrature of the defining integral, which uses
none of this machinery, at six $(\nbare,b)$ points heavy and light, to $1.6$--$8.6\times10^{-12}$; the
relation \cref{eq:uf:threeftwo} itself to $4\times10^{-31}$ at $\epsilon=10^{-2}$; and
\cref{eq:app:Yclosed} against the series \cref{eq:app:Y} at real and complex $s$
(\cref{app:numerics}).

\section{The triple-exchange moment in closed form}
\label{app:triple}

We evaluate the weak-mixing moment $\Tcal_b$ of \cref{eq:sq:Hdef} exactly, for every mass. Write
$d\equiv\tfrac12-b\nbare$, $h\equiv1+\tfrac d2$, $s\equiv d+2$ and $c\equiv\Scal(d-\tfrac12)=\Scal(-b\nbare)$,
so that $\Tcal_b=\int_0^\infty\dd\xi\,\xi^{d+1}e^{-\xi}v(\xi)^2$; for a heavy field $\Real\,d=\tfrac12$.

\paragraph{A Legendre representation of the one-insertion leg.} The Bessel form
\cref{eq:sq:vsplit} is unsuited to squaring: its $I_{\pm\nbare}$ and ${}_2F_2$ pieces grow
exponentially while their sum decays. Instead, with the ordinary Legendre functions on the cut
$z>1$,
\begin{equation}
  \label{eq:tri:legendre}
  Q_{d-1}(1+2u)=\frac{\Gamma(d)^2}{2\Gamma(2d)}\,(1+u)^{-d}\,\hyp\Big(d,d;2d;\frac1{1+u}\Big),
  \qquad
  P_{d-1}(1+2u)=\hyp(d,1-d;1;-u),
\end{equation}
the leg is the Laplace transform
\begin{equation}
  \label{eq:tri:leg}
  v(\xi)=\int_0^\infty\!\dd u\;e^{-\xi u}\Big[Q_{d-1}(1+2u)+c\,P_{d-1}(1+2u)\Big] .
\end{equation}
To see it, let $g$ be the parameter-derivative integrand of \cref{app:V2},
$g=\tfrac{\ln u}{1+u}F+2\partial_c\hyp(d,1-d;c;-u)|_{c=1}$ with $F=P_{d-1}(1+2u)$: the Gauss
operator $\mathcal{P}$ of \cref{app:V2} gives $\mathcal{P}F=0$, $\mathcal{P}[\partial_cF_c]=-F'$
and $\mathcal{P}[F\ln u/(1+u)]=2F'$, hence $\mathcal{P}g=0$, and matching the two local
coefficients at $u\to0$, where $g=\ln u+\mathcal{O}(u\ln u)$ and
$Q_{d-1}=-\tfrac12\ln u-\gamma-\psi(d)+\ldots$, fixes $g=-2Q_{d-1}-2[\gamma+\psi(d)]P_{d-1}$. With
$\Vv_2=\gamma\Vv_1+\mathcal{L}g$ and $v=-\tfrac12\Vv_2-\ii\varrho_1\Vv_1$ this is \cref{eq:tri:leg},
the bracket $\gamma/2+\psi(d)-\ii\varrho_1$ being $\Scal(d-\tfrac12)$ by \cref{eq:uf:digamma}, and
$\mathcal{L}P_{d-1}=\Vv_1$. Both weights are logarithmic at $u\to0$ and bounded by $u^{-1/2}\ln u$
at infinity, so every moment below converges separately.

\paragraph{Three moments.} By bilinearity, with $p=\Vv_1$ and $q_d=\mathcal{L}Q_{d-1}$,
\begin{equation}
  \label{eq:tri:split}
  \Tcal_b=\mathsf{Q}(d)+2c\,\mathsf{R}(d)+c^2\,\mathsf{P}(d),
  \qquad
  (\mathsf{P},\mathsf{R},\mathsf{Q})=\int_0^\infty\!\dd\xi\,\xi^{d+1}e^{-\xi}\,(p^2,\,p\,q_d,\,q_d^2) .
\end{equation}
$\mathsf{P}$ is a Mellin moment of $K_{\nbare}^2$, $\int_0^\infty\dd y\,y^{r-1}K_{\nbare}(y)^2
=2^{r-3}\Gamma(r/2)^2\Gamma(r/2+\nbare)\Gamma(r/2-\nbare)/\Gamma(r)$ at $r=d+1$,
\begin{equation}
  \label{eq:tri:P}
  \mathsf{P}(d)=2^{d-1}\sqrt\pi\;\frac{\Gamma(\tfrac{d+1}2)\,\Gamma(\tfrac{3d}2)\,\Gamma(1-\tfrac d2)}{\Gamma(1+\tfrac d2)} .
\end{equation}
For the other two, doing the $\xi$-integral first turns the product of two Laplace transforms into
$\Gamma(s)(1+u+w)^{-s}$; with $t=u/(1+u)$, $z=w/(1+w)$ one has $1+u+w=(1-tz)/[(1-t)(1-z)]$, and
the binomial expansion of $(1-tz)^{-s}$ separates the two weights,
$\int\xi^{d+1}e^{-\xi}(\mathcal{L}W)(\mathcal{L}Z)=\Gamma(s)\sum_{k\ge0}\tfrac{(s)_k}{k!}W_kZ_k$ with
$W_k=\int_0^\infty\dd u\,(1+u)^{-s}t^kW(1+2u)$. The vertex exponent is correlated with the
Legendre degree in exactly the way that makes the $Q$ moments Gauss-summable: at $s=d+2$,
$Q_k=\tfrac{\Gamma(d)^2}{2\Gamma(2d)}B(k+1,2d+1)\,{}_3F_2(d,d,2d+1;2d,2d+k+2;1)$, a Gauss function
plus its derivative at unit argument, and
\begin{equation}
  \label{eq:tri:Qk}
  Q_k=\frac{d\,\Gamma(d)^2\,\Gamma(k+1)^2\,(k+h)}{\Gamma(k+d+2)^2}
  \quad\Longrightarrow\quad
  \mathsf{Q}(d)=\frac{h^2\,\Gamma(d)}{d\,(d+1)^3}\;
  {}_6F_5\Big(\begin{matrix}1,1,1,1,h+1,h+1\\ d+2,d+2,d+2,h,h\end{matrix};1\Big) .
\end{equation}
For the mixed moment, Pfaff and Euler transformations give
$P_k=\int_0^1\dd t\,t^k(1-t)\,\hyp(1-d,1-d;1;t)$, and summing the $k$-series with \cref{eq:tri:Qk},
\begin{equation}
  \label{eq:tri:R}
  \begin{gathered}
  \mathsf{R}(d)=\Gamma(d)\Big[J_1(d)-\frac{d}{2(d+1)}J_2(d)\Big],\\[2pt]
  J_j(d)=\int_0^1\!\dd t\,(1-t)\,\hyp(1-d,1-d;1;t)\,\hyp(1,1;d+j;t) ,
  \end{gathered}
\end{equation}
by the contiguous relation $(t\partial_t+h)\hyp(1,1;d+2;t)=(d+1)\hyp(1,1;d+1;t)-\tfrac d2\hyp(1,1;d+2;t)$.
The connection formula at $t=1$ for the second Gauss function and termwise Euler integrals reduce
each $J_j$ to ordinary hypergeometric functions at unit argument,
\begin{equation}
  \label{eq:tri:J}
  \begin{gathered}
  J_j=\frac{d+j-1}{d+j-2}\,\frac{\Gamma(1+2d)}{\Gamma(d+2)^2}\;
  {}_4F_3\Big(\begin{matrix}1,1,2,1+2d\\ 3-d-j,d+2,d+2\end{matrix};1\Big)\\[3pt]
  +\frac{\Gamma(d+j)^2\Gamma(2-d-j)\Gamma(3d+j-1)}{\Gamma(2d+j)^2}\;
  {}_3F_2\Big(\begin{matrix}d+j-1,d+j,3d+j-1\\ 2d+j,2d+j\end{matrix};1\Big),
  \end{gathered}
\end{equation}
convergent as written for $j=1$ (parametric excess $1-d$). For $j=2$ the excess is $-d$ and the
series are to be read through their analytic continuation in the parameters, which the finite
summation-by-parts identity
${}_pF_{p-1}(a;b;1)=\frac{1}{e\prod_kb_k}\sum_{r=0}^{p-2}D_r\frac{\prod_k(a_k)_r}{\prod_k(b_k+1)_r}
\,{}_pF_{p-1}(a+r;b+1+r;1)$ provides, with $e=\sum b_k-\sum a_k$ and $D_r$ the falling-power
coefficients of the polynomial $D(z)=\prod_k(z+a_k)+(e-z)\prod_k(z+b_k)$; every shifted function on
the right converges, the smallest excess being $\tfrac12$ for any mass. This completes
\cref{eq:sq:Texact}: a finite combination of $\Gamma$ functions, one ${}_6F_5$, four ${}_4F_3$ and
three ${}_3F_2$ at unit argument, all absolutely convergent, with no large-mass expansion.

\paragraph{The observable.} On the physical line $\Vv_1$ and $\Vv_2$ are real and
$\Real\,\varrho_1=\pi/2$, so that $v^*=v+\ii\pi\Vv_1$. Using the representation \cref{eq:tri:leg}
with the same $d_+=\tfrac12-\nbare$ for $v$ and $v^*$, the two branches of \cref{eq:sq:clock} carry
the coefficients $c$ and $c+\ii\pi$ of $p$, and with $a=\Scal(\nbare)=-\tfrac\pi2(\chi+\ii)$,
$c=\Scal(-\nbare)=-\tfrac\pi2(\chi^{-1}+\ii)$, $\chi\equiv\cot(\tfrac\pi4+\tfrac{\pi\nbare}2)$, the
three coefficients of $\mathsf{Q}$, $\mathsf{R}$, $\mathsf{P}$ in $a\Tcal_++(a+\ii\pi)\Tcal_-^*$ are
$-\pi\chi$, $0$ and $\pi^3/(2\cos\pi\nbare)$: the mixed moment cancels identically, which is
\cref{eq:sq:A3exact}. Its large-mass limit follows from ${}_6F_5\to1+\mathcal{O}(\mu^{-3})$,
$|\mathsf{P}|=\mathcal{O}(e^{-\pi\mu})$ and Stirling's formula.

\section{Recovering the perturbative literature}
\label{app:dictionaries}

This appendix shows that the fixed-order results of
Refs.~\cite{Arkani-Hamed:2015bza,Qin:2023ejc,Pinol:2021aun,Aoki:2024uyi} are particular cases of
the weak-mixing coefficients of \cref{eq:sq:weaktable}. For each reference we give the map between its conventions
and ours, which is needed only to make the comparison, and the identities that turn its result into
ours. The comparisons are made in $\langle\pic^3\rangle'$, the object the references quote. Undoing
the shape definition with $\Dzo=H^2/(2\pi\fpi^2)$, $\Dz=R^{1/2}\Dzo$ and
$\zeta=-H\pic/\fpi^2$, every factor of $R$ cancels and, for the single-exchange channels at the
symmetry point $\Lambda^{-1}=\Lambda_1^{-1}=\rho/\fpi^2$, i.e.\ $1/\Lambda=\la H/\fpi^2$,
\begin{equation}
  \label{eq:app:pic3sq}
  \langle\pic^3\rangle'\;\simeq_{\kappa\ll1}\;
  -\frac{H^5\la}{32\fpi^2}\,\frac{1}{k_s^{3/2}k_h^{9/2}}\;
  \Real\Big[\big(\Ccal^{\rm LI}_++\Ccal^{{\rm LI}*}_-\big)\,\kappa^{-\ii\mu}\Big],
  \qquad \kappa=\frac{k_s}{k_h},
\end{equation}
$k_s$ the soft and $k_h$ the two hard momenta, with $\Ccal^{\rm LI}_b\propto\la$ so that the whole
is $\propto\rho^2$. In each of the four references the mass parameter is the bare
$\sqrt{m^2/H^2-9/4}$, the $\mu$ of \cref{sec:sq:weak}, and we write $\mu$ for it in every
transcribed equation below; at the fixed order in $\la$ at which the comparisons are made it
coincides with $\muf$ up to the relative $\mathcal{O}(\la^2)$ dropped throughout.

\paragraph{Single exchange, Ref.~\cite{Arkani-Hamed:2015bza}.} This reference uses a single
coupling, denoted $\lambda_{\rm A}$ here, which multiplies the operator
$\int\sqrt{-g}\,(\nabla\phi)^2\sigma$, with $\phi=\phi_0+\xi$ and $\zeta=-(H/\dot\phi_0)\xi$. Since
$\fpi^4=\dot\phi_0^2$ and $\pic=\fpi^2\pi$ with $\zeta=-H\pi$,
\begin{equation}
  \label{eq:app:AHMdict}
  \xi=\pic,\qquad \rho=-2\lambda_{\rm A}\fpi^2,\qquad \Lambda^{-1}=-2\lambda_{\rm A}=\frac{\rho}{\fpi^2},
\end{equation}
the last equality not being an input: their one coupling generates both the mixing and the cubic
vertex, and the ratio it forces is our symmetry-fixed $\Lambda_1^{-1}=\rho/\fpi^2$. Their squeezing
variable is $k_3/k_1$ with $k_3$ soft, our $\kappa$; they set $H=1$, and dimensional analysis
restores a unique $H^3$. Their Eq.~(6.130) reads
$\langle\xi^3\rangle'=[\dot\phi_0\lambda_{\rm A}^2H^3/(2k_1^3k_3^3)]\,J(\mu,k_3/k_1)$, whose prefactor
is $\rho^2/(8\fpi^2)=H^2\la^2/(8\fpi^2)$, and their Eq.~(6.131) is
\begin{equation}
  \label{eq:app:AHM}
  J=\frac{\pi^2}{4\sqrt\pi\cosh^2\pi\mu}\Big(\frac{k_3}{k_1}\Big)^{3/2}
  \bigg[\Big(\frac{k_3}{4k_1}\Big)^{-\ii\mu}\!\!(1-\ii\sinh\pi\mu)
  (\tfrac52-\ii\mu)(\tfrac32-\ii\mu)\frac{\Gamma(\ii\mu)}{\Gamma(\tfrac12+\ii\mu)}+{\rm c.c.}\bigg].
\end{equation}
With $\kappa^{3/2}/(k_1^3k_3^3)=1/(k_s^{3/2}k_h^{9/2})$ and
$1/\Gamma(\tfrac12+\ii\mu)=\Gamma(\tfrac12-\ii\mu)\cosh\pi\mu/\pi$, their result becomes
$(H^5\la^2\sqrt\pi/(16\fpi^2\cosh\pi\mu))(k_s^{3/2}k_h^{9/2})^{-1}\Real[4^{\ii\mu}\Gamma(\ii\mu)\Gamma(\tfrac12-\ii\mu)(\tfrac52-\ii\mu)(\tfrac32-\ii\mu)(1-\ii\sinh\pi\mu)\kappa^{-\ii\mu}]$,
which is \cref{eq:app:pic3sq} with \cref{eq:sq:Gclosed} inserted, term by term. One reading detail
matters: the denominator of \cref{eq:app:AHM} carries $\cosh^2\pi\mu$, not $\cosh\pi\mu$; with a
single power the amplitude of $J$ would grow like $\mu^{3/2}$ instead of decaying like
$\mu^{3/2}e^{-\pi\mu}$, contradicting the reference's own statement of an exponential suppression,
and a text extraction that drops the exponent manufactures a spurious discrepancy of exactly
$1/\cosh\pi\mu$.

\paragraph{Single exchange, Ref.~\cite{Qin:2023ejc}.} Its \S2.2.1 uses
$O_{\phi\phi\sigma}=\tfrac12(-\tau)^{-2}\eta^{\mu\nu}\partial_\mu\phi\partial_\nu\phi\,\sigma$ and
$O_{\phi\sigma}=(-\tau)^{-3}\phi'\sigma$ with unit couplings, $H=1$, mostly-plus signature,
$\tilde\nu=\sqrt{m^2-9/4}$, and bulk-to-boundary functions
$(1-\ii\mathsf{a}k\tau)e^{\ii\mathsf{a}k\tau}/(2k^3)$, $\mathsf{a}=\pm$ their SK index. Writing our
two operators in their normalisation,
\begin{equation}
  \label{eq:app:QXdict}
  c_{\phi\phi\sigma}=-\frac1\Lambda,\qquad c_{\phi\sigma}=+\rho,\qquad \phi=\pic,\qquad \tilde\nu=\mu,
  \qquad u_{\rm QX}=\frac{2k_3}{k_{123}}\to\kappa\ (k_3\ {\rm soft}),
\end{equation}
and writing their kinematic ratios $k_1/k_3$, $k_2/k_3$, $k_{12}/k_3$, which tend to $1/\kappa$,
$1/\kappa$, $2/\kappa$. Their Eq.~(18) writes the correlator as three seed integrals
$\tilde I^{p_1,-2}_{\mathsf{ab}}(u_{\rm QX},1)$, $p_1=0,-1,-2$, whose coefficients tend in the
squeezed limit to $4/\kappa^2$, $-4\ii\mathsf{a}/\kappa$ and $-2$, while the non-analytic part of
their homogeneous solution, their Eq.~(91), is
$Y^{p}_{\pm}(u)|_{\rm signal}=2^{-5/2-p\mp2\ii\mu}\Gamma(\tfrac52+p\pm\ii\mu)\Gamma(\mp\ii\mu)u^{5/2+p\pm\ii\mu}$;
the three seeds thus land at the same order $\kappa^{\frac12\pm\ii\mu}$. With their
Eqs.~(89)--(90) for the signal parts of $\tilde I_{\pm\pm}$ and $\tilde I_{\pm\mp}$, whose common
constant at $p_2=-2$ is $\sqrt\pi/(2^{3/2}\cosh\pi\mu)$, the sum over the four SK branches gives for
the coefficient $A^{\rm QX}_-$ of $\kappa^{\frac12-\ii\mu}$ in $16(k_1k_2k_3)^2\langle\phi^3\rangle'$
\begin{equation}
  \label{eq:app:QXmatch}
  A^{\rm QX}_-=\frac{\Ccal^{\rm LI}_++\Ccal^{{\rm LI}*}_-}{4\la},
\end{equation}
modulus and phase, at every $\mu$, with $A^{\rm QX}_-=(A^{\rm QX}_+)^*$ so that their expression is
real; restoring the couplings, $c_{\phi\phi\sigma}c_{\phi\sigma}=-\rho/\Lambda$ reproduces the
prefactor of \cref{eq:app:pic3sq} with no leftover constant. The two permutations that put the
massive line on a hard momentum contribute no clock.

\paragraph{Single exchange, Ref.~\cite{Pinol:2021aun}.} This reference couples $N_{\rm flavour}$
scalars to $\dot\zeta$ through one portal combination; at $N_{\rm flavour}=1$ the portal is
$\sigma$ itself, its mixing angle is trivial, $(O^1{}_1)^2=1$, and the theory is ours at the symmetry
point. Its quadratic mixing is $2a^3\sqrt{2\epsilon}\mpl\,\omega\,\sigma\dot\zeta$, their Eq.~(1)
with the $4$ inside the overall $\tfrac{a^3}{2}$, and its cubic vertex
$-\tfrac{a^3}{H}\sqrt{2\epsilon}\mpl\,\omega\,\sigma[\dot\zeta^2-(\partial\zeta)^2/a^2]$, their
Eq.~(4). With $\zeta=-H\pic/\fpi^2$ and $\sqrt{2\epsilon}\mpl=\fpi^2/H$ these are
$-2a^3\omega\,\sigma\dot\pic$ and $-a^3(\omega/\fpi^2)\,\sigma[\dot\pic^2-(\partial\pic)^2/a^2]$, so that
\begin{equation}
  \label{eq:app:flavdict}
  \rho=-2\omega,\qquad \la^2=\frac{4\omega^2}{H^2},\qquad
  \Lambda^{-1}=-\frac{2\omega}{\fpi^2}=\frac{\rho}{\fpi^2},
\end{equation}
the last equality being the symmetry point: the vertex is fixed by the mixing, as it is for the
single coupling of Ref.~\cite{Arkani-Hamed:2015bza}. Their shape function is defined through
the correlator $(2\pi)^7\delta^3(\textstyle\sum\vec k)\,(\mathcal{P}^{(0)}_\zeta)^2\,S/(k_1k_2k_3)^2$,
with the unmixed spectrum, so that it equals ours at $R\to1$, and their squeezing ratio is our
$\kappa$.
Their heavy-field result, Sec.~IV\,B, is $S\simeq-\tfrac\pi2(\omega^2/H^2)\,\mathcal{S}_1$ with
$\mathcal{S}_1=-\sqrt\kappa\,\mathcal{A}(\mu)\,\sin[\mu\ln\kappa-\varphi(\mu)]$ and
\begin{equation}
  \label{eq:app:flavA}
  \mathcal{A}(\mu)e^{\ii\varphi(\mu)}=
  \frac{\sqrt\pi\,2^{-1+2\ii\mu}\tanh(\pi\mu)\,\Gamma(\tfrac72-\ii\mu)}{(2\mu+\ii)\,\Gamma(1-\ii\mu)}\,
  e^{-\pi\mu}\big(\coth\pi\mu+\ii\,\csch\pi\mu+1\big)^2 .
\end{equation}
Ours, from \cref{eq:sq:clock} with $\Ncal_{\rm LI}=(64\pi R^{3/2}\Dz)^{-1}H/\Lambda=\la/(32R^2)$ at
the symmetry point, is $S^{\rm LI}\simeq\tfrac{\la}{32}\sqrt\kappa\,\Acal_{\rm LI}\sin[\mu\ln\kappa-\delta_{\rm LI}]$
with $\Acal_{\rm LI}e^{\ii\delta_{\rm LI}}=-\ii(\Ccal^{\rm LI}_++\Ccal^{{\rm LI}*}_-)$; equating the two
with \cref{eq:app:flavdict} requires
\begin{equation}
  \label{eq:app:flavmatch}
  \mathcal{A}(\mu)e^{\ii\varphi(\mu)}=\frac{\ii}{4\pi}\times\Big[-\frac{\Ccal^{\rm LI}_++\Ccal^{{\rm LI}*}_-}{\la}\Big]
  =\frac{\ii}{4\pi}\times\text{r.h.s.\ of \cref{eq:sq:Gclosed}},
\end{equation}
which holds identically in $\mu$, modulus and phase, and was checked to $10^{-31}$ at
$\mu=0.7,1.7,2.5,4$ (\cref{app:numerics}). Since Ref.~\cite{Pinol:2021aun} recovers
Ref.~\cite{Arkani-Hamed:2015bza} at one flavour, this is the same anchor reached by a third route;
what it adds is the check of the flavour-basis prefactor, $-\tfrac\pi2(\omega/H)^2$, against ours.

\paragraph{Double exchange, Ref.~\cite{Aoki:2024uyi}.} This reference works with an external field
$\varphi=-(\fpi^2/H)\zeta$, i.e.\ $\varphi=\pic$ identically; a quadratic mixing
$\int\dd\tau\dd^3x\,\rho\,a^3\sigma\varphi'$, their (2.11), which is our $\rho$; a cubic coupling
$\int\dd\tau\dd^3x\,\lambda_\star a^3\sigma^2\varphi'$, their (2.12), so that $\alpha=-2\lambda_\star$,
the $\tfrac12$ being our explicit symmetry factor and theirs being carried inside their SK rule; the
bare mass variable $\sqrt{m^2/H^2-9/4}=\sqrt{\muf^2-\la^2}$, our $\mu$; the correlator
$\langle\varphi^3\rangle'=-[H^5/(\fpi^2(k_1k_2k_3)^2)]S$, their (5.4), whose $S$ coincides with ours
once $R\to1$; the squeezing variables $r_1=2k_1/k_{1234}\to\kappa$ and $r_3\to1$ at $k_4\to0$; and
the clock packaging of their (5.17),
$S=(\rho/H)^2\lambda_\star(2\pi\Dz)^{-1}\Real[\kappa^{\frac12+\ii\mu}A_{\rm DE}e^{\ii\delta_{\rm DE}}]$.
From \cref{eq:bisp:master} at $R=1$,
\begin{equation}
  \label{eq:app:Jfrak}
  S^{(2)}\;\simeq_{\kappa\ll1}\;\Big(\frac\rho H\Big)^{2}\lambda_\star\,(2\pi\Dz)^{-1}\,
  \Real\big[\,\mathcal{J}\,\kappa^{\frac12+\ii\mu}\big],
  \qquad
  \mathcal{J}\;\equiv\;-\frac{\Ccal^{(2)*}_++\Ccal^{(2)}_-}{16\la^2}
  \;=\;\frac{\ii}{16\la^2}\big(\Acal_2e^{\ii\delta_2}\big)^{*},
\end{equation}
which is the object to be compared with $A_{\rm DE}e^{\ii\delta_{\rm DE}}$. Their mass parameter is
our $\mu$, and the comparison is made at the relative $\mathcal{O}(\la^2)$ dropped throughout; this
is the only approximation of the comparison. Before comparing coefficients, their
prefactor was rebuilt from their own \S2: from their (2.14)--(2.15) at $p_{123}=-6$, the branch sum
is $2\Real[I_{+++}+I_{++-}+I_{+-+}+I_{-++}]$, in the squeezed limit
$(4k_1k_2^4k_3)^{-1}\kappa^{-\frac12+\ii\mu}=[4(k_1k_2k_3)^2]^{-1}\kappa^{\frac12+\ii\mu}$, and
exactly two of their three permutations carry a soft massive line and give equal clocks; collecting
$\tfrac14\times2\Real\times2$ reproduces their (4.20) prefactor $-\rho^2\lambda_\star H/(k_1k_2k_3)^2$
with no leftover constant, so that the dictionary is derived and independently confirmed. Their third
permutation puts the cubic vertex on the soft leg and contributes only the analytic background: it
is our dressed-leg clock, which at their fixed order does not exist. A second confirmation tests the
dictionary rather than the result: pushing our single-exchange coefficient \cref{eq:sq:Gclosed},
anchored above, through the same shape-function, $\kappa$ and $(\rho/H)^2$ entries against their
(5.18) for the Lorentz-covariant single-exchange channel returns ratio $1$.

Their squeezed limit is the sum of four seed integrals, their (4.16)--(4.19), each with its partner
under the reflection of the mass parameter, and their (4.20) and (5.14) quote the coefficient at
leading order in $e^{-\pi\mu}$ as
\begin{equation}
  \label{eq:sq:Cmu}
  C(\mu)=\frac{\sqrt\pi\,(2\ii\mu+5)}{2^{4+2\ii\mu}(2\mu-3\ii)}\,
  \Gamma(\tfrac12+\ii\mu)\Gamma(-\ii\mu)\big(1+\tanh\pi\mu\big) .
\end{equation}
We compare the complete sum, not its leading order. By \cref{eq:app:Jfrak,eq:sq:C2obs} our
$\mathcal{J}=\tfrac{\ii}{16\la^2}(\Acal_2e^{\ii\delta_2})^*$ is, in closed form,
$\mathcal{J}=C(\mu)\,2(\mathsf{t}-\ii\mathsf{s})/(1+\mathsf{t})$ with $\mathsf{t}\equiv\tanh\pi\mu$ and
$\mathsf{s}\equiv\sech\pi\mu$, the two identities $2^{4+2\ii\mu}=16\cdot4^{\ii\mu}$ and
$(2\ii\mu+5)/(2\mu-3\ii)=\ii(\tfrac52+\ii\mu)/(\tfrac32+\ii\mu)$ turning \cref{eq:sq:Cmu} into the
conjugate of the $\Gamma$-part of \cref{eq:sq:C2obs} term by term. On their side, the mirror of $C$
obeys $C(-\mu)^*=-C(\mu)(1-\mathsf{t})/(1+\mathsf{t})$; the homogeneous piece of their $I_{+++}$,
the whole of $I_{++-}$ and $I_{+-+}$ and the $\sech^2$ piece of $I_{-++}$ sum to
\begin{equation}
  \label{eq:app:threeway}
  \frac{\pi^2}{3+2\ii\mu}\Big\{\big[1-\mathsf{t}\coth2\pi\mu-\tfrac12\mathsf{s}^2\big]
  +\ii\big[\csch(2\pi\mu)\,\mathsf{s}-\tfrac12\mathsf{s}^2\,\csch(\pi\mu)\big]\Big\}=0,
\end{equation}
both brackets vanishing identically by $\mathsf{t}\coth2\pi\mu=\tfrac12(1+\mathsf{t}^2)$,
$\mathsf{t}^2+\mathsf{s}^2=1$ and $\csch2x=\tfrac12\csch x\sech x$; and the sole survivor, the
$\Gamma$-type particular solution of their $I_{-++}$, contributes $-2\ii\mathsf{s}C(\mu)/(1+\mathsf{t})$.
Adding,
\begin{equation}
  \label{eq:app:stage2}
  \mathcal{J}_{\rm DE}=C(\mu)\,\frac{2(\mathsf{t}-\ii\,\mathsf{s})}{1+\mathsf{t}}=\mathcal{J}
  \qquad\text{identically in }\mu .
\end{equation}
The agreement is analytic and exact, in amplitude and phase and at every $\mu$, not only at
leading order in $e^{-\pi\mu}$; the unit-modulus bracket $\mathsf{t}+\ii\mathsf{s}$ of
\cref{eq:sq:C2obs} is the $\Gamma$-type solution of their $I_{-++}$ riding on the mirror factor
$2\mathsf{t}/(1+\mathsf{t})$ of their $I_{+++}$.

\section{Numerical verification}
\label{app:numerics}

Every analytic statement of this paper was checked numerically, and every numerical statement up to
$\la=8$ was produced by at least two engines sharing no code beyond the definitions. The kernel
engines use the finite-interval representation \cref{eq:lin:pfaff}, in which the two endpoint powers
are explicit and removed by Taylor subtraction: an arbitrary-precision engine (mpmath, adaptive
tanh--sinh quadrature, library ${}_2F_1$) and a double-precision one (fixed composite
Gauss--Legendre panels, ${}_2F_1$ summed as a power series with the connection formula near $t=1$),
with a third, fixed-node engine in $w=\ln u$ for the $\xi$-quadratures of the exchange channels. The
strong-mixing observable of \cref{sec:sq:strong} is a cancellation, of the two equalised branches
of the clock, on top of the $e^{0.6\pi\la}$ cancellation inside every kernel quadrature, and double
precision is exhausted beyond $\la=8$ and, for the $b=-$ hard integral, for $\muf\gtrsim5$; beyond it a fourth engine, the
Gauss--Legendre panels carried out in ball arithmetic (arb) at $40$--$64$ digits, carries the
results on its own, agreeing with the mpmath engine to $10^{-15}$ at $\la=9$ and with itself, at
two unrelated node sets, to $10^{-25}$ at $\la=17$. Every clock number above $\la=8$ rests on this
one engine; reproducing them with an implementation sharing no definitions with ours, of the kind
deferred to \cite{PPRW}, is the check that would satisfy the two-engine rule. Two lessons from
building it: the junction between the two quadrature regions of \cref{eq:lin:pfaff}, at
$t=\tfrac12$, must be the same point to all digits in both, the unsubtracted integrand there being
of size $e^{\pi\la/4}$; and the lower cut of the $\xi$-quadrature must be $10^{-18}$ rather than
$10^{-8}$, since the truncation error at the latter is exactly the size of the cancelled combination
$|\Ccal_++\Ccal_-^*|/|\Ccal_+|$ at fixed $\muf$ and $\la\simeq12$, and moves fitted powers by
$0.1$--$0.2$. Three free anchors are used throughout: the identity
$\Wn^a_2=\Wn^a_0+4\Vv^a$, blind to any consistent misnormalisation of $\omega_a$; the boundary value
$\Wn^a_0(0)=r_a$ against \cref{eq:lin:raclosed}, which fixes it; and the exact channel-summed soft
leg \cref{eq:lin:Wplane}, which the dressed leg must reproduce through
$\bW_2\simeq4\sum_b\Wcal_b\beta^{-1/2-b\n}$ at small $\beta$. The table gives the tolerance behind each
number quoted in the text; tolerances are relative, ``digits'' means significant figures, and the
complete list of checks, with per-point tolerances, is distributed with the
implementation~\cite{Pinol:2026code}.

{\small
\begin{longtable}{@{}p{0.56\textwidth}p{0.41\textwidth}@{}}
\toprule
statement & achieved tolerance \\
\midrule
\endfirsthead
\toprule
statement (continued) & achieved tolerance \\
\midrule
\endhead
\multicolumn{2}{@{}l}{\emph{Linear theory (\cref{sec:linear})}}\\
kernels: mpmath vs float64 on $\{\Wn_0,\Wn_2,\Vv\}$, 54 points; $\Wn^a_0(0)=r_a$; $\Wn^a_2=\Wn^a_0+4\Vv^a$ & $3.4\times10^{-15}$; $2.3\times10^{-31}$ (mpmath), $6.7\times10^{-15}$ (float64); $1.6\times10^{-16}$\\
residuals of \cref{eq:lin:P,eq:lin:S} on $(\Ma^\pi_a,\Ma^\sigma_a)$; the sign of \cref{eq:lin:Msig} & $2.7\times10^{-25}$; the wrong sign gives $\mathcal{O}(1)$\\
$R$: weak mixing \cref{eq:lin:Rweak} at $\la=10^{-5}$; fixed bare mass \cref{eq:lin:Rstrongbare} at $\la=10^7$; fixed $\n$ \cref{eq:lin:Rstrongeff}, residual $\propto\la^{-2}$ & $1.3\times10^{-9}$; $1.9\times10^{-7}$ (at $\nbare=0$ the next order is $-7/64$ to all digits); residual ratio $16.01$ between $\la=50,200$\\
$\tfrac14$ theorem, 12 points; tail coefficients $\Wcal^a_b$ vs fit; exact modulus \cref{eq:lin:Wcalmod}, 20 random points; the per-channel ratio below it & $5\times10^{-10}$; $10^{-12}$--$10^{-9}$; $2.8\times10^{-14}$ (float64), $2\times10^{-48}$ (arb); $\le8\times10^{-30}$\\
the plane form \cref{eq:lin:Rplane} vs \cref{eq:lin:R}; the channel-summed soft leg \cref{eq:lin:Wplane} vs the sum \cref{eq:lin:dressedtails}, and $|\Wcal_+/\Wcal_-|e^{\pi\muf}-1$, at $(\la,\muf)=(0.3,1.7),(3,2.5),(8,2),(12,12),(5,20),(30,4)$, 40 digits & $\le8\times10^{-41}$; $\le3\times10^{-40}$ ($5\times10^{-21}$ at $(5,20)$, the precision of $e^{\pm\pi\la}$ there); $\le2\times10^{-40}$\\
ridge value \cref{eq:lin:ridge}, $R^{-3/2}|\Wcal_-|$ at $\la=\muf=10,30,100,300$ vs $\Gamma(\tfrac34)^2/\sqrt2\pi=0.3379891$ & $0.33836$, $0.33803$, $0.337993$, $0.337990$\\
a hard leg normalised to the spectrum, $R^{-1/2}|\bW_2(\beta)|$ of \cref{fig:lin:ingredients}b: its large-argument limit $\sqrt2\,e^{\pi\la/4}$ at $\beta=50$, $\muf=2.5$, $\la=2,4,6,8,10$ (the plane-wave regime sets in only at $\beta=\mathcal{O}(\la)$); and, at unit argument on the grid $\la\le8$, $\muf\le5$, the slope of its logarithm in $\la$ at $\muf=1$ against the $\pi/2$ of $e^{\pi\delta/2}$ & $6.59,33.6,180,1.01\times10^3,5.90\times10^3$ against $6.80,32.7,158,758,3.65\times10^3$; $1.52$ per unit $\la$ between $\la=4$ and $8$ against $1.57$\\
$c_2$ of \cref{eq:lin:Rstrongbare}: extracted from $R$ at $\la=400,800,1600$ along $\nbare^2=0,\tfrac94,6,-3$ vs the closed form & $\le5\times10^{-7}$ (the extraction, not the identity, limits this)\\
two-parity structure of $\bV$, \cref{eq:lin:sigmapower}, 4 points & $1.5\times10^{-6}$; odd$\times$odd alone off by $15$--$108\%$\\
$\Vv_2$ closed form \cref{eq:lin:V2closed} vs defining integral, heavy and light; recursion \cref{eq:app:crec} vs \texttt{hyp2f2}; $\mathcal{L}\Vv_2-1$ on the closed form & $10^{-34}$--$10^{-51}$; $\le5\times10^{-61}$; $\le3\times10^{-61}$ ($\beta\le5$)\\
$-\tfrac12\langle\Vv_2\rangle_s$ from \cref{eq:lin:V2closed} vs the even$\times$even half of \cref{eq:sq:H}, $\muf=1.7,3$ and $\n=0.35,0.6,0.8$ & $10^{-28}$--$10^{-31}$\\
\midrule
\multicolumn{2}{@{}l}{\emph{Contour rotation (\cref{sec:rules:contour})}}\\
rays $\theta=0.45,1.0,\pi/2$ at $\vec k=(0.9,1,1.1)$, $\la=1.1$, $\n=1.8\ii$, $N=1,2,3$, three leg sets; $N=0$; $N=2$ by an independent engine & $10^{-13}$--$10^{-15}$; disagree at $\mathcal{O}(1)$; $8\times10^{-16}$\\
\midrule
\multicolumn{2}{@{}l}{\emph{Weak mixing and the anchors (\cref{sec:sq:weak})}}\\
end-to-end squeezed fits of \cref{eq:bisp:shapes} vs \cref{eq:sq:family} at $\la=0.3$, $\n=1.7\ii$: single, double, triple exchange; alternative powers and frequencies ($2\muf$, $3\muf$, $\muf/2$, $\kappa^{1\mp\n}$) & single $3.2\times10^{-9}$ ($b=-$), $7.5\times10^{-8}$ ($b=+$); double $4.2\times10^{-5}$ ($b=+$), $5.2\times10^{-7}$ ($b=-$); triple $1.6\times10^{-5}$ ($b=+$), $3.6\times10^{-7}$ ($b=-$); alternatives worse by $5$--$6$ orders\\
reduction to Ref.~\cite{Arkani-Hamed:2015bza}, Eqs.~(6.130)--(6.131), and Ref.~\cite{Qin:2023ejc}, Eqs.~(18), (89)--(91), $\muf=0.25\ldots6$; from the tabulated kernels, modulus and phase & analytic identity, $10^{-31}$; $3.3\times10^{-8}$, $1.0\times10^{-7}$~rad\\
$\Hcal_b$ \cref{eq:sq:H}: Richardson in $\la^2$ from the exact channel sums; direct two-stage quadrature, six $(\n,b)$ points; two-term relation \cref{eq:app:image} at $\epsilon=10^{-2}$; $\Ycal$ \cref{eq:app:Yclosed} vs its series & $1\times10^{-9}$; $1.6$--$8.6\times10^{-12}$; $4\times10^{-31}$; $10^{-10}$\\
reduction to Ref.~\cite{Aoki:2024uyi}, Eqs.~(4.16)--(4.20), (5.14), (5.17) & analytic identity, $|{\rm ratio}-1|\lesssim4\times10^{-41}$\\
weak-mixing clocks of all six channels vs \cref{eq:sq:weaktable} from the pipeline of \cref{sec:implementation}; $\Ccal^{(2)}_b$ exact vs \cref{eq:sq:weaktable} at $\la=0.10,0.05,0.02$ & $\le9\times10^{-7}$; $9.2\times10^{-3},2.3\times10^{-3},3.6\times10^{-4}$ (clean $\mathcal{O}(\la^2)$)\\
$\Tcal_b$ of \cref{eq:sq:Hdef}: Richardson from the exact legs vs direct quadrature with $\Vv_1$ closed and $\Vv_2$ from \cref{eq:lin:V2closed} & $4.3\times10^{-6}$ ($b=+$), $1.6\times10^{-6}$ ($b=-$)\\
the contact approximation of \cref{eq:sq:Tclosed} and its large-mass limit for the triple-exchange clock against the exact $\Acal_3e^{\ii\delta_3}$ at $\muf=1,1.5,2,2.5,4,6,8$, from two quadratures of $v$ sharing no nodes (the $\xi$-rule of the pipeline and an independent Gauss--Legendre rule in $\ln\xi$, $330$ and $880$ nodes) & contact approximation: modulus ratio $1.81$, $1.52$, $1.36$, $1.23$, $1.08$, $1.03$, $1.02$, phase $0.25$, $0.28$, $0.19$, $0.13$, $0.05$, $0.02$, $0.01$~rad; large-mass limit: modulus $1.01$, $0.95$, $0.94$, $0.93$, $0.93$, $0.96$, $0.97$, phase off by $2.05$, $1.64$, $1.25$, $0.98$, $0.57$, $0.36$, $0.26$~rad ($\simeq2/\muf$); exact over the large-mass law: $1.28$, $1.27$, $1.21$, $1.18$, $1.12$, $1.07$, $1.04$; the two rules agree to $10^{-10}$\\
the closed form \cref{eq:sq:Texact} of $\Tcal_b$, both branches, against the quadrature of \cref{eq:sq:Hdef} at $\mu=0,0.25,0.5,1,1.5,2,2.5,4,6$ (40- against 35-digit arithmetic), and the one-function observable \cref{eq:sq:A3exact} & $\le3\times10^{-32}$; the mixed moment cancels exactly in the observable\\
the identity \cref{eq:sq:Tclosed} (bilinearity) and $\Hcal_b$ from \cref{eq:sq:H} vs the direct moment of $v$; the remainder $J^{(2)}_b[v+2/(1-4\n^2),v+2/(1-4\n^2)]/\Tcal_b$ on $b=-$ at $\muf=1.5,2,4$ & $|\Hcal_b^{\rm quad}/\Hcal_b-1|\le3\times10^{-11}$; remainder $0.67,0.42,0.094$, i.e.\ $1.5,1.7,1.5$ times $\muf^{-2}$\\
$\Hcal_-\to-2\Gamma(\tfrac52+\ii\muf)/(1+4\muf^2)$ and the double-exchange law $\Acal_2/\la^2\to4\pi^{3/2}\muf^{-1/2}e^{-\pi\muf}$ from \cref{eq:sq:H}, $\muf=2,4,8,12$ & ratios $1.06,1.04,1.01,1.007$ and $1.28,1.10,1.03,1.014$\\
\cref{eq:sq:C2obs} vs $-(\Ccal^{(2)}_++\Ccal^{(2)*}_-)/\la^2$ assembled from \cref{eq:sq:weaktable,eq:sq:H}, $\muf=0.4,1.7,2.5,5$; unit modulus of its bracket over $\muf\in[0.05,10]$ & identity, $10^{-31}$; $4\times10^{-61}$\\
reduction to Ref.~\cite{Pinol:2021aun} at $N_{\rm flavour}=1$, \cref{eq:app:flavmatch}, $\muf=0.7,1.7,2.5,4$ & identity, $|{\rm ratio}-1|\le2\times10^{-31}$\\
\midrule
\multicolumn{2}{@{}l}{\emph{Strong mixing in the plane (\cref{sec:sq:strong})}}\\
proportionality \cref{eq:bisp:prop} at $\la\gg\muf$, equilateral, $\n=1.7\ii$: imaginary part at $\la=5,7,9$; $(\text{ratio}-1)\la^2$ for double exchange; cross-channel $3{:}2$ ratio at $\la=3,5,7,9$ and Richardson; the same ratio from the pipeline of \cref{sec:implementation} at $\la=10$ and extrapolated & $10^{-5},10^{-8},10^{-11}$; $6.2,5.5,5.3$; $1.952,1.600,1.544,1.524$, $1.488$; $1.529$, $|{\rm ratio}-\tfrac32|=7\times10^{-3}$\\
branch equalisation: $|J^{(2)}_+/J^{(2)}_-|e^{-\pi\muf}$ at $\muf=2$, $\la=3,\dots,8$ (float64); at $\muf=2.5$, $\la=8,\dots,20$ (arb), minus $1$ & $0.634$, $0.950$, $0.99924$, $1.00012$, $1.000012$, $1.0000007$; $2\times10^{-6}$ falling to $4\times10^{-16}$\\
the grid of \cref{fig:sq:plane}: $\la\in[0.5,8]$, $\muf\in[1,5]$, step $0.25$, six channels, float64; range of $\pi^{-1}\ln(R^{-3/2}\Acal^{(0)})$ and of its residual after $\tfrac12(\la-2\muf)$; local slopes of $\pi^{-1}\ln(R^{-3/2}\Acal^{(0)})$ at $\la=8$ in $\la$ and in $\muf$ ($\muf=1\to5$); slope of the amplitude contours $\dd\la/\dd\muf$ at $\la=4,6,8$ & $[-3.9,+5.5]$ against $[0.33,2.53]$; $0.64$--$0.59$ against the asymptotic $0.5$; $-1.37\to-1.10$ against $-1$; $1.6$, $1.9$, $2.1$ against the asymptotic $2$ (the tilt of the contours in \cref{fig:sq:plane}a)\\
the residual along the arb lines: local slope in $\la$ of $\pi^{-1}\ln(R^{-3/2}\Acal^{(0)})-\tfrac12(\la-2\muf)$ at fixed $\muf=2.5$ ($2$), $\la=9\to19$; at fixed $m/H=2.5$ ($3.5$), $\la=9\to23$ & $0.122\to0.058$ ($0.121\to0.058$), $\propto1/\la$ with coefficient $3.5$; $0.081\to0.034$ ($0.060\to0.038$), $\propto1/\la$ with coefficient $2.3$\\
the hard-leg factor $\Qcal$ of \cref{eq:sq:planeB} from the arb lines: along fixed $m/H=2.5$ ($3.5$) at $\la=8\to24$ ($10\to22$), and its limit; along fixed $\muf=2.5$ ($2$), the fit $\ln\Qcal=c+p\ln\la$ on $\la\in[8,20]$ and $\Qcal/\la^{5/2}$ & $1.996\to1.876$ ($2.037\to1.898$), $\Qcal_\infty=1.86$ from both masses; $p=2.503$ ($2.493$), rms $4\times10^{-3}$ ($7\times10^{-3}$), $\Qcal/\la^{5/2}=0.098$ ($0.199$) constant to $1\%$\\
the diagonal law \cref{eq:sq:diagonal} vs the arb clock along fixed $m/H=2.5$ and $3.5$ & ratio $1.09\to1.01$ over $\la=8\to24$ and $1.11\to1.02$ over $\la=10\to22$\\
weak-mixing corner of \cref{eq:sq:planeexp}: $|\Ccal_++\Ccal_-^*|/(\la^2\muf^{3/2}e^{-\pi\muf})$ from the first row of \cref{eq:sq:weaktable} at $\muf=2,3,5,8,12$ & $14.3,12.6,11.7,11.4,11.2\to2\pi^{3/2}=11.14$\\
arb vs mpmath kernels at $\la=9$, $17$; arb at two node sets at $\la=12$, $17$; arb clock vs float64 scan at $\la=6$, $8$ & $10^{-15}$, $10^{-10}$; $10^{-26}$, $10^{-25}$; $3\times10^{-9}$, $4\times10^{-3}$ (the float64 error)\\
\bottomrule
\end{longtable}}

Every squeezed fit above uses the basis $\{\kappa^{\frac12-\n},\kappa^{\frac12+\n},\kappa,\kappa^2\}$,
since a fit failure and a physics disagreement look identical unless the basis contains the
expected non-analytic powers, and alternative powers are rejected on the same basis and the same
points. Ratios of complex brackets are conditioned by their own moduli and are reliable to
$\la\simeq20$ in double precision, while any real shape or clock amplitude carries a Boltzmann
cancellation and is exhausted near $\la\simeq10$ and $\la\simeq9$ respectively; every clock entry
beyond $\la=8$ is from the arb engine alone. Not done: an arbitrary-precision scan of the plane away
from the four lines, which would fix the powers of \cref{eq:sq:planeexp}; the light-field
continuation of \cref{sec:rules:contour}; and a
reproduction of the shapes by an implementation sharing no definitions with ours, such as a
transport or flow-equation solver, which is deferred to \cite{PPRW}.

\section{Statement on the use of AI assistance}
\label{app:ai}

This project was carried out with substantial assistance from a large language model, used as an
interactive collaborator rather than as a text-editing tool. We state here what it was used for,
what it was not used for, and what was done to make the results independent of it, both as a matter
of research ethics and because the reader is entitled to know how the calculations in this paper
were produced.

\paragraph{What was used.} A large language model, accessed through an agentic coding interface
with tool access to a shell, a Python environment and the published literature, between July and
September~2026. It was used in the following tasks.
\begin{itemize}\setlength\itemsep{1pt}
\item \emph{Derivations.} Proposing and carrying out algebraic steps, principally for the exchange
  channels: the weak-mixing expansions and closed forms of the double- and triple-exchange
  coefficients of \cref{sec:sq:weak}, the closed form of $\Vv_2$ (\cref{app:V2}), the
  double-exchange moment through the two-term relation \cref{eq:uf:threeftwo} (\cref{app:double}),
  the closed form of the triple-exchange moment through the Legendre representation of the
  one-insertion leg (\cref{app:triple}), and the analysis of the hard-leg factor of the
  strong-mixing clock in \cref{sec:sq:strong}.
\item \emph{Numerical implementation.} Writing, testing and documenting the production engine, the
  independent mpmath engine used to certify it, and the arbitrary-precision (arb) engine that the
  strong-mixing cancellations require, together with the quadrature rules, the tail splices and the
  self-tests tabulated in \cref{app:numerics}; and producing every figure of this paper.
\item \emph{Comparison with the perturbative literature.} From the references chosen by the author,
  building the dictionaries between conventions of \cref{app:dictionaries} and performing the
  explicit checks.
\item \emph{Adversarial checking.} Re-deriving results along structurally different routes,
  attempting to break them by limits, symmetries and dimensional audits, and reading successive
  drafts as a hostile referee would.
\item \emph{Writing.} Drafting and restructuring the manuscript, its figures' captions and this
  appendix.
\end{itemize}

\paragraph{What was not delegated.} The physical questions, the choice of what to compute, the
conventions, the interpretation of every result, the decisions about what is established and what is
not, and the final text are the author's. No number, equation or statement entered the paper on the
model's assertion alone.

\paragraph{How the results are made independent of it.} A language model is a fluent source of
plausible errors, and the safeguard used throughout was to make every claim checkable by something
that is not a language model. Every equation that enters the paper is verified numerically, at the
level recorded in \cref{app:numerics}, against at least one route that shares no step with its
derivation, and wherever possible against an implementation sharing no code --- except where the
text says otherwise, and it says so in two places: above $\la=8$ the two-engine rule is not
met and every strong-mixing number rests on one engine, and the contour rotation is tested at one
kinematic point. Those exceptions are listed in \cref{app:numerics} and are the places where a
reader should be most sceptical. The measured statements
of \cref{sec:sq:strong} are quoted as local slopes over the ranges on which they were measured, not
as fitted exponents. Two classes of error found this way during the project are worth recording, because they are
the kinds of error this way of working produces: a plausible but wrong asymptotic argument, which
the numerics contradicted, and silent convention drift between our normalisation and a published
one, which only an explicit dictionary and a limit check caught.

\bibliographystyle{JHEP}
\bibliography{biblio}

\providecommand{\href}[2]{#2}\begingroup\raggedright\begin{thebibliography}{10}

\bibitem{Planck:2018jri}
{Planck Collaboration}, \emph{{Planck 2018 results. X. Constraints on
  inflation}}, {\emph{Astron. Astrophys.} {\bfseries 641} (2020) A10}
  [\href{https://arxiv.org/abs/1807.06211}{{\ttfamily 1807.06211}}].

\bibitem{Baumann:2014nda}
D.~Baumann and L.~McAllister, \emph{{Inflation and String Theory}}. Cambridge
  University Press, 2015, [\href{https://arxiv.org/abs/1404.2601}{{\ttfamily
  1404.2601}}].

\bibitem{Cheung:2007st}
C.~Cheung, P.~Creminelli, A.~L. Fitzpatrick, J.~Kaplan and L.~Senatore,
  \emph{{The Effective Field Theory of Inflation}},
  \href{https://doi.org/10.1088/1126-6708/2008/03/014}{\emph{JHEP} {\bfseries
  03} (2008) 014} [\href{https://arxiv.org/abs/0709.0293}{{\ttfamily
  0709.0293}}].

\bibitem{Senatore:2010wk}
L.~Senatore and M.~Zaldarriaga, \emph{{The Effective Field Theory of Multifield
  Inflation}}, \href{https://doi.org/10.1007/JHEP04(2012)024}{\emph{JHEP}
  {\bfseries 04} (2012) 024} [\href{https://arxiv.org/abs/1009.2093}{{\ttfamily
  1009.2093}}].

\bibitem{Noumi:2012vr}
T.~Noumi, M.~Yamaguchi and D.~Yokoyama, \emph{{Effective field theory approach
  to quasi-single field inflation and effects of heavy fields}},
  \href{https://doi.org/10.1007/JHEP06(2013)051}{\emph{JHEP} {\bfseries 06}
  (2013) 051} [\href{https://arxiv.org/abs/1211.1624}{{\ttfamily 1211.1624}}].

\bibitem{Pinol:2024arz}
L.~Pinol, \emph{{Effective field theory of multifield inflationary
  fluctuations}},
  \href{https://doi.org/10.1103/PhysRevD.110.L041302}{\emph{Phys. Rev. D}
  {\bfseries 110} (2024) L041302}
  [\href{https://arxiv.org/abs/2405.02190}{{\ttfamily 2405.02190}}].

\bibitem{Achucarro:2010jv}
A.~Ach\'ucarro, J.-O. Gong, S.~Hardeman, G.~A. Palma and S.~P. Patil,
  \emph{{Mass hierarchies and non-decoupling in multi-scalar field dynamics}},
  \href{https://doi.org/10.1103/PhysRevD.84.043502}{\emph{Phys. Rev. D}
  {\bfseries 84} (2011) 043502}
  [\href{https://arxiv.org/abs/1005.3848}{{\ttfamily 1005.3848}}].

\bibitem{Achucarro:2010da}
A.~Ach\'ucarro, J.-O. Gong, S.~Hardeman, G.~A. Palma and S.~P. Patil,
  \emph{{Features of heavy physics in the CMB power spectrum}}, {\emph{JCAP}
  {\bfseries 01} (2011) 030} [\href{https://arxiv.org/abs/1010.3693}{{\ttfamily
  1010.3693}}].

\bibitem{Cespedes:2012hu}
S.~Cespedes, V.~Atal and G.~A. Palma, \emph{{On the importance of heavy fields
  during inflation}}, {\emph{JCAP} {\bfseries 05} (2012) 008}
  [\href{https://arxiv.org/abs/1201.4848}{{\ttfamily 1201.4848}}].

\bibitem{Garcia-Saenz:2018ifx}
S.~Garcia-Saenz, S.~Renaux-Petel and J.~Ronayne, \emph{{Primordial fluctuations
  and non-Gaussianities in sidetracked inflation}},
  \href{https://doi.org/10.1088/1475-7516/2018/07/057}{\emph{JCAP} {\bfseries
  07} (2018) 057} [\href{https://arxiv.org/abs/1804.11279}{{\ttfamily
  1804.11279}}].

\bibitem{Maldacena:2002vr}
J.~M. Maldacena, \emph{{Non-Gaussian features of primordial fluctuations in
  single field inflationary models}},
  \href{https://doi.org/10.1088/1126-6708/2003/05/013}{\emph{JHEP} {\bfseries
  05} (2003) 013} [\href{https://arxiv.org/abs/astro-ph/0210603}{{\ttfamily
  astro-ph/0210603}}].

\bibitem{Chen:2010xka}
X.~Chen, \emph{{Primordial Non-Gaussianities from Inflation Models}},
  \href{https://doi.org/10.1155/2010/638979}{\emph{Adv. Astron.} {\bfseries
  2010} (2010) 638979} [\href{https://arxiv.org/abs/1002.1416}{{\ttfamily
  1002.1416}}].

\bibitem{Meerburg:2019qqi}
P.~D. Meerburg et~al., \emph{{Primordial Non-Gaussianity}}, {\emph{Bull. Am.
  Astron. Soc.} {\bfseries 51} (2019) 107}
  [\href{https://arxiv.org/abs/1903.04409}{{\ttfamily 1903.04409}}].

\bibitem{Achucarro:2022qrl}
A.~Ach\'ucarro et~al., \emph{{Inflation: Theory and Observations}},
  \href{https://arxiv.org/abs/2203.08128}{{\ttfamily 2203.08128}}.

\bibitem{Creminelli:2004yq}
P.~Creminelli and M.~Zaldarriaga, \emph{{Single field consistency relation for
  the 3-point function}},
  \href{https://doi.org/10.1088/1475-7516/2004/10/006}{\emph{JCAP} {\bfseries
  10} (2004) 006} [\href{https://arxiv.org/abs/astro-ph/0407059}{{\ttfamily
  astro-ph/0407059}}].

\bibitem{Chen:2009we}
X.~Chen and Y.~Wang, \emph{{Large non-Gaussianities with Intermediate Shapes
  from Quasi-Single Field Inflation}},
  \href{https://doi.org/10.1103/PhysRevD.81.063511}{\emph{Phys. Rev. D}
  {\bfseries 81} (2010) 063511}
  [\href{https://arxiv.org/abs/0909.0496}{{\ttfamily 0909.0496}}].

\bibitem{Chen:2009zp}
X.~Chen and Y.~Wang, \emph{{Quasi-Single Field Inflation and
  Non-Gaussianities}},
  \href{https://doi.org/10.1088/1475-7516/2010/04/027}{\emph{JCAP} {\bfseries
  04} (2010) 027} [\href{https://arxiv.org/abs/0911.3380}{{\ttfamily
  0911.3380}}].

\bibitem{Chen:2012ge}
X.~Chen and Y.~Wang, \emph{{Quasi-Single Field Inflation with Large Mass}},
  \href{https://doi.org/10.1088/1475-7516/2012/09/021}{\emph{JCAP} {\bfseries
  09} (2012) 021} [\href{https://arxiv.org/abs/1205.0160}{{\ttfamily
  1205.0160}}].

\bibitem{Baumann:2011nk}
D.~Baumann and D.~Green, \emph{{Signatures of Supersymmetry from the Early
  Universe}}, {\emph{Phys. Rev. D} {\bfseries 85} (2012) 103520}
  [\href{https://arxiv.org/abs/1109.0292}{{\ttfamily 1109.0292}}].

\bibitem{Arkani-Hamed:2015bza}
N.~Arkani-Hamed and J.~Maldacena, \emph{{Cosmological Collider Physics}},
  \href{https://arxiv.org/abs/1503.08043}{{\ttfamily 1503.08043}}.

\bibitem{Lee:2016vti}
H.~Lee, D.~Baumann and G.~L. Pimentel, \emph{{Non-Gaussianity as a Particle
  Detector}}, \href{https://doi.org/10.1007/JHEP12(2016)040}{\emph{JHEP}
  {\bfseries 12} (2016) 040}
  [\href{https://arxiv.org/abs/1607.03735}{{\ttfamily 1607.03735}}].

\bibitem{Arkani-Hamed:2018kmz}
N.~Arkani-Hamed, D.~Baumann, H.~Lee and G.~L. Pimentel, \emph{{The Cosmological
  Bootstrap: Inflationary Correlators from Symmetries and Singularities}},
  \href{https://doi.org/10.1007/JHEP04(2020)105}{\emph{JHEP} {\bfseries 04}
  (2020) 105} [\href{https://arxiv.org/abs/1811.00024}{{\ttfamily
  1811.00024}}].

\bibitem{Baumann:2019oyu}
D.~Baumann, C.~Duaso~Pueyo, A.~Joyce, H.~Lee and G.~L. Pimentel, \emph{{The
  Cosmological Bootstrap: Weight-Shifting Operators and Scalar Seeds}},
  {\emph{JHEP} {\bfseries 12} (2020) 204}
  [\href{https://arxiv.org/abs/1910.14051}{{\ttfamily 1910.14051}}].

\bibitem{Pimentel:2022fsc}
G.~L. Pimentel and D.-G. Wang, \emph{{Boostless cosmological collider
  bootstrap}}, {\emph{JHEP} {\bfseries 10} (2022) 177}
  [\href{https://arxiv.org/abs/2205.00013}{{\ttfamily 2205.00013}}].

\bibitem{Chen:2017ryl}
X.~Chen, Y.~Wang and Z.-Z. Xianyu, \emph{{Schwinger-Keldysh Diagrammatics for
  Primordial Perturbations}},
  \href{https://doi.org/10.1088/1475-7516/2017/12/006}{\emph{JCAP} {\bfseries
  12} (2017) 006} [\href{https://arxiv.org/abs/1703.10166}{{\ttfamily
  1703.10166}}].

\bibitem{Qin:2022fbv}
Z.~Qin and Z.-Z. Xianyu, \emph{{Helical inflation correlators: partial
  Mellin-Barnes and bootstrap equations}}, {\emph{JHEP} {\bfseries 04} (2023)
  059} [\href{https://arxiv.org/abs/2208.13790}{{\ttfamily 2208.13790}}].

\bibitem{Qin:2023ejc}
Z.~Qin and Z.-Z. Xianyu, \emph{{Closed-form formulae for inflation
  correlators}}, \href{https://doi.org/10.1007/JHEP07(2023)001}{\emph{JHEP}
  {\bfseries 07} (2023) 001}
  [\href{https://arxiv.org/abs/2301.07047}{{\ttfamily 2301.07047}}].

\bibitem{Liu:2024str}
H.~Liu and Z.-Z. Xianyu, \emph{{Massive inflationary amplitudes: differential
  equations and complete solutions for general trees}},
  \href{https://doi.org/10.1007/JHEP09(2025)183}{\emph{JHEP} {\bfseries 09}
  (2025) 183} [\href{https://arxiv.org/abs/2412.07843}{{\ttfamily
  2412.07843}}].

\bibitem{Liu:2024xyi}
H.~Liu, Z.~Qin and Z.-Z. Xianyu, \emph{{Dispersive bootstrap of massive
  inflation correlators}},
  \href{https://doi.org/10.1007/JHEP02(2025)101}{\emph{JHEP} {\bfseries 02}
  (2025) 101} [\href{https://arxiv.org/abs/2407.12299}{{\ttfamily
  2407.12299}}].

\bibitem{Ema:2024hkj}
Y.~Ema and K.~Mukaida, \emph{{Cutting rule for in-in correlators and
  cosmological collider}},
  \href{https://doi.org/10.1007/JHEP12(2024)194}{\emph{JHEP} {\bfseries 12}
  (2024) 194} [\href{https://arxiv.org/abs/2409.07521}{{\ttfamily
  2409.07521}}].

\bibitem{Pinol:2021aun}
L.~Pinol, S.~Aoki, S.~Renaux-Petel and M.~Yamaguchi, \emph{{Inflationary flavor
  oscillations and the cosmic spectroscopy}},
  \href{https://doi.org/10.1103/PhysRevD.107.L021301}{\emph{Phys. Rev. D}
  {\bfseries 107} (2023) L021301}
  [\href{https://arxiv.org/abs/2112.05710}{{\ttfamily 2112.05710}}].

\bibitem{Aoki:2024jha}
S.~Aoki, A.~Ghoshal and A.~Strumia, \emph{{Cosmological collider
  non-Gaussianity from multiple scalars and R$^{2}$ gravity}},
  \href{https://doi.org/10.1007/JHEP11(2024)009}{\emph{JHEP} {\bfseries 11}
  (2024) 009} [\href{https://arxiv.org/abs/2408.07069}{{\ttfamily
  2408.07069}}].

\bibitem{Aoki:2024uyi}
S.~Aoki, L.~Pinol, F.~Sano, M.~Yamaguchi and Y.~Zhu, \emph{{Cosmological
  correlators with double massive exchanges: bootstrap equation and
  phenomenology}}, \href{https://doi.org/10.1007/JHEP09(2024)176}{\emph{JHEP}
  {\bfseries 09} (2024) 176}
  [\href{https://arxiv.org/abs/2404.09547}{{\ttfamily 2404.09547}}].

\bibitem{Jazayeri:2022kjy}
S.~Jazayeri and S.~Renaux-Petel, \emph{{Cosmological Bootstrap in Slow
  Motion}}, \href{https://doi.org/10.1007/JHEP12(2022)137}{\emph{JHEP}
  {\bfseries 12} (2022) 137}
  [\href{https://arxiv.org/abs/2205.10340}{{\ttfamily 2205.10340}}].

\bibitem{Jazayeri:2023xcj}
S.~Jazayeri, S.~Renaux-Petel and D.~Werth, \emph{{Shapes of the Cosmological
  Low-Speed Collider}},
  \href{https://doi.org/10.1088/1475-7516/2023/12/035}{\emph{JCAP} {\bfseries
  12} (2023) 035} [\href{https://arxiv.org/abs/2307.01751}{{\ttfamily
  2307.01751}}].

\bibitem{Qin:2025xct}
Z.~Qin, S.~Renaux-Petel, X.~Tong, D.~Werth and Y.~Zhu, \emph{{The exact and
  approximate tales of boost-breaking cosmological correlators}},
  \href{https://doi.org/10.1088/1475-7516/2025/09/058}{\emph{JCAP} {\bfseries
  09} (2025) 058} [\href{https://arxiv.org/abs/2506.01555}{{\ttfamily
  2506.01555}}].

\bibitem{Bordin:2018pca}
L.~Bordin, P.~Creminelli, A.~Khmelnitsky and L.~Senatore, \emph{{Light
  Particles with Spin in Inflation}},
  \href{https://doi.org/10.1088/1475-7516/2018/10/013}{\emph{JCAP} {\bfseries
  10} (2018) 013} [\href{https://arxiv.org/abs/1806.10587}{{\ttfamily
  1806.10587}}].

\bibitem{Wu:2024wti}
Y.-P. Wu, \emph{{The cosmological collider in R $^{2}$ inflation}},
  \href{https://doi.org/10.1088/1475-7516/2024/07/010}{\emph{JCAP} {\bfseries
  07} (2024) 010} [\href{https://arxiv.org/abs/2404.05031}{{\ttfamily
  2404.05031}}].

\bibitem{McCulloch:2024hiz}
C.~McCulloch, E.~Pajer and X.~Tong, \emph{{A Cosmological Tachyon Collider:
  Enhancing the Long-Short Scale Coupling}},
  \href{https://doi.org/10.1007/JHEP05(2024)262}{\emph{JHEP} {\bfseries 05}
  (2024) 262} [\href{https://arxiv.org/abs/2401.11009}{{\ttfamily
  2401.11009}}].

\bibitem{Aoki:2026qea}
S.~Aoki, D.~Roest and D.~Werth, \emph{{Universal Non-Gaussian Signatures from
  Transient Instabilities}},
  \href{https://arxiv.org/abs/2604.01035}{{\ttfamily 2604.01035}}.

\bibitem{Chen:2016uwp}
X.~Chen, Y.~Wang and Z.-Z. Xianyu, \emph{{Standard Model Background of the
  Cosmological Collider}}, {\emph{Phys. Rev. Lett.} {\bfseries 118} (2017)
  261302} [\href{https://arxiv.org/abs/1610.06597}{{\ttfamily 1610.06597}}].

\bibitem{Kumar:2017ecc}
S.~Kumar and R.~Sundrum, \emph{{Heavy-Lifting of Gauge Theories By Cosmic
  Inflation}}, {\emph{JHEP} {\bfseries 05} (2018) 011}
  [\href{https://arxiv.org/abs/1711.03988}{{\ttfamily 1711.03988}}].

\bibitem{Wang:2019gbi}
L.-T. Wang and Z.-Z. Xianyu, \emph{{In Search of Large Signals at the
  Cosmological Collider}}, {\emph{JHEP} {\bfseries 02} (2020) 044}
  [\href{https://arxiv.org/abs/1910.12876}{{\ttfamily 1910.12876}}].

\bibitem{Bodas:2020yho}
A.~Bodas, S.~Kumar and R.~Sundrum, \emph{{The Scalar Chemical Potential in
  Cosmological Collider Physics}},
  \href{https://doi.org/10.1007/JHEP02(2021)079}{\emph{JHEP} {\bfseries 02}
  (2021) 079} [\href{https://arxiv.org/abs/2010.04727}{{\ttfamily
  2010.04727}}].

\bibitem{Sou:2021juh}
C.~M. Sou, X.~Tong and Y.~Wang, \emph{{Chemical-potential-assisted particle
  production in FRW spacetimes}},
  \href{https://doi.org/10.1007/JHEP06(2021)129}{\emph{JHEP} {\bfseries 06}
  (2021) 129} [\href{https://arxiv.org/abs/2104.08772}{{\ttfamily
  2104.08772}}].

\bibitem{An:2017hlx}
H.~An, M.~McAneny, A.~K. Ridgway and M.~B. Wise, \emph{{Quasi Single Field
  Inflation in the non-perturbative regime}},
  \href{https://doi.org/10.1007/JHEP06(2018)105}{\emph{JHEP} {\bfseries 06}
  (2018) 105} [\href{https://arxiv.org/abs/1706.09971}{{\ttfamily
  1706.09971}}].

\bibitem{Iyer:2017qzw}
A.~V. Iyer, S.~Pi, Y.~Wang, Z.~Wang and S.~Zhou, \emph{{Strongly Coupled
  Quasi-Single Field Inflation}},
  \href{https://doi.org/10.1088/1475-7516/2018/01/041}{\emph{JCAP} {\bfseries
  01} (2018) 041} [\href{https://arxiv.org/abs/1710.03054}{{\ttfamily
  1710.03054}}].

\bibitem{Cremonini:2010ua}
S.~Cremonini, Z.~Lalak and K.~Turzynski, \emph{{Strongly Coupled Perturbations
  in Two-Field Inflationary Models}},
  \href{https://doi.org/10.1088/1475-7516/2011/03/016}{\emph{JCAP} {\bfseries
  03} (2011) 016} [\href{https://arxiv.org/abs/1010.3021}{{\ttfamily
  1010.3021}}].

\bibitem{Aoki:2025ywt}
S.~Aoki, H.~Otsuka and R.~Yanagita, \emph{{Heavy field effects on inflationary
  models in light of ACT data}},
  \href{https://doi.org/10.1088/1475-7516/2025/11/088}{\emph{JCAP} {\bfseries
  11} (2025) 088} [\href{https://arxiv.org/abs/2509.06739}{{\ttfamily
  2509.06739}}].

\bibitem{Sefusatti:2012ye}
E.~Sefusatti, J.~R. Fergusson, X.~Chen and E.~P.~S. Shellard, \emph{{Effects
  and Detectability of Quasi-Single Field Inflation in the Large-Scale
  Structure and Cosmic Microwave Background}}, {\emph{JCAP} {\bfseries 08}
  (2012) 033} [\href{https://arxiv.org/abs/1204.6318}{{\ttfamily 1204.6318}}].

\bibitem{Planck:2019kim}
{\scshape Planck} collaboration, \emph{{Planck 2018 results. IX. Constraints on
  primordial non-Gaussianity}},
  \href{https://doi.org/10.1051/0004-6361/201935891}{\emph{Astron. Astrophys.}
  {\bfseries 641} (2020) A9}
  [\href{https://arxiv.org/abs/1905.05697}{{\ttfamily 1905.05697}}].

\bibitem{Sohn:2024xzd}
W.~Sohn, D.-G. Wang, J.~R. Fergusson and E.~P.~S. Shellard, \emph{{Searching
  for Cosmological Collider in the Planck CMB Data}},
  \href{https://doi.org/10.1088/1475-7516/2024/09/016}{\emph{JCAP} {\bfseries
  09} (2024) 016} [\href{https://arxiv.org/abs/2404.07203}{{\ttfamily
  2404.07203}}].

\bibitem{Suman:2025vuf}
P.~Suman, D.-G. Wang, W.~Sohn, J.~R. Fergusson and E.~P.~S. Shellard,
  \emph{{How Significant are Cosmological Collider Signals in the Planck
  Data?}},  \href{https://arxiv.org/abs/2511.17500}{{\ttfamily 2511.17500}}.

\bibitem{Cabass:2024wob}
G.~Cabass, O.~H.~E. Philcox, M.~M. Ivanov, K.~Akitsu, S.-F. Chen,
  M.~Simonovi{\'c} et~al., \emph{{BOSS constraints on massive particles during
  inflation: The cosmological collider in action}},
  \href{https://doi.org/10.1103/PhysRevD.111.063510}{\emph{Phys. Rev. D}
  {\bfseries 111} (2025) 063510}
  [\href{https://arxiv.org/abs/2404.01894}{{\ttfamily 2404.01894}}].

\bibitem{Kumar:2026ogn}
S.~Kumar, Q.~Lu, Z.-Z. Xianyu and Y.~Zhang, \emph{{Cosmological Collider
  Searches beyond the Hubble Scale with Planck Data}},
  \href{https://arxiv.org/abs/2603.15728}{{\ttfamily 2603.15728}}.

\bibitem{Pinol:2023oux}
L.~Pinol, S.~Renaux-Petel and D.~Werth, \emph{{The Cosmological Flow: A
  Systematic Approach to Primordial Correlators}},
  \href{https://doi.org/10.1088/1475-7516/2025/02/019}{\emph{JCAP} {\bfseries
  02} (2025) 019} [\href{https://arxiv.org/abs/2312.06559}{{\ttfamily
  2312.06559}}].

\bibitem{Philcox:2026njr}
O.~H.~E. Philcox, \emph{{What Shape is the Inflationary Bispectrum?}},
  \href{https://arxiv.org/abs/2603.17004}{{\ttfamily 2603.17004}}.

\bibitem{Werth:2023pfl}
D.~Werth, L.~Pinol and S.~Renaux-Petel, \emph{{Cosmological Flow of Primordial
  Correlators}},
  \href{https://doi.org/10.1103/PhysRevLett.133.141002}{\emph{Phys. Rev. Lett.}
  {\bfseries 133} (2024) 141002}
  [\href{https://arxiv.org/abs/2302.00655}{{\ttfamily 2302.00655}}].

\bibitem{Werth:2024aui}
D.~Werth, L.~Pinol and S.~Renaux-Petel, \emph{{CosmoFlow: Python Package for
  Cosmological Correlators}},
  \href{https://doi.org/10.1088/1361-6382/ad6740}{\emph{Class. Quant. Grav.}
  {\bfseries 41} (2024) 175015}
  [\href{https://arxiv.org/abs/2402.03693}{{\ttfamily 2402.03693}}].

\bibitem{Philcox:2026tjj}
O.~H.~E. Philcox, \emph{{Dissecting the Scalar Cosmological Collider with the
  Cosmic Microwave Background}},
  \href{https://arxiv.org/abs/2607.18369}{{\ttfamily 2607.18369}}.

\bibitem{Pinol:2026code}
L.~Pinol, ``{exact-collider: explorer, data and code for exact bispectra in
  strongly mixed multifield inflation}.''
  \href{https://doi.org/10.5281/zenodo.XXXXXXX}{doi:10.5281/zenodo.XXXXXXX},
  2026.

\bibitem{Huenupi:2026abj}
J.~Huenupi, C.~Mu{\~n}oz, G.~A. Palma and S.~Sypsas, \emph{{Pushing the
  Primordial Frontier: Exact Linear Solutions in Multifield Inflation}},
  \href{https://arxiv.org/abs/2606.18248}{{\ttfamily 2606.18248}}.

\bibitem{Pinol:2026xnl}
L.~Pinol, \emph{{New exact bispectrum shapes in multifield inflation}},
  \href{https://arxiv.org/abs/2607.15251}{{\ttfamily 2607.15251}}.

\bibitem{Huenupi:2026aqc}
J.~Huenupi, C.~Mu{\~n}oz, G.~A. Palma and S.~Sypsas, \emph{{Pushing the
  Primordial Frontier: Cosmological Collider Signatures at Strong Mixing}},
  \href{https://arxiv.org/abs/2607.14529}{{\ttfamily 2607.14529}}.

\bibitem{Wang:2026lff}
X.~Wang, Y.~Wang and Y.~Zhao, \emph{{Cosmological Collider Signals at Strong
  Mixing}},  \href{https://arxiv.org/abs/2607.14891}{{\ttfamily 2607.14891}}.

\bibitem{Belrhali:2026uxn}
N.~Belrhali, A.~Poisson and S.~Renaux-Petel, \emph{{Analytical Cosmological
  Collider at Strong Mixing in Laplace Space}},
  \href{https://arxiv.org/abs/2608.23243}{{\ttfamily 2608.23243}}.

\bibitem{Planck:2018vyg}
{Planck Collaboration}, \emph{{Planck 2018 results. VI. Cosmological
  parameters}}, {\emph{Astron. Astrophys.} {\bfseries 641} (2020) A6}
  [\href{https://arxiv.org/abs/1807.06209}{{\ttfamily 1807.06209}}].

\bibitem{Werth:2024mjg}
D.~Werth, \emph{{Spectral representation of cosmological correlators}},
  \href{https://doi.org/10.1007/JHEP12(2024)017}{\emph{JHEP} {\bfseries 12}
  (2024) 017} [\href{https://arxiv.org/abs/2409.02072}{{\ttfamily
  2409.02072}}].

\bibitem{Babich:2004gb}
D.~Babich, P.~Creminelli and M.~Zaldarriaga, \emph{{The Shape of
  non-Gaussianities}},
  \href{https://doi.org/10.1088/1475-7516/2004/08/009}{\emph{JCAP} {\bfseries
  08} (2004) 009} [\href{https://arxiv.org/abs/astro-ph/0405356}{{\ttfamily
  astro-ph/0405356}}].

\bibitem{Jung:2025nss}
G.~Jung, M.~Citran, B.~van Tent, L.~Dumilly and N.~Aghanim, \emph{{Constraints
  on primordial non-Gaussianity from Planck PR4 data}},
  \href{https://doi.org/10.1051/0004-6361/202555283}{\emph{Astron. Astrophys.}
  {\bfseries 702} (2025) A204}
  [\href{https://arxiv.org/abs/2504.00884}{{\ttfamily 2504.00884}}].

\bibitem{PPRW}
O.~H.~E. Philcox, L.~Pinol, D.~Roest and D.~Werth, \emph{{in preparation}}, .

\bibitem{Belrhali:2026ygh}
N.~Belrhali, A.~Poisson and S.~Renaux-Petel, \emph{{Laplace Space for
  Cosmological Correlators}},
  \href{https://arxiv.org/abs/2606.27309}{{\ttfamily 2606.27309}}.

\bibitem{Belrhali:2026jqe}
N.~Belrhali, A.~Poisson and S.~Renaux-Petel, \emph{{Massive Cosmological
  Correlators from Flat Space: a Laplace-Space Approach}},
  \href{https://arxiv.org/abs/2606.27311}{{\ttfamily 2606.27311}}.

\bibitem{DLMF}
``{NIST Digital Library of Mathematical Functions}.''
  \url{https://dlmf.nist.gov/}, Release 1.2.4 of 2025-03-15.

\end{thebibliography}\endgroup
\end{document}